\documentclass[12pt]{report}

\usepackage[linesnumbered,ruled]{algorithm2e}
\usepackage{algpseudocode}
\usepackage{amsfonts}
\usepackage{amsmath}
\usepackage{amssymb}
\usepackage{amsthm}
\usepackage{authblk}
\usepackage[font=footnotesize,labelfont=bf]{caption}
\usepackage{enumitem}
\usepackage{float} % For [H] option in figure
\usepackage{epsfig}
\usepackage{gensymb}
\usepackage[margin=2.5cm]{geometry}
\usepackage{graphicx}
\usepackage[colorlinks=true, allcolors=blue]{hyperref}
\usepackage{latexsym}
\usepackage{lineno}
\usepackage{listings}
\usepackage{mathrsfs}
\usepackage{parskip}
\usepackage{placeins}
\usepackage{relsize}
\usepackage{setspace}
\usepackage{soul}
\usepackage{subcaption}
\usepackage[most]{tcolorbox}
\usepackage{textcomp}
\usepackage[cc]{titlepic} % To add figure on first page
\usepackage{xcolor}

\definecolor{codegreen}{rgb}{0,0.6,0}
\definecolor{codegray}{rgb}{0.5,0.5,0.5}
\definecolor{codepurple}{rgb}{0.58,0,0.82}
\definecolor{backcolour}{rgb}{0.95,0.95,0.92}
\definecolor{backterminal}{rgb}{0.0,0.0,0.0}
\definecolor{codeblue}{rgb}{0.13, 0.67, 0.8}
\definecolor{applegreen}{rgb}{0.24, 0.82, 0.44}
\definecolor{davysgrey}{rgb}{0.21, 0.21, 0.21}

\newcounter{codeCnt}[section]
\lstdefinestyle{PythonStyle}{
    backgroundcolor=\color{backcolour},
    commentstyle=\color{codegreen},
    frame=single,
    keywordstyle=\color{magenta},
    numberstyle=\tiny\color{codegray},
    stringstyle=\color{codepurple},
    basicstyle=\ttfamily\footnotesize,
    breakatwhitespace=false,
    breaklines=true,
    captionpos=b,
    numbers=left,                    
    numbersep=5pt
}
\lstnewenvironment{Code}[1][]{
  \setcounter{lstlisting}{\value{codeCnt}}
  \lstset{caption={#1}, label={#1}, style=PythonStyle, #1}
}{\addtocounter{codeCnt}{1}}

\newcounter{terminalCnt}[chapter]
\lstdefinestyle{TerminalStyle}{
    basicstyle=\linespread{1.1}\color{white}\ttfamily\small,
    captionpos=b,
    frame = none,
    literate={tilde} {{{\color{codeblue} $\sim$}}}1,
    classoffset=0,
    morekeywords={user@mypc},
    keywordstyle={\bf \color{applegreen}},
    breaklines=true
}
\lstnewenvironment{Terminal}[1][]{
  \setcounter{lstlisting}{\value{terminalCnt}}
  \lstset{language=bash, caption={#1}, label={#1}, escapechar=ä,numbers=none, framerule=0.0pt, style=TerminalStyle, #1}
}{\addtocounter{terminalCnt}{1}}

\title{Indoor Geometry Generator (IGG)}

\author[a,*]{Laetitia Mottet}

\affil[a]{Department of Earth Science $\&$ Engineering, Imperial College London, UK}
\affil[*]{laetitia.mottet@gmail.com}

\titlepic{\includegraphics[width=\textwidth]{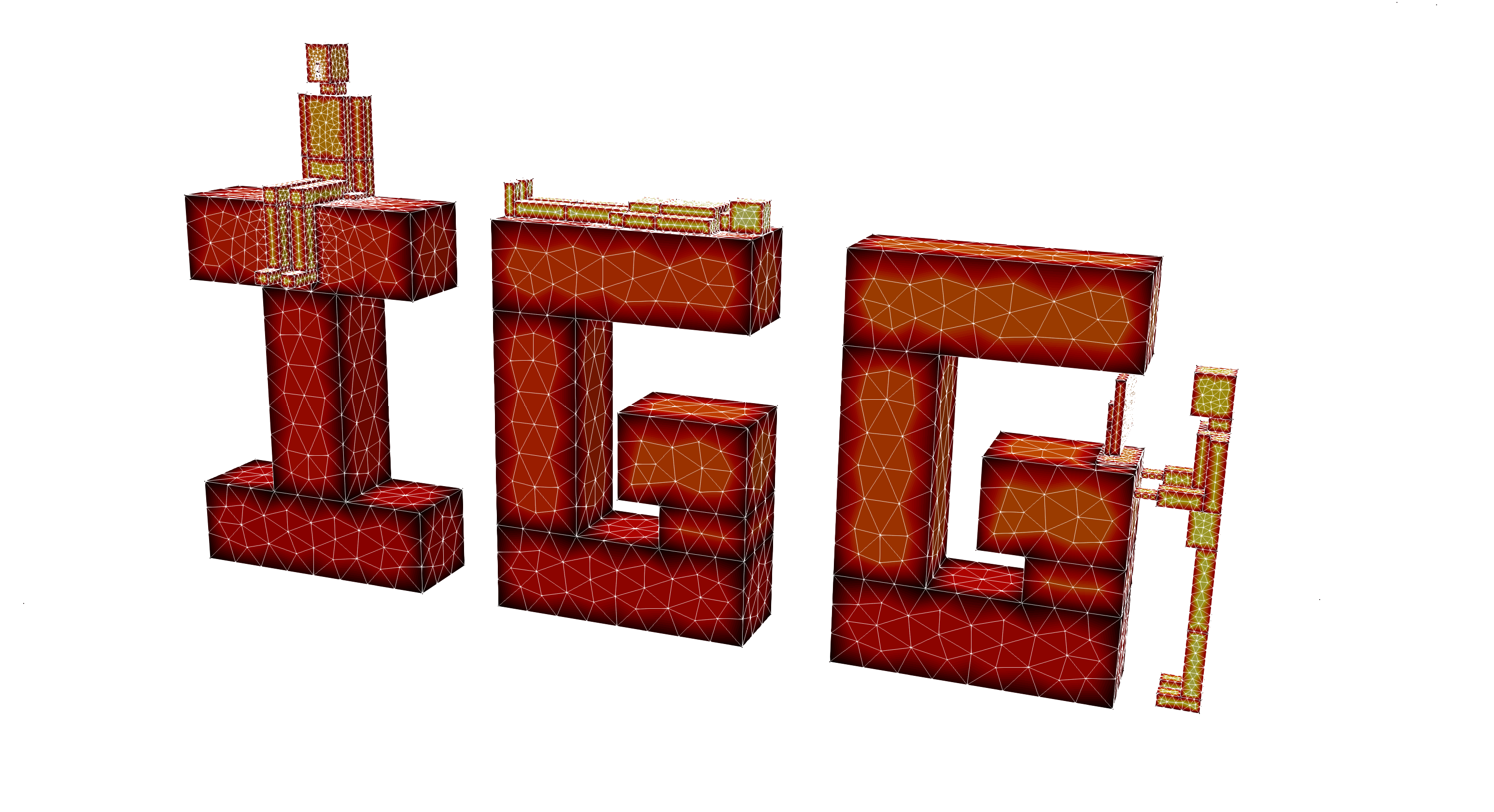}}

\begin{document}

    \begin{sloppypar}
        \maketitle
        
        %------------------------------------%
        %-- ABSTRACT                       --%
        %------------------------------------%
        \begin{abstract}
            %% Text of abstract
            The Indoor Geometry Generator (\textbf{IGG}) is able to generate automatically simplified indoor geometry and its associated unstructured mesh for Computational Fluid Dynamics (CFD) simulations purposes given very simple user inputs. A large number of indoor features are supported by \textbf{IGG} such as (non-exhaustive list): shelves, tills, tables, chairs, seats... Smaller features such as laptops, computer towers and screens can also be included. In addition, doors/windows as well as ventilation inlet and outlet can be taken into account. Finally, the adding of simplified shape of humans standing, sitting or lying are also supported by \textbf{IGG}. 
            \textbf{IGG} allows the user to generate without much effort indoor geometries such that shop, train, bus, plane, school, open spaces..., while being fully consistent with CFD requirements. The geometry and the mesh are outputted in \textbf{GMSH} format supported by both \textbf{Fluidity} and \textbf{IC-FERST} open-source finite-element CFD software.
        \end{abstract}

        %------------------------------------%
        %-- Table of Contents              --%
        %------------------------------------%
        \tableofcontents{}

        %------------------------------------%
        %-- Manuscript                     --%
        %------------------------------------% 
        %----------------------------------------------------------%
%-- Generalities                                         --%
%----------------------------------------------------------%
\chapter{Generalities}\label{Sec:Generalities}

%-----------------------
%-- Brief description
%-----------------------
\section{Brief description}
The Indoor Geometry Generator (\textbf{IGG}) is able to generate automatically simplified indoor geometry and its associated unstructured mesh for Computational Fluid Dynamics (CFD) simulations purposes given very simple user inputs. \textbf{IGG} supports a large number of features allowing the user to generate without much effort indoor geometries such that supermarket, train, bus, plane, school, open spaces..., while being fully consistent with CFD requirements. The geometry and the mesh are outputted in \textbf{GMSH} format supported by both \textbf{Fluidity} and \textbf{IC-FERST} open-source finite-element CFD software~\cite{AMCG2015}. \textbf{IGG} is designed such that different IDs are assigned to different surfaces to allow the user to specify various boundary conditions if wanted such as thermal fluxes for example.

%-----------------------
%-- Dependencies and structure
%-----------------------
\section{Getting and running IGG}

\textbf{IGG} is available on GitHub under request, just send an email to Dr Laetitia Mottet - \url{laetitia.mottet@gmail.com}. Python3, shapely, numpy, descartes, \textbf{Fluidity} (or \textbf{IC-FERST}) and \textbf{GMSH} need to be installed on the machine as detailed in the following sections. \textbf{Fluidity} is (in some cases) optional if the tool \texttt{checkmesh} is not used.

%------------------------
%-- Dependencies
\subsection{External libraries and software}
%-- Python libraries
\textbf{IGG} is written in python v3 - it is therefore required to have python3 install on your machine. In addition, the three following python libraries are needed:
\begin{itemize}
    \item \textbf{numpy} mathematical library
    \item \textbf{shapely} manipulation and analysis of geometric objects
    \item \textbf{descartes} plot using matplotlib of shapely objects (for debug mode in \textbf{IGG})
\end{itemize}

To install them using \texttt{pip}:
\begin{Terminal}[caption={Installation of python libraries used in IGG.}]
ä\colorbox{davysgrey}{
\parbox{435pt}{
\color{applegreen} \textbf{user@mypc}\color{white}\textbf{:}\color{codeblue}$\sim$
\color{white}\$ pip install numpy
\newline
\color{applegreen} \textbf{user@mypc}\color{white}\textbf{:}\color{codeblue}$\sim$
\color{white}\$ pip install shapely
\newline
\color{applegreen} \textbf{user@mypc}\color{white}\textbf{:}\color{codeblue}$\sim$
\color{white}\$ pip install descartes
}}
\end{Terminal}

%-- Fluidity and GMSH libraries
In addition, it is assumed that the CFD software \textbf{Fluidity} as well as the mesh generator \textbf{GMSH} are installed.
\begin{itemize}
    \item \textbf{Fluidity} an open-source finite-element mesh-adaptive CFD software available at \newline \url{https://github.com/FluidityProject/fluidity}
    \item \textbf{GMSH} an open source 3D finite element mesh generator available at \url{gmsh.info/}
\end{itemize}

To install them use:
\begin{Terminal}[caption={Installation of external software used by IGG: Fluidity and GMSH.}]
ä\colorbox{davysgrey}{
\parbox{435pt}{
\color{applegreen} \textbf{user@mypc}\color{white}\textbf{:}\color{codeblue}$\sim$
\color{white}\$ pip install gmsh
\newline
\color{applegreen} \textbf{user@mypc}\color{white}\textbf{:}\color{codeblue}$\sim$
\color{white}\$ sudo apt-add-repository -y ppa:fluidity-core/ppa
\newline
\color{applegreen} \textbf{user@mypc}\color{white}\textbf{:}\color{codeblue}$\sim$
\color{white}\$ sudo apt-get update
\newline
\color{applegreen} \textbf{user@mypc}\color{white}\textbf{:}\color{codeblue}$\sim$
\color{white}\$ sudo apt-get -y install fluidity-dev
}}
\end{Terminal}

\textbf{Remark about GMSH versions:} It is highly recommended to use \textbf{GMSH v3.x} to enjoy all the capabilities of \textbf{IGG}, especially concerning the mesh refinement and more particularly the sphere and box options (see Section~\ref{Sec:Mesh}) which are only supported from \textbf{GMSH} v3.x and higher. \textbf{GMSH} v2.x will work but the mesh refinement constraint specified by the user will be ignored (see Section~\ref{Sec:Mesh} and Section~\ref{Sec:InputFiles_Mesh}). \textbf{GMSH} v4.x will also work (the mesh will be well outputted), however the migration of \textbf{Fluidity} to \textbf{GMSH} v4.x is not completed yet (at the time of the manual is written!), hence it is not recommended to use it neither. 

%-- Test that IGG is working using makefile
To test if all the dependencies are correctly installed, in the folder \texttt{IGG/code/}, the user can run:
\begin{Terminal}[caption={Test if all IGG dependencies are installed.}]
ä\colorbox{davysgrey}{
\parbox{435pt}{
\color{applegreen} \textbf{user@mypc}\color{white}\textbf{:}\color{codeblue}$\sim$/IGG/code
\color{white}\$ make test
}}
\end{Terminal}

%------------------------
%-- Running IGG
\subsection{Running IGG}
%-- Creating an alias is recommended
The main code is \textit{IGG.py} and the functions are in the \texttt{src/} forder. To use \textbf{IGG} from anywhere in your machine without copy/paste the code all the time, it is recommended to add an alias in your \texttt{$\sim$/.bashrc}: 
\begin{Code}[language=Python, firstnumber=1, caption={Alias for IGG to be added into .bashrc user file.}]
alias IGG = "python3 path_of_IGG/IGG/code/IGG.py"
\end{Code}

%-- Running IGG
The following 6 input files need to be in the folder the user wants to work in:
\begin{enumerate}
    \item \textit{IGG\_Numerics.dat}
    \item \textit{IGG\_Domain.dat}
    \item \textit{IGG\_Furniture.dat}
    \item \textit{IGG\_Accessories.dat}
    \item \textit{IGG\_Humans.dat}
    \item \textit{IGG\_Mesh.dat}
\end{enumerate}

Examples of the input files can be found in the folder \texttt{examples/}. In addition, they are detailed in Section~\ref{Sec:InputFiles}.

Finally, in a terminal, where the input files are located - here for example \texttt{MyFolder/} -, assuming the alias has been added into \texttt{$\sim$/.bashrc}, run the command \texttt{IGG}:
\begin{Terminal}[caption={Running IGG.}]
ä\colorbox{davysgrey}{
\parbox{435pt}{
\color{applegreen} \textbf{user@mypc}\color{white}\textbf{:}\color{codeblue}$\sim$/MyFolder
\color{white}\$ IGG
}}
\end{Terminal}

%-----------------------
%-- Geometry
%-----------------------
\section{Geometry and convention}
In this document and in \textbf{IGG} more generally, the word ``\textit{feature}" is used to refer to furniture, accessories, windows, doors, inlet and outlet of ventilation.

%-- Geometry
The features into \textbf{IGG} can be either 2D or 3D:
\begin{itemize}
    \item \textbf{2D features} are planar rectangles (ex: inlet and outlet of ventilation, windows...);
    \item \textbf{3D features} are a compound of one or several rectangular boxes (ex: shelf, chair...).
\end{itemize}

All 3D features are a compound of one or several rectangular boxes, with the exception of ``Bevel" which has one of its surface with an inclination that meets the other surfaces at any angle but 90\degree (see Section~\ref{Sec:Feature_Bevel} for more details). Moreover, humans are 3D feature made by rectangular boxes and one bevel for the nose.

\begin{figure}
    \centering
    \includegraphics[scale=0.8]{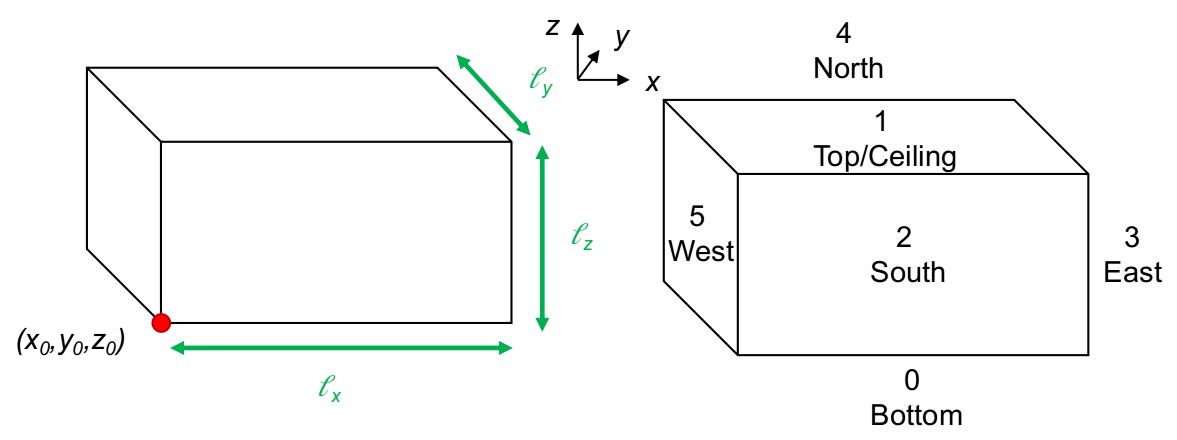}
    \caption{Rectangular box representation in \textbf{IGG}. Size: $(\ell_{x}, \ell_{y}, \ell_{z})$. Reference point: $(x_{0}, y_{0}, z_{0})$.}
    \label{Fig:Generalities_Convention}
\end{figure}

%-- Convention - Naming of surfaces, size and reference points
The 3D feature surfaces are named Bottom (id 0), Top/Ceiling (id 1), South (id 2), East (id 3), North (id 4) and West (id 5) as shown in Figure~\ref{Fig:Generalities_Convention}. Each rectangular box is defined by its size $(\ell_{x}, \ell_{y}, \ell_{z})$ and its reference point $(x_{0}, y_{0}, z_{0})$ as shown in Figure~\ref{Fig:Generalities_Convention}. By convention, the reference point is chosen to be always the \textbf{minimum} $x$, $y$ and $z$ of the feature.

\textbf{Remark:} The reference point of the domain itself is $(0, 0, 0)$. This implies that all the reference points of features are 3-tuples of positive values.

\begin{figure}
    \centering
    \includegraphics[scale=0.55]{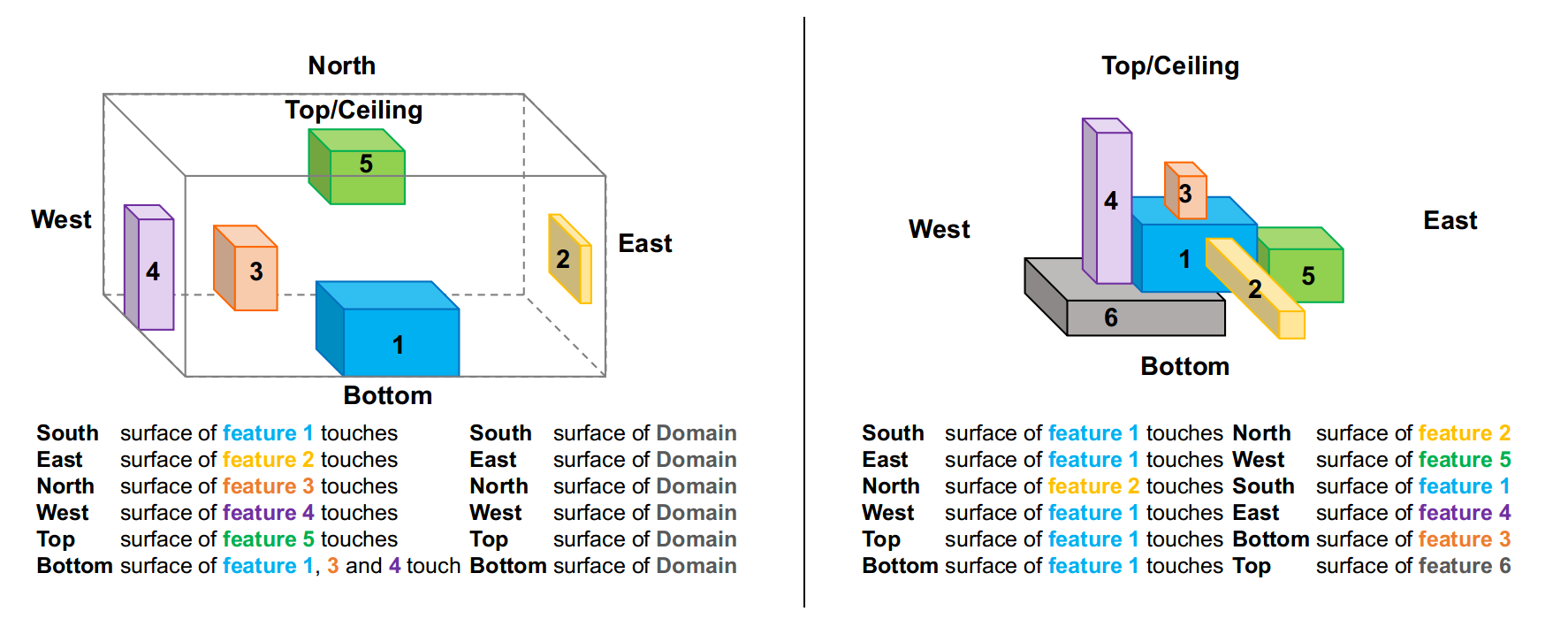}
    \caption{Convention for features touching each others.}
    \label{Fig:Generalities_Surfaces}
\end{figure}

All these conventions imply that features will always touch each others following the same rules as summarised in Figure~\ref{Fig:Generalities_Surfaces}

%-- Figure Convention for rectangular boxes

%-- Figure Rectangular boxes and surfaces

%-----------------------
%-- Capabilities of IGG
%-----------------------
\section{Capabilities of IGG}
\subsection{What IGG can do...}
By combining 2D features, generic rectangular boxes and bevels, furniture, accessories and humans, it is possible to generate very complex and realistic indoor environments: just be imaginative!

The \textbf{2D features} supported by \textbf{IGG} are (Section~\ref{Sec:Feature_2D}):
\begin{itemize}
    \item Inlet of ventilation/heating
    \item Outlet of ventilation/heating
    \item Doors
    \item Windows
\end{itemize}

The \textbf{furniture} (3D features) supported by \textbf{IGG} are (Section~\ref{Sec:Feature_Furniture}):
\begin{itemize}
    \item Box: rectangular box shape
    \item Normal Shelf: rectangular box shape
    \item Refrigerated Shelf: L-shape
    \item Manual Till: rectangular box shape
    \item Automatic Till: L-shape
    \item Chair: complex shape
    \item Seat: L-shape
    \item Stool: complex shape
    \item Table: complex shape
    \item Bed: complex shape
    \item Bevel: rectangular box shape with one inclined surface
\end{itemize}

The \textbf{accessories} (3D features) supported by \textbf{IGG} are (Section~\ref{Sec:Feature_Accessories}):
\begin{itemize}
    \item Screen separator: rectangular shape
    \item Laptop: L-shape
    \item Computer screen: complex shape
    \item Computer tower: rectangular shape
    \item Ceiling diffuser (exception: 2D feature)
    \item Ceiling extractor (exception: 2D feature)
\end{itemize}

In addition, \textbf{humans} can be added into the geometry - see Section~\ref{Sec:Humans} for more details.

Finally, the \textbf{mesh} generated can be refined or coarsen in specific area if required (Section~\ref{Sec:Mesh}).

%-----------------------
%-- Limitations
%-----------------------
\subsection{... and cannot do (limitations)}
The limitations and known issues of \textbf{IGG} are the followings:
\begin{itemize}
    \item The geometry can only be based on rectangular or bevel shapes. Curved surfaces are not supported.
    \item Features need to be aligned with cartesian coordinates: they cannot be rotated, i.e. rotated by an angle $]0,\pi/2[$.
    \item The output of \textbf{IGG} is the geometry and the mesh in \textbf{GMSH} format (\textit{*.geo} and \textit{*.msh}) suitable to run the CFD software \textbf{Fluidity}. \textit{*.stl} format can be also generated if wanted by running \texttt{gmsh -2 MyGeom.geo -o MyGeom.stl} where \texttt{MyGeom.geo} is the output of \textbf{IGG}. Indeed, most of the commercial CFD software support \textit{*.stl} format as mesh input.
    \item If two features have only one edge in common (one single line - features attached by a corner): \textbf{Fluidity} mesh adaptivity capabilities will complain and most of the time crash. Even though this situation is supported and can be generated by \textbf{IGG}, it is recommended to avoid it to avoid future issues with \textbf{Fluidity}.
\end{itemize}

        %----------------------------------------------------------%
%-- Input files                                          --%
%----------------------------------------------------------%
\chapter{Input files}\label{Sec:InputFiles}

The user needs to define the following 6 input files:
\begin{itemize}
    \item Section~\ref{Sec:InputFiles_Numerics}: \textit{IGG\_Numerics.dat}
    \item Section~\ref{Sec:InputFiles_Domain}: \textit{IGG\_Domain.dat}
    \item Section~\ref{Sec:InputFiles_Furniture}: \textit{IGG\_Furniture.dat}
    \item Section~\ref{Sec:InputFiles_Accessories}: \textit{IGG\_Accessories.dat}
    \item Section~\ref{Sec:InputFiles_Humans}: \textit{IGG\_Humans.dat}
    \item Section~\ref{Sec:InputFiles_Mesh}: \textit{IGG\_Mesh.dat}
\end{itemize}

%-----------------------
%-- IGG_Numerics.dat
%-----------------------
\section{IGG\_Numerics.dat: general options}\label{Sec:InputFiles_Numerics}
The input file \textit{IGG\_Numerics.dat} (see Code~\ref{Lst:InputNumerics}) allows the user to specify:
\begin{itemize}
    \item \textbf{Numerical variables} To avoid rounded errors \texttt{Round} and \texttt{Epsilon} are used into \textbf{IGG}. \texttt{Round} is the number of digits kept after rounded a number and \texttt{Epsilon} is the tolerance used when comparing numbers (equality). It is recommended to keep them equal to 7 and $1e-4$, respectively as shown in Code~\ref{Lst:InputNumerics}.
    \item \textbf{File name} is the root name the output files (geometry, mesh...) will have. In the example shown in Code~\ref{Lst:InputNumerics}, the root filename will be \texttt{MyGeom}.
    \item \textbf{GMSH path} The user can specify where \textbf{GMSH} is located on the machine. \texttt{gmsh} will call the version installed by default. In case of several \textbf{GMSH} versions, the absolute path can be specified. It is re-called that \textbf{GMSH} v3.x is recommended.
    \item \textbf{Checkmesh path} The user can specify where the checkmesh tool is located on the machine. \texttt{checkmesh} will use the checkmesh version installed by default on your machine (this should have been installed alongside \textbf{Fluidity}), otherwise its absolute path can be specified.
    \item \textbf{Outputs} The user can choose to output or not the \texttt{Mesh} in \textbf{GMSH} format (\textit{*.msh}) and/or the geometry/mesh in ParaView \texttt{vtk} format (\textit{*.vtk}). \texttt{T} stands for True (they are outputted) and \texttt{F} for False (they are not). \textbf{IGG} always outputs the geometry in \textbf{GMSH} format (\textit{*.geo}). If the mesh is outputted, the ParaView file \textit{*.vtk} will show the 3D mesh, otherwise it will show the 2D mesh.
    \item \textbf{Check mesh consistency} Once the mesh is created, the user can use two tools to check if the mesh is consistent to run CFD simulations. A tool coming alongside \textbf{Fluidity} \texttt{ToolFluidity} - \texttt{checkmesh} - and a tool coming alongside \textbf{GMSH} \texttt{ToolGmsh}. \texttt{T} stands for True (the mesh consistency is checked) and \texttt{F} for False (the mesh consistency is not checked). 
    \item \textbf{Debug} mode can be used if wanted - this can be useful for \textbf{IGG} developers only: if turned on, it is plotting each 2D surfaces of every feature. It is recommended to let it turned off.
\end{itemize}

%-- Example IGG_Numerics.dat
\begin{Code}[language=Python, firstnumber=1, caption={Example of \textit{IGG\_Numerics.dat} input file.}, label={Lst:InputNumerics}]
# NUMERICAL VARIABLES
Round     7
Epsilon   1e-4

# FILE NAME
MyGeom

# GMSH PATH
gmsh

# CHECKMESH PATH (Fluidity tool)
checkmesh

# OUTPUTS
Mesh T
vtk  T

# CHECK MESH TOOL
ToolFluidity T
ToolGmsh     T

# DEBUG MODE
Debug F
\end{Code}

%-----------------------
%-- IGG_Domain.dat
%-----------------------
\section{IGG\_Domain.dat: setup of the domain}\label{Sec:InputFiles_Domain}
The input file \textit{IGG\_Domain.dat} (see Code~\ref{Lst:InputDomain}) allows the user to specify:
\begin{itemize}
    \item \textbf{Domain size} is specified by a 3-tuple $(\ell_{x}, \ell_{y}, \ell_{z})$. In the example Code~\ref{Lst:InputDomain}, the domain is 20 m $\times$ 20 m $\times$ 3 m.
    \item \textbf{Periodicity} can be used in $x-$ (\texttt{PerioX}) and $y-$ (\texttt{PerioY}) direction if wanted. \texttt{T} stands for True (periodicity) and \texttt{F} for False (no periodicity). If \texttt{PerioX} is \texttt{T}, the East and West surface of the domain are the ``same". If \texttt{PerioY} is \texttt{T}, the North and South surface of the domain are the ``same".
    \item \textbf{Inlets, outlets, doors and windows} need to be specified in this order. For each feature type, the number of the feature should be given first: in the example in Code~\ref{Lst:InputDomain}, there are 2 inlets, 1 outlet, 1 door and 2 windows. Then, the location $(x_{0}, y_{0}, z_{0})$ and the size $(\ell_{x}, \ell_{y}, \ell_{z})$ of each feature is given. 
\end{itemize}

\textbf{Remark:} The inlets, outlets, doors and windows are 2D features: this implies that one of their three size is equal to $0$. In Code~\ref{Lst:InputDomain} for example, the size of Inlet 1 is $(0.0, 1.0, 0.5)$: its size along $x$, i.e. $\ell_{x}$, is equal to zero, i.e. Inlet 1 is on a $(yOz)$-plane.

%-- Example IGG_Domain.dat
\begin{Code}[language=Python, firstnumber=1, caption={Example of \textit{IGG\_Domain.dat} input file.}, label={Lst:InputDomain}]
#DOMAIN
20.0 20.0 3.0
PerioX F
PerioY F

#INLETS VENTILATION
2

#Inlet 1
0.0 2.0 0.5
0.0 1.0 0.5

#Inlet 2
0.0 4.0 0.5
0.0 1.0 0.5

#OUTLETS VENTILATION
1

#Outlet 1
3.0 20.0 0.5
1.0  0.0 0.5

#DOORS
1

#Door 1
5.0 0.0 0.0
1.0 0.0 2.0

#WINDOWS
2

#Window 1
20.0 2.0 1.25
0.0  1.0 1.00

#Window 2
8.5 8.5 3.0
1.5 1.5 0.0
\end{Code}

%-----------------------
%-- IGG_Furniture.dat
%-----------------------
\section{IGG\_Furniture.dat: furniture into the domain}\label{Sec:InputFiles_Furniture}
The input file \textit{IGG\_Furniture.dat} (see Code~\ref{Lst:InputFurniture}) allows the user to specify the number, the type, the location and the size of the furniture into the domain.
\begin{itemize}
    \item The \textbf{total number of furniture} into the domain is given at the first line. In Code~\ref{Lst:InputFurniture} for example, there is 10 furniture.
    \item Each \textbf{Furniture} is then defined: the first line specifies the furniture type, then following lines provide the position and the size of the furniture. The furniture type keywords list is: \texttt{Box}, \texttt{NShelf}, \texttt{RShelf}, \texttt{MTill}, \texttt{ATill}, \texttt{Bed}, \texttt{Stool}, \texttt{Table}, \texttt{Seat}, \texttt{Chair}, \texttt{Bevel}. See Section~\ref{Sec:Feature_Furniture} for more details about how to define each furniture type.
\end{itemize}

%-- Example IGG_Furniture.dat
\begin{Code}[language=Python, firstnumber=1, caption={Example of \textit{IGG\_Furniture.dat} input file.}, label={Lst:InputFurniture}]
11

#FURNITURE 1
Chair
1.00 1.00 0.40
0.40 0.40 0.05
1.00 1.00 0.60
0.40 0.05 0.20
0.0

#FURNITURE 2
Seat
5.00 1.00 0.30
0.40 0.40 0.10
5.00 1.00 0.40
0.10 0.40 0.40

#FURNITURE 3
Stool
9.00 1.00 0.60
0.40 0.40 0.05
0.05

#FURNITURE 4
Table
1.00 3.00 0.75
1.50 0.75 0.03
0.05

#FURNITURE 5
Bed
1.00 5.00 0.20
1.90 0.90 0.20
0.10

#FURNITURE 6
NShelf
6.0 7.5 0.0
1.0 2.0 1.5

#FURNITURE 7
RShelf
8.0 5.0 0.0
2.0 0.5 0.5
8.0 5.5 0.0
2.0 0.5 1.5

#FURNITURE 8
MTill
6.0 15.0 0.00
1.0  2.0 0.75

#FURNITURE 9
ATill
9.00 2.00 0.00
0.40 0.40 0.50
9.00 2.00 0.50
0.10 0.40 0.50

#FURNITURE 10
Bevel
Top North
11.0 5.0 0.0
1.0  1.5 1.5
0.5

#FURNITURE 11
Box
15.0 17.5 1.0
1.0   0.5 1.5
\end{Code}

%-----------------------
%-- IGG_Accessories.dat
%-----------------------
\section{IGG\_Accessories.dat: accessories into the domain}\label{Sec:InputFiles_Accessories}
The input file \textit{IGG\_Accessories.dat} (see Code~\ref{Lst:InputAccessories}) allows the user to specify the number, the type, the location and the size of the accessories into the domain.
\begin{itemize}
    \item The \textbf{total number of accessories} into the domain is given at the first line. In Code~\ref{Lst:InputAccessories} for example, there is 4 accessories.
    \item Each \textbf{Accessories} is then defined: the first line specifies the furniture type, then following lines provide the position and the size of the accessory. The accessory keywords list is: \texttt{Tower}, \texttt{Screen}, \texttt{Laptop}, \texttt{Separator}, \texttt{Diffuser}, \texttt{Extractor}. See Section~\ref{Sec:Feature_Accessories} for more details about how to define each accessories type. NB: Note that \texttt{Diffuser} and \texttt{Extractor} are 2D features.
\end{itemize}

%-- Example IGG_Accessories.dat
\begin{Code}[language=Python, firstnumber=1, caption={Example of \textit{IGG\_Accessories.dat} input file.}, label={Lst:InputAccessories}]
4

#ACCESSORY 1
Tower
6.0 3.00 0.0
0.25 0.5 0.5

#ACCESSORY 2
Separator
5.5 4.50 0.7
2.0 0.05 1.0

#ACCESSORY 3
Screen
Centre          4.7 3.2
ScreenSize      32
Feature         18
ScreenDirection South

#ACCESSORY 4
Laptop
Centre          7.7 3.2
ScreenSize      16
Furniture       19
ScreenDirection South

#ACCESSORY 5
Diffuser
Direction  North South East West
Centre     2.0 2.0
Size       0.5 0.5

#ACCESSORY 6
Extractor
Direction  North East
Centre     1.0 2.0
Size       0.5 0.5
\end{Code}

%-----------------------
%-- IGG_Humans.dat
%-----------------------
\section{IGG\_Humans.dat: people into the domain}\label{Sec:InputFiles_Humans}
The input file \textit{IGG\_Humans.dat} (see Code~\ref{Lst:InputHumans}) allows the user to specify the number, the height, the weight profile, the activity, the location... of people into the domain.
\begin{itemize}
    \item The \textbf{total number of humans} is given at the first line. In Code~\ref{Lst:InputHumans} for example, there are 3 humans.
    \item \textbf{Human details} allow the user to specify which part of the human is being represented into the geometry. Humans are divided into 10 parts: \texttt{Head}, \texttt{Neck}, \texttt{Arms}, \texttt{Forearms}, \texttt{Hands}, \texttt{Trunk}, \texttt{Pelvis}, \texttt{Thighs}, \texttt{Legs} and \texttt{Feet}. \texttt{T} stands for True and \texttt{F} for False.
    \item Each \textbf{Human} is then defined. The keywords list defining a human is (no specific order required): \texttt{Activity} (mandatory), \texttt{Height} (optional), \texttt{Weight} (optional), \texttt{Forearm} (optional), \texttt{FaceDirection} (optional), \texttt{Centre} (mandatory if \texttt{Activity} is \texttt{Standing}, optional otherwise), \texttt{Furniture} (mandatory if \texttt{Activity} is \texttt{Sitting} or \texttt{Lying}), \texttt{LyingPosition} (mandatory if \texttt{Activity} is \texttt{Lying}), \texttt{NoseModel} (optional), \texttt{MouthModel} (optional), \texttt{NoseArea} (optional), \texttt{MouthArea} (optional), \texttt{NoseAngle} (optional), \texttt{HeatingPart} (optional). See Section~\ref{Sec:Humans} for more details about how to define humans.
\end{itemize}

%-- Example IGG_Humans.dat
\begin{Code}[language=Python, firstnumber=1, caption={Example of \textit{IGG\_Humans.dat} file.}, label={Lst:InputHumans}]
3

#Human Details
Head     T
Neck     T
Arms     T
Forearms T
Hands    T
Trunk    T
Pelvis   T
Thighs   T
Legs     T
Feet     T

#HUMAN 1
Activity      Standing   
Forearm       Down
Height        170
Weight        Normal
NoseModel     T
FaceDirection West
HeatingPart   Head Hands Legs Forearms
Centre        8 5 0

#HUMAN 2
Activity      Sitting   
Furniture     5
Forearm       Up
Weight        Thin
MouthModel    T

#HUMAN 3
Activity      Lying   
Furniture     22
FaceDirection North
LyingPosition West
\end{Code}

%-----------------------
%-- IGG_Mesh.dat
%-----------------------
\section{IGG\_Mesh.dat: mesh characteristics}\label{Sec:InputFiles_Mesh}
The input file \textit{IGG\_Mesh.dat} (see Code~\ref{Lst:InputMesh}) allows the user to specify the mesh characteristics.
\begin{itemize}
    \item \textbf{General mesh sizes} are firstly defined. \texttt{NMinElement} is the minimum number of elements on one surface. \texttt{MinLength} and \texttt{MaxLength} are the minimum and maximum edge length of the mesh, respectively.
    \item \textbf{Constraints} can be added to the mesh. First, the number of constraints is defined, e.g. 7 constraints are specified in Code~\ref{Lst:InputMesh}; then the constraints are specified. See Section~\ref{Sec:Mesh} for more details.
    \begin{itemize}
        \item Feature-specific constraint: the feature type to refine is specified, then the edge length wanted given.
        \item Non-feature-specific constraint: the user can refined/coarsen the mesh in some areas if wanted. The area to be refined/coarsen can be defined by a box or a sphere.
    \end{itemize}
\end{itemize}

%-- Example IGG_Mesh.dat
\begin{Code}[language=Python, firstnumber=1, caption={Example of \textit{IGG\_Mesh.dat} file.}, label={Lst:InputMesh}]
# GENERAL VARIABLES
NMinElement 2
MinLength   0.001
MaxLength   2.0

# CONSTRAINTS
7

#CONSTRAINT 1
Inlet 1
0.1

#CONSTRAINT 2
Domain Top
0.8

#CONSTRAINT 3
Furniture 1
Legs
0.1

#CONSTRAINT 4
Accessory 1
All
0.3

#CONSTRAINT 5
Human 1
Hand
0.01

#CONSTRAINT 6
Sphere
16.0 1.0 1.5
0.5
0.05

#CONSTRAINT 7
Box
16.0 3.0 1.5
1.5  1.0 0.5
0.05
\end{Code}
        %----------------------------------------------------------%
%-- Output files                                          --%
%----------------------------------------------------------%
\chapter{Output files}\label{Sec:OutputFiles}

Assuming that the \texttt{File Name} defined in \textit{IGG\_Numerics.dat} is \texttt{MyGeom}, the output files generated are:
\begin{itemize}
    \item Section~\ref{Sec:OutputFiles_geo}: \textit{MyGeom.geo}
    \item Section~\ref{Sec:OutputFiles_msh}: \textit{MyGeom.msh}
    \item Section~\ref{Sec:OutputFiles_vtk}: \textit{MyGeom.vtk}
    \item Section~\ref{Sec:OutputFiles_humans}: \textit{MyGeom\_HumansStates.dat}
    \item Section~\ref{Sec:OutputFiles_physicalIDs}: \textit{MyGeom\_PhysicalIDs.dat}
\end{itemize}

%-----------------------
%-- MyGeom.geo
%-----------------------
\section{*.geo file: GMSH geometry}\label{Sec:OutputFiles_geo}
\begin{figure}
    \centering
    \includegraphics[width=0.5\textwidth]{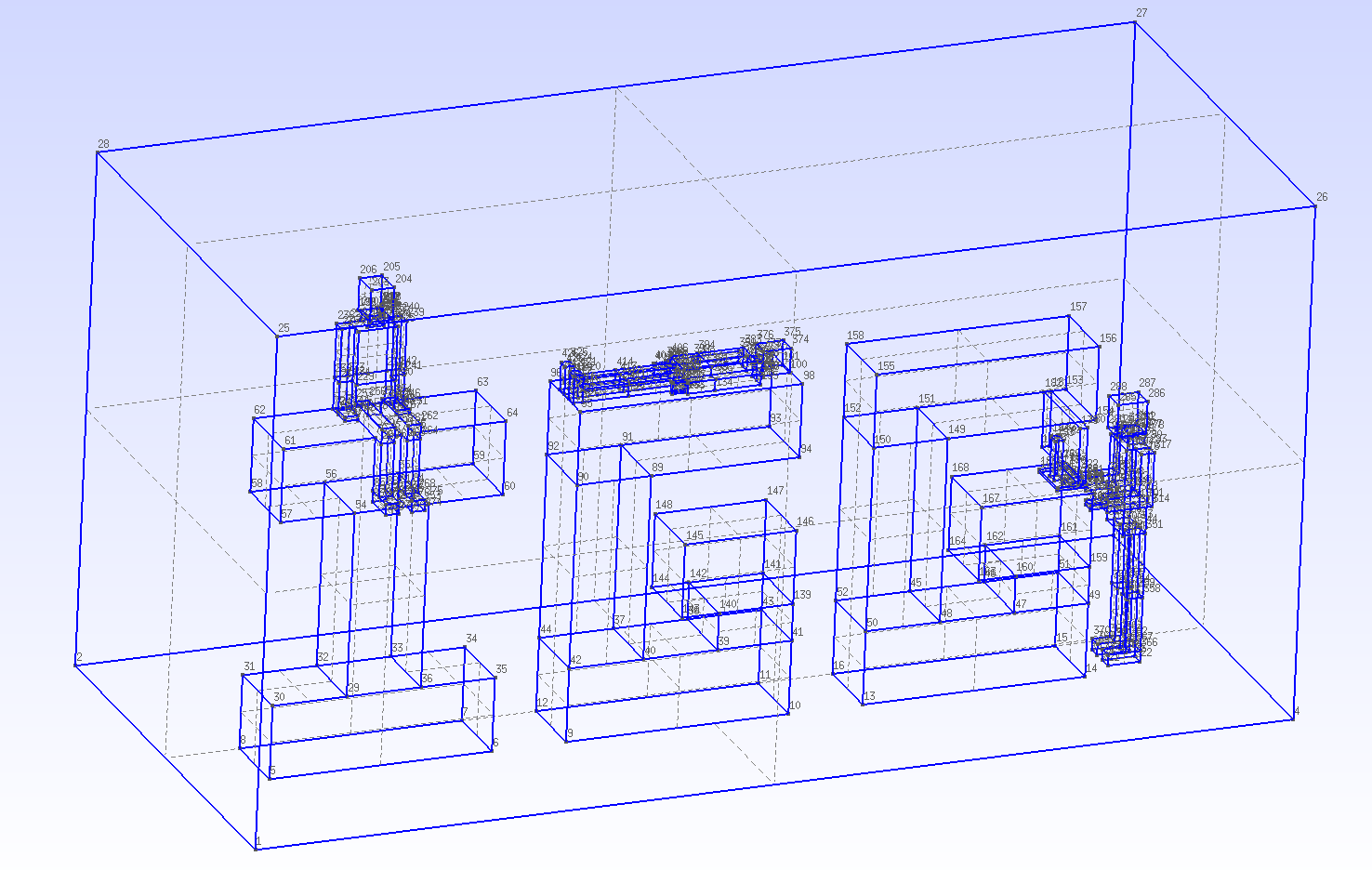}
    \caption{Visualising the geometry \textit{MyGeom.geo} in \textbf{GMSH}. Points IDs are also shown in this example.}
    \label{Fig:OutputFile_VisualiseGeo}
\end{figure}

Assuming that the \texttt{File Name} defined in \textit{IGG\_Numerics.dat} is \texttt{MyGeom}, the geometry in \textbf{GMSH} format outputted is \textit{MyGeom.geo} and can be visualised  using the graphical interface of \textbf{GMSH} by running Command~\ref{Lst:OutputFile_VisualiseGeo} in a terminal. The IDs of points, lines, surfaces... can be seen on the graphical interface of \textbf{GMSH} for convenience. In \textbf{GMSH}, under \texttt{Tools/Options/Geometry/Visibility/}, the \texttt{Point labels}, \texttt{Line labels} and \texttt{Surface labels} can be easily displayed. An example of \textit{*.geo} file visualisation in \textbf{GMSH} is shown in Figure~\ref{Fig:OutputFile_VisualiseGeo}.

%-- CODE: To visualise the geometry
\begin{Terminal}[caption={Visualising the geometry \textit{MyGeom.geo} in \textbf{GMSH}.}, label=Lst:OutputFile_VisualiseGeo]
    ä\colorbox{davysgrey}{
    \parbox{435pt}{
    \color{applegreen} \textbf{user@mypc}\color{white}\textbf{:}\color{codeblue}$\sim$
    \color{white}\$ gmsh MyGeom.geo
}}
\end{Terminal}

%-- Figure - example of MyGeom.geo

%-----------------------
%-- MyGeom.msh
%-----------------------
\section{*.msh file: GMSH mesh}\label{Sec:OutputFiles_msh}
\begin{figure}
    \centering
    \includegraphics[width=0.5\textwidth]{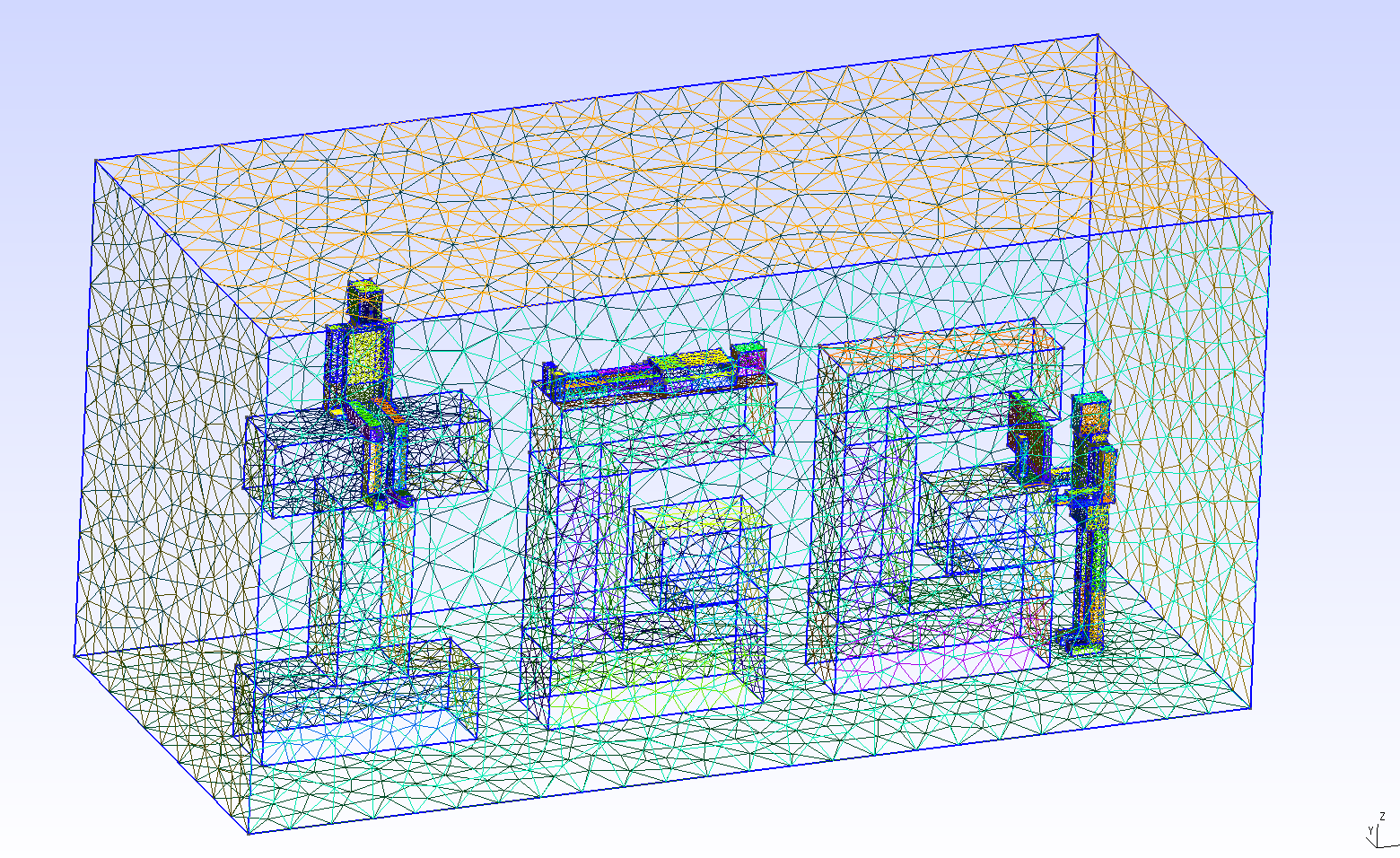}
    \caption{Visualising the mesh \textit{MyGeom.msh} in \textbf{GMSH}.}
    \label{Fig:OutputFile_VisualiseMsh}
\end{figure}

The mesh is outputted if the option \texttt{Mesh} in \textit{IGG\_Numerics.dat} is True \texttt{T}. Assuming that the \texttt{File Name} defined in \textit{IGG\_Numerics.dat} is \texttt{MyGeom}, the mesh in \textbf{GMSH} format outputted is \textit{MyGeom.msh} and can be visualised using the graphical interface of \textbf{GMSH} by running Command~\ref{Lst:OutputFile_VisualiseMsh} in a terminal. An example of \textit{*.msh} file visualisation in \textbf{GMSH} is shown in Figure~\ref{Fig:OutputFile_VisualiseMsh}.

%-- CODE: To visualise the mesh
\begin{Terminal}[caption={Visualising the mesh \textit{MyGeom.msh} in \textbf{GMSH}.}, label=Lst:OutputFile_VisualiseMsh]
    ä\colorbox{davysgrey}{
    \parbox{435pt}{
    \color{applegreen} \textbf{user@mypc}\color{white}\textbf{:}\color{codeblue}$\sim$
    \color{white}\$ gmsh MyGeom.msh
}}
\end{Terminal}

%-- Figure - example of MyGeom.msh

%-----------------------
%-- MyGeom.vtk
%-----------------------
\section{*vtk: ParaView file}\label{Sec:OutputFiles_vtk}
The \textbf{ParaView} file is outputted if the option \texttt{vtk} in \textit{IGG\_Numerics.dat} is True \texttt{T}. If \texttt{Mesh} is True \texttt{T}, the user will be able to see the 3D mesh into the \textit{ParaView} file, if \texttt{Mesh} is False \texttt{F}, the user will be able to see the 2D mesh only into the \textit{vtk} file.

\begin{figure}
    \centering
    \includegraphics[width=\textwidth]{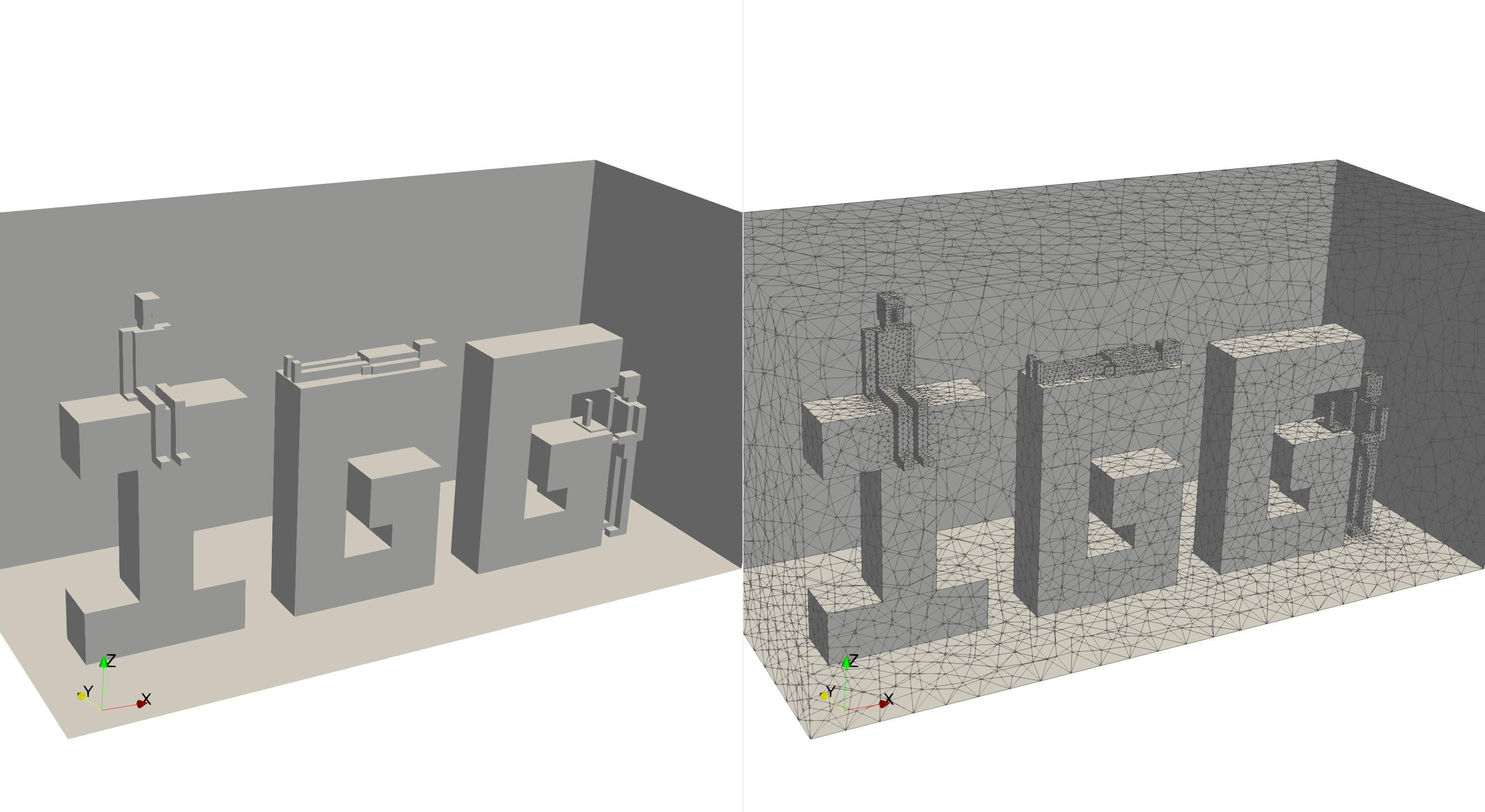}
    \caption{Visualising the geometry/mesh \textit{MyGeom.vtk} in \textbf{ParaView}.}
    \label{Fig:OutputFile_VisualiseVtk}
\end{figure}

Assuming that the \texttt{File Name} defined in \textit{IGG\_Numerics.dat} is \texttt{MyGeom}, the geometry/mesh in \textbf{ParaView} format outputted is \textit{MyGeom.vtk} and can be visualised  using the graphical interface of ParaView by running Command~\ref{Lst:OutputFile_VisualiseVtk} in a terminal. An example of \textit{*.vtk} file visualisation in \textbf{ParaView} is shown in Figure~\ref{Fig:OutputFile_VisualiseVtk}

%-- CODE: To visualise paraview file
\begin{Terminal}[caption={Visualising the geometry/mesh \textit{MyGeom.vtk} in \textbf{ParaView}.}, label=Lst:OutputFile_VisualiseVtk]
    ä\colorbox{davysgrey}{
    \parbox{435pt}{
    \color{applegreen} \textbf{user@mypc}\color{white}\textbf{:}\color{codeblue}$\sim$
    \color{white}\$ paraview MyGeom.vtk
}}
\end{Terminal}

%-- Figure - example of MyGeom.msh

%-----------------------
%-- MyGeom_HumansStates.dat
%-----------------------
\section{*\_HumansStates.dat: summary of humans state}\label{Sec:OutputFiles_humans}
Assuming that the \texttt{File Name} defined in \textit{IGG\_Numerics.dat} is \texttt{MyGeom}, a summary of the human state is given in \textit{MyGeom\_HumansStates.dat} and can be open with a text editor. It consists in 3 part:
\begin{itemize}
    \item At the beginning of the file, it is re-called how the physical IDs of humans are defined. This is also explained in this manual in Section~\ref{Sec:PhysicalIDsBC_IDHuman}.
    \item As the user can let \textbf{IGG} uses the default values for, amongst other, the \texttt{Height}, the \texttt{Weight}..., \textit{MyGeom\_HumansStates.dat} summarises the values and characteristics used for each human.
    \item The area of each body part are summarised at the end of \textit{MyGeom\_HumansStates.dat}.
\end{itemize}

%-----------------------
%-- MyGeom_PhysicalIDs.dat
%-----------------------
\section{*\_PhysicalIDs.dat: summary of the physical IDs}\label{Sec:OutputFiles_physicalIDs}
Assuming that \texttt{File Name} defined in \textit{IGG\_Numerics.dat} is \texttt{MyGeom}, \textit{MyGeom\_PhysicalIDs.dat} summarises the Physical IDs assigned to surfaces. These Physical IDs are needed by \textbf{Fluidity} to apply boundary conditions. In the first part of the file, the Physical IDs by feature types (furniture, accessories, humans...) are summarised. In a second part, an attempt is done to group the Physical IDs by boundary conditions types, i.e. walls, heating, inlet, outlet, heat generation... The Physical IDs are grouped by common boundary conditions used in CFD - but the user can decide to use different boundary conditions than the ones suggested. More details about the Physical IDs are given in Section~\ref{Sec:PhysicalIDsBC}.
        %----------------------------------------------------------%
%-- Description of features                              --%
%----------------------------------------------------------%
\chapter{Features supported by IGG}\label{Sec:Feature}
Features are divided into 2D features defined in \textit{IGG\_Domain.dat}, furniture defined in \textit{IGG\_Furniture.dat} and accessories defined in \textit{IGG\_Accessories.dat}. Each feature types and how to define them in their respective input files are described in this chapter.

%%%%%%%%%%%%%%%%%%%%%%%%%%%%%%%%%%%
%-- 2D features
%%%%%%%%%%%%%%%%%%%%%%%%%%%%%%%%%%%
\section{2D features: inlets, outlets, doors and windows}\label{Sec:Feature_2D}
In \textit{IGG\_Domain.dat}, the inlets, outlets, doors and windows can be added, and need to be defined in this specific order: inlets, outlets, doors then windows. These 4 features are 2D and are defined by their location $(x_{0}, y_{0}, z_{0})$ and then their size $(\ell_{x}, \ell_{y}, \ell_{z})$. Code~\ref{Lst:InputDomain_2Dfeatures} shows how to define these features. For each feature type, the following need to be defined:
\begin{itemize}
    \item First the number of feature is given;
    \item Then each feature of this type is defined by giving in this order:
    \begin{itemize}
        \item its location $(x_{0}, y_{0}, z_{0})$ and
        \item its size $(\ell_{x}, \ell_{y}, \ell_{z})$. 
    \end{itemize}
\end{itemize}

%-- Example IGG_Domain.dat
\begin{Code}[language=Python, firstnumber=1, caption={How to define inlets, outlets, doors and windows in \textit{IGG\_Domain.dat}.}, label={Lst:InputDomain_2Dfeatures}]
#INLETS VENTILATION
Nbr_Inlets

#Inlet 1
x0 y0 z0
lx ly lz

...

#Inlet Nbr_Inlets
x0 y0 z0
lx ly lz

#OUTLETS VENTILATION
Nbr_Outlets

#Outlet 1
x0 y0 z0
lx ly lz

...

#Outlet Nbr_Outlets
x0 y0 z0
lx ly lz

#DOORS
Nbr_Doors

#Door 1
x0 y0 z0
lx ly lz

...

#Door Nbr_Doors
x0 y0 z0
lx ly lz

#WINDOWS
Nbr_Windows

#Window 1
x0 y0 z0
lx ly lz

...

#Window Nbr_Windows
x0 y0 z0
lx ly lz
\end{Code}

\textbf{Remark 1:} By convention, the location, i.e. the reference point, $(x_{0}, y_{0}, z_{0})$ is chosen to be always the \textbf{minimum} $x$, $y$ and $z$ of the feature.

\textbf{Remark 2:} The inlets, outlets, doors and windows are 2D features: this implies that one of their three sizes $(\ell_{x}, \ell_{y}, \ell_{z})$ is null:
\begin{itemize}
    \item If $\ell_{x}$ is equal to zero, the 2D feature is defined on a $(yOz)$-plane; 
    \item if $\ell_{y}$ is equal to zero, the 2D feature is defined on a $(xOz)$-plane; 
    \item if $\ell_{z}$ is equal to zero, the 2D feature is defined on a $(xOy)$-plane. 
\end{itemize}

\textbf{Remark 3:} If one or several of these 2D features are not intended to be in the geometry, set the number of the feature type equal to 0 as shown in Code~\ref{Lst:InputDomain_No2Dfeatures}.

\begin{Code}[language=Python, firstnumber=1, caption={Example of \textit{IGG\_Domain.dat} if the geometry does not include any 2D feature.}, label={Lst:InputDomain_No2Dfeatures}]
#INLETS VENTILATION
0

#OUTLETS VENTILATION
0

#DOORS
0

#WINDOWS
0
\end{Code}

\textbf{Remark 4:} These 2D features can be located on any surfaces of the domain, furniture or accessories; even humans (!) if wanted.

The physical IDs of these 2D features have the form \textbf{xyy}, where:
\begin{itemize}
    \item \textbf{x} is the integer:
    \begin{itemize}
        \item \textbf{1} for an inlet
        \item \textbf{2} for an outlet
        \item \textbf{3} for a door
        \item \textbf{4} for a window
    \end{itemize}
    \item \textbf{yy} is an integer referring to the feature number from \textbf{01} to \textbf{99} as defined in \textit{IGG\_Domain.dat}.
\end{itemize}

\textbf{Example:} the physical ID \textbf{102} refers to the inlet number 2 defined in \textit{IGG\_Domain.dat}; the physical ID \textbf{410} refers to the window number 10 defined in \textit{IGG\_Domain.dat}.

\textbf{Ceiling diffuser or extractor}

Note that it exists two other 2D feature defined under \textit{IGG\_Accessories.dat}: \texttt{Diffuser} and \texttt{Extractor}. A diffuser/extractor is a 2D feature located at ceiling height (only for now!). The user needs to defined its centre $(x_{C}, y_{C})$ as well as its size $(\ell_{x}, \ell_{y})$ as shown in Code~\ref{Lst:InputDomain_DiffuserExtractor}. The diffuser/extractor is divided into sub-shaped depending the in/out air direction desired, i.e. inlet/outlet of the flow towards \texttt{North}, \texttt{South}, \texttt{East} and/or \texttt{West}. This allows the user to generate 1-way, 2-way, 3-way or 4-way diffusers/extractors. 

%-- Code - Diffuser Extractor
\begin{Code}[language=Python, firstnumber=1, caption={Example of \texttt{Diffuser}.}, label={Lst:InputDomain_DiffuserExtractor}]
#ACCESSORY 1
Diffuser
Direction  North South East West
Centre     xC yC
Size       lx ly

#ACCESSORY 2
Extractor
Direction  North South East West
Centre     xC yC
Size       lx ly
\end{Code}

The physical IDs defining the diffuser are:
\begin{itemize}
    \item \textbf{631} for the surface where the air is released towards the \texttt{South}
    \item \textbf{632} for the surface where the air is released towards the \texttt{East}
    \item \textbf{633} for the surface where the air is released towards the \texttt{North}
    \item \textbf{634} for the surface where the air is released towards the \texttt{West}
\end{itemize}

The physical IDs defining the extractor are:
\begin{itemize}
    \item \textbf{641} for the surface where the air is extracted towards the \texttt{South}
    \item \textbf{642} for the surface where the air is extracted towards the \texttt{East}
    \item \textbf{643} for the surface where the air is extracted towards the \texttt{North}
    \item \textbf{644} for the surface where the air is extracted towards the \texttt{West}
\end{itemize}

\begin{figure}
    \centering
    \includegraphics[width=\textwidth]{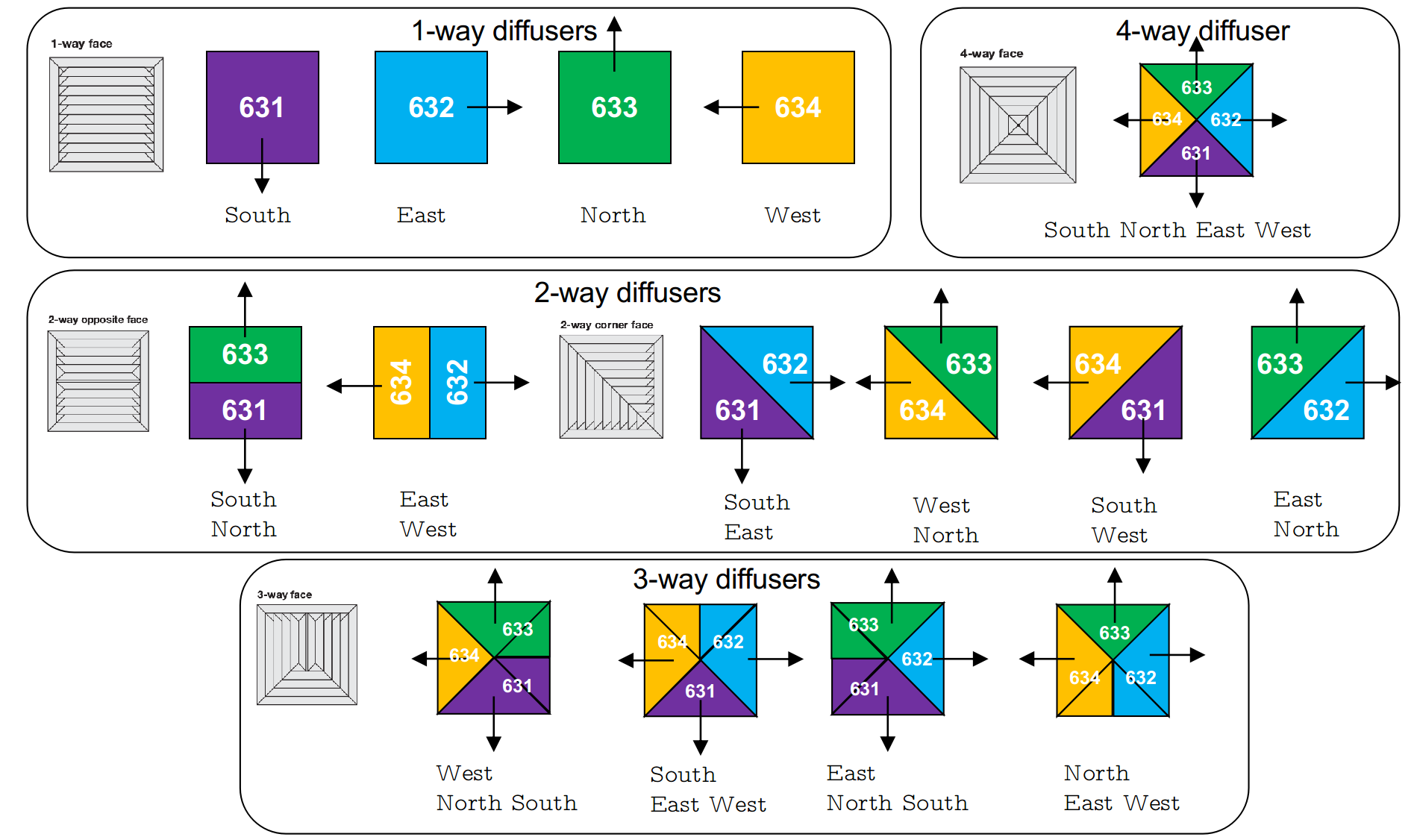}
    \caption{Different shape of ceiling \texttt{Diffuser}. \textbf{631}, \textbf{632}, \textbf{633} and \textbf{634} are the physical IDs of the \texttt{Diffuser} surfaces. \texttt{North}, \texttt{South}, \texttt{East} and/or \texttt{West} are the keywords to be used in \textit{IGG\_Accessories.dat}. Note that ceiling \texttt{Extractor} is represented the same way, excepted it behaves as an outlet.}
    \label{Fig:Diffuser_Notation}
\end{figure}

Schematic of the diffusers as represented in \textbf{IGG} are shown in Figure~\ref{Fig:Diffuser_Notation}. Extractor are represented as the same, excepted the air is extracted, i.e. going out of the domain.

%-- Figure - Diffuser
%%%%%%%%%%%%%%%%%%%%%%%%%%%%%%%%%%%
%-- Furniture
%%%%%%%%%%%%%%%%%%%%%%%%%%%%%%%%%%%
\section{Furniture}\label{Sec:Feature_Furniture}
The furniture, which are defined by the user in \textit{IGG\_Furniture.dat}, supported by \textbf{IGG} are summarised in Table~\ref{Tab:Feature_Furniture}.

\begin{table}
    \centering
    \begin{tabular}{ |p{3.4cm}||p{3.75cm}|p{1.9cm}|p{4.0cm}|p{1.5cm}|  }
         \hline
         \textbf{Furniture type} & \textbf{Shape} & \textbf{Keyword} & \textbf{Physical ID} & \textbf{Section}  \\\hline
         Box                & Rectangular Box      & \texttt{Box}    & 40 & \ref{Sec:Feature_Box} \\\hline
         Normal Shelf       & Rectangular box      & \texttt{NShelf} & 30  & \ref{Sec:Feature_NShelf} \\\hline
         Refrigerated Shelf & L-shape              & \texttt{RShelf} & 311 cold surfaces \newline 312 otherwise & \ref{Sec:Feature_RShelf} \\\hline
         Manual Till        & Rectangular box      & \texttt{MTill}  & 32  & \ref{Sec:Feature_MTill} \\\hline
         Automatic Till     & L-shape              & \texttt{ATill}  & 33  & \ref{Sec:Feature_ATill} \\\hline
         Table              & 5 rectangular boxes  & \texttt{Table}  & 341 top surface \newline 342 otherwise & \ref{Sec:Feature_Table} \\\hline
         Chair              & 8 rectangular boxes  & \texttt{Chair}  & 351 in-contact w/ people surfaces \newline 352 otherwise & \ref{Sec:Feature_Chair} \\\hline
         Seat               & L-shape              & \texttt{Seat}   & 361 in-contact w/ people surfaces \newline 362 otherwise & \ref{Sec:Feature_Seat} \\\hline
         Stool              & 5 rectangular boxes  & \texttt{Stool}  & 371 top surface \newline 372 otherwise & \ref{Sec:Feature_Stool} \\\hline
         Bed                & 11 rectangular boxes & \texttt{Bed}    & 381 mattress top surface \newline 382 otherwise & \ref{Sec:Feature_Bed} \\\hline
         Bevel              & Bevel                & \texttt{Bevel}  & 39x: x=bevel ID & \ref{Sec:Feature_Bevel} \\\hline

    \end{tabular}
    \caption{\label{Tab:Feature_Furniture}Summary of the furniture supported by \textbf{IGG}. }
\end{table}

%--------------------------------------------------
%-----------------------
%-- Generic box
%-----------------------
\subsection{Box}\label{Sec:Feature_Box}
\begin{figure}
    \centering
    \includegraphics[width=0.5\textwidth]{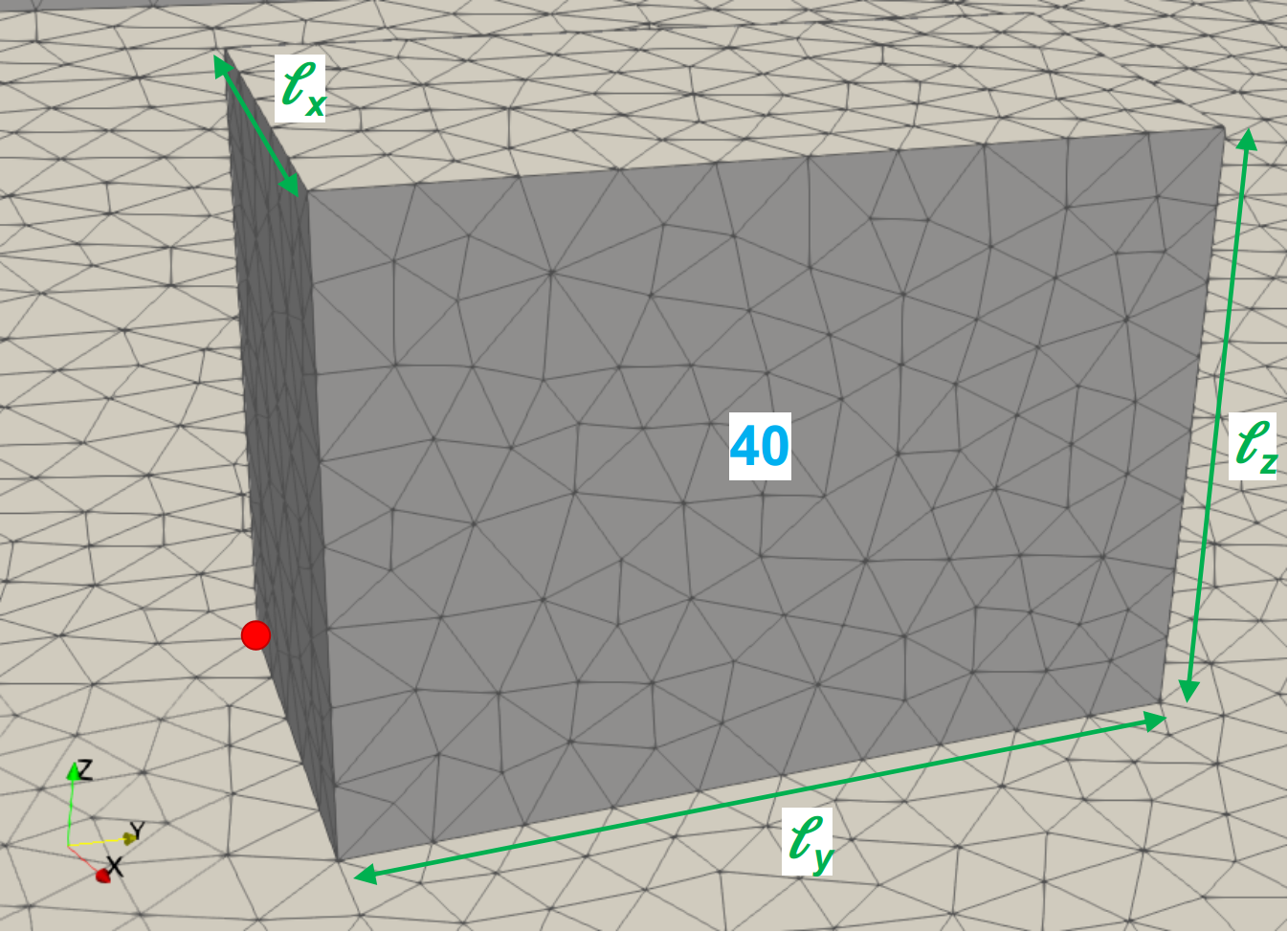}
    \caption{Visualisation of a generic box \texttt{Box} in \textbf{ParaView}. \textbf{40} is the physical ID of boxes.}
    \label{Fig:Box}
\end{figure}

A generic box (keyword: \texttt{Box}) is a rectangular box object. A box is defined, in this order, by:
\begin{itemize}
    \item its keyword \texttt{Box};
    \item its location $(x_{0}, y_{0}, z_{0})$ and
    \item its size $(\ell_{x}, \ell_{y}, \ell_{z})$,
\end{itemize}

as shown in Code~\ref{Lst:InputFurniture_Box} and in Figure~\ref{Fig:Box}. The physical surface ID of every box is \textbf{40}.

%-- Code - Box
\begin{Code}[language=Python, firstnumber=1, caption={How to define a generic box in \textit{IGG\_Furniture.dat}.}, label={Lst:InputFurniture_Box}]
# FURNITURE 1
Box
x0 y0 z0
lx ly lz
\end{Code}

%-- Figure - Box

%-----------------------
%-- Normal Shelf
%-----------------------
\subsection{Normal Shelf}\label{Sec:Feature_NShelf}
\begin{figure}
    \centering
    \includegraphics[width=0.5\textwidth]{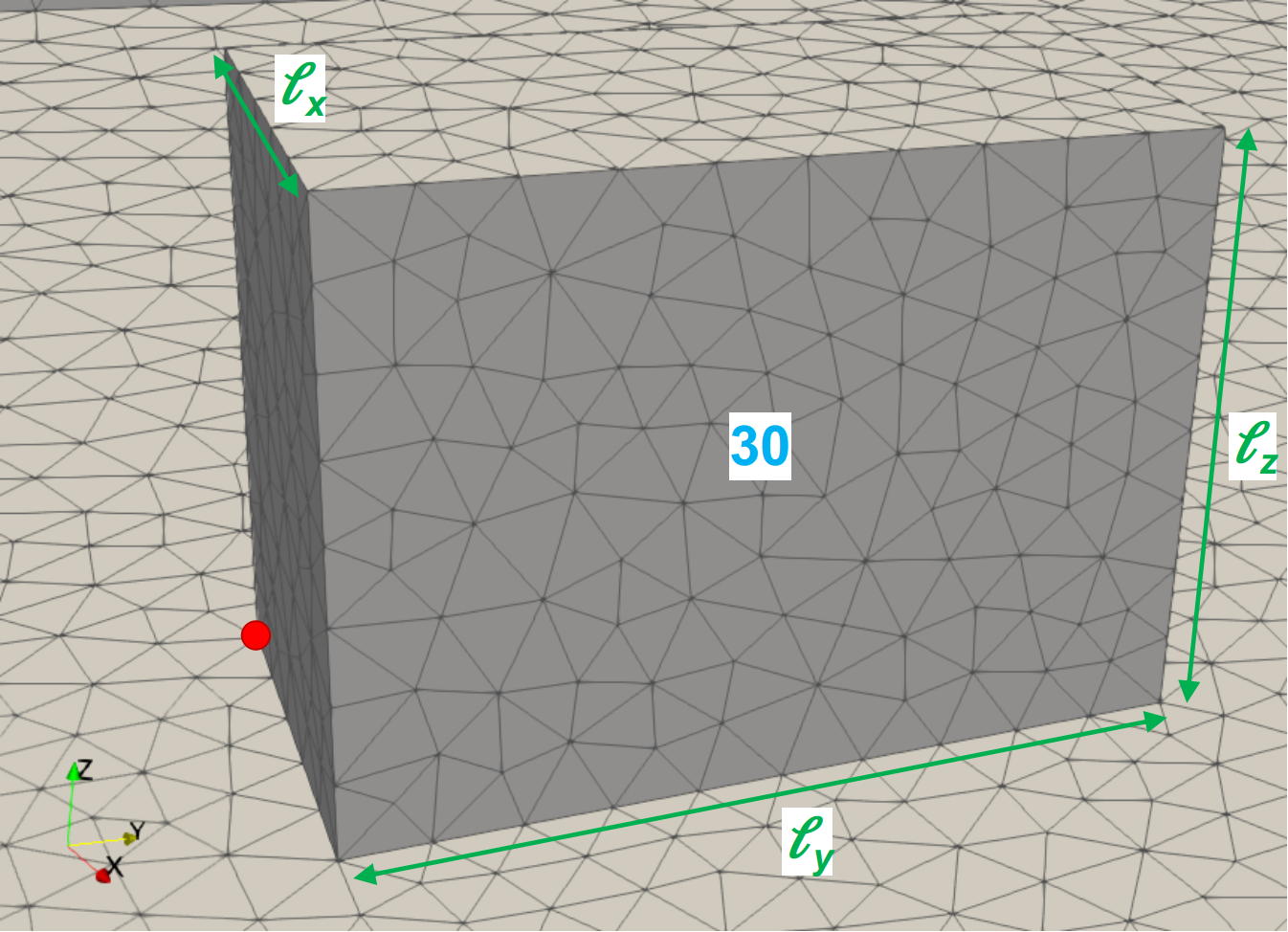}
    \caption{Visualisation of a normal shelf \texttt{NShelf} in \textbf{ParaView}. \textbf{30} is the physical ID of normal shelves.}
    \label{Fig:NShelf}
\end{figure}

A normal shelf (keyword: \texttt{NShelf}) is a rectangular box object. A normal shelf is defined, in this order, by:
\begin{itemize}
    \item its keyword \texttt{NShelf};
    \item its location $(x_{0}, y_{0}, z_{0})$ and
    \item its size $(\ell_{x}, \ell_{y}, \ell_{z})$,
\end{itemize}

as shown in Code~\ref{Lst:InputFurniture_NShelf} and in Figure~\ref{Fig:NShelf}. The physical surface ID of every normal shelf is \textbf{30}.

%-- Code - NShelf
\begin{Code}[language=Python, firstnumber=1, caption={How to define normal shelf in \textit{IGG\_Furniture.dat}.}, label={Lst:InputFurniture_NShelf}]
# FURNITURE 1
NShelf
x0 y0 z0
lx ly lz
\end{Code}

%-- Figure - NShelf

%--------------------------------------------------
%-----------------------
%-- Refrigerated Shelf
%-----------------------
\subsection{Refrigerated shelf}\label{Sec:Feature_RShelf}
\begin{figure}
    \centering
    \begin{subfigure}[t]{\textwidth}
        \includegraphics[width=\textwidth]{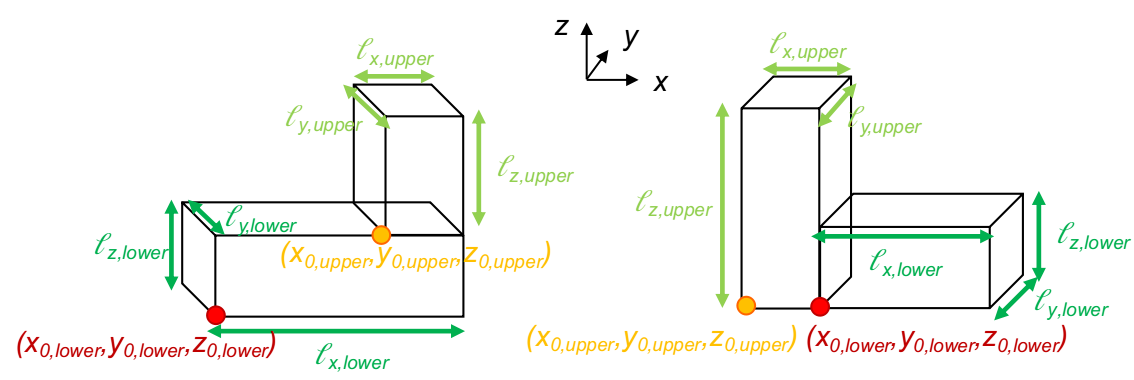}
        \caption{}
        \label{Fig:RShelf_Notation}
    \end{subfigure}
    \begin{subfigure}[t]{0.5\textwidth}
        \includegraphics[width=\textwidth]{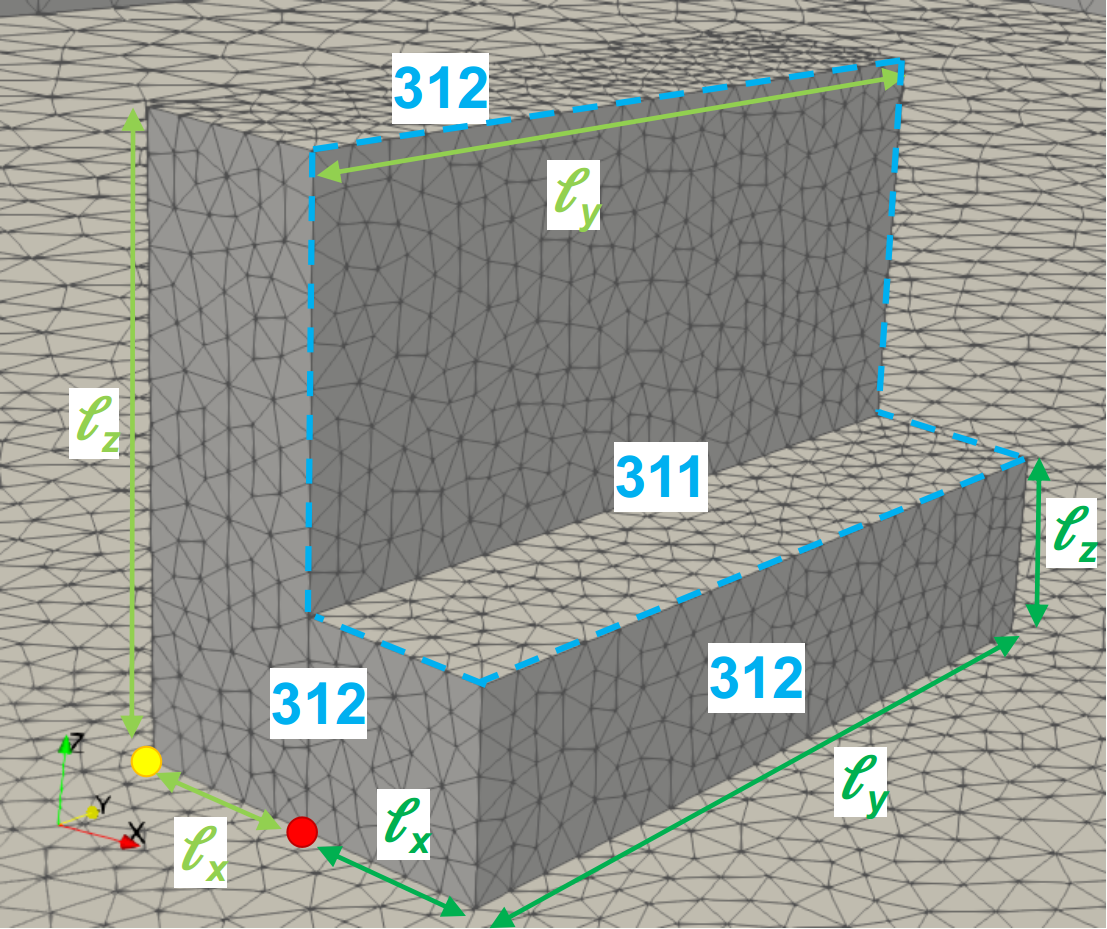}
        \caption{}
        \label{Fig:RShelf_ParaView}
    \end{subfigure}
    \caption{(a) Different way of representing L-shape in \textbf{IGG} and (b) visualisation of a refrigerated shelf \texttt{RShelf} in \textbf{ParaView}. \textbf{311} is the physical ID of ``cold" surfaces and \textbf{312} is the physical ID of the other surfaces.}
    \label{Fig:RShelf}
\end{figure}

A refrigerated shelf (keyword: \texttt{RShelf}) is a L-shape object: in other words, refrigerated shelf is a compound of 2 rectangular boxes. A refrigerated shelf is defined, in this order, by:
\begin{itemize}
    \item its keyword \texttt{RShelf};
    \item its lower part location $(x_{0,lower}, y_{0,lower}, z_{0,lower})$,
    \item its lower part size $(\ell_{x,lower}, \ell_{y,lower}, \ell_{z,lower})$,
    \item its upper part location $(x_{0,upper}, y_{0,upper}, z_{0,upper})$ and
    \item its upper part size $(\ell_{x,upper}, \ell_{y,upper}, \ell_{z,upper})$,
\end{itemize}

as shown in Code~\ref{Lst:InputFurniture_RShelf} and in Figure~\ref{Fig:RShelf}. It exists two ways to represent refrigerated shelf, and more generally L-shape object, as shown in Figure~\ref{Fig:RShelf_Notation}: both are supported by \textbf{IGG}, and the user can use one or the other. The physical surface IDs of every refrigerated shelf is \textbf{31x}, where \textbf{x} is equal to \textbf{1} for the ``cold" surfaces or \textbf{2} otherwise. A refrigerated shelf has 2 ``cold" surfaces being the top surface of the lower part and the adjacent lateral surface of the upper part as shown in Figure~\ref{Fig:RShelf_ParaView}.

%-- Code - RShelf
\begin{Code}[language=Python, firstnumber=1, caption={How to define refrigerated shelf in \textit{IGG\_Furniture.dat}.}, label={Lst:InputFurniture_RShelf}]
# FURNITURE 1
RShelf
x0_lower y0_lower z0_lower
lx_lower ly_lower lz_lower
x0_upper y0_upper z0_upper
lx_upper ly_upper lz_upper
\end{Code}

% Figure -- RShelf

%--------------------------------------------------
%-----------------------
%-- Manual Till
%-----------------------
\subsection{Manual Till}\label{Sec:Feature_MTill}
\begin{figure}
    \centering
    \includegraphics[width=0.5\textwidth]{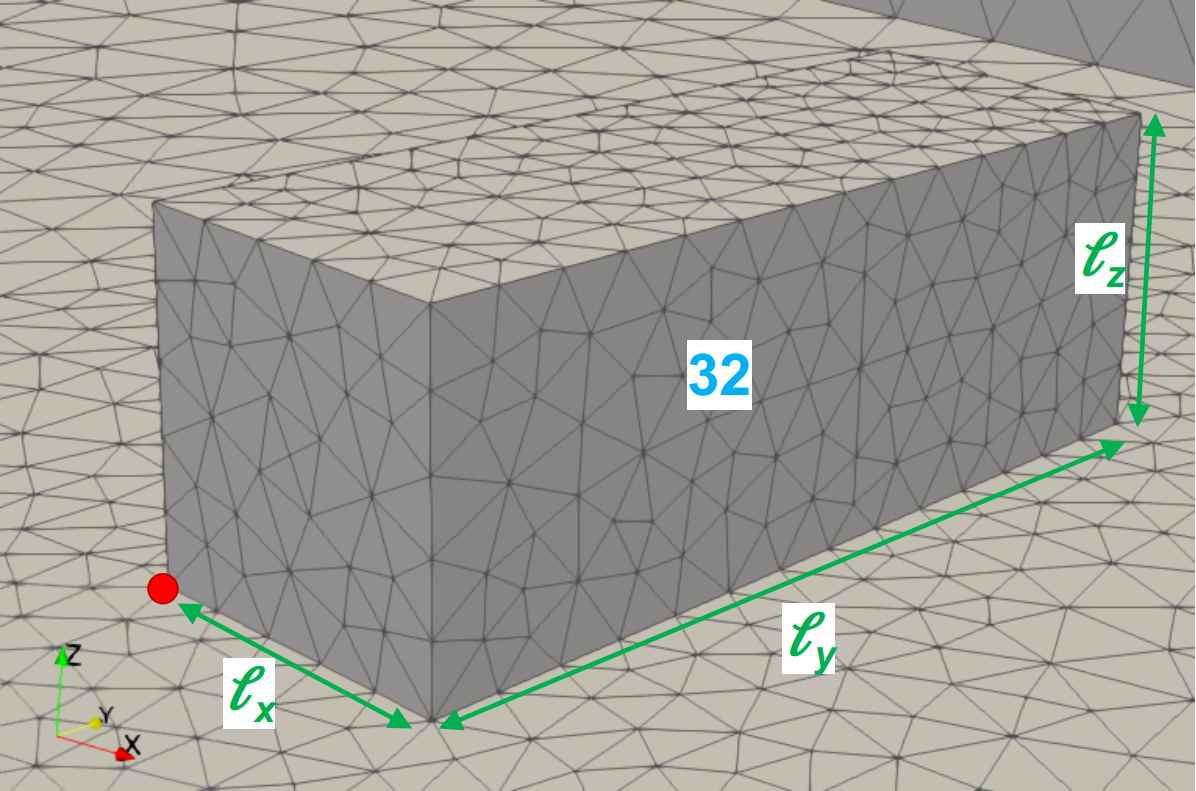}
    \caption{Visualisation of a manual till \texttt{MTill} in \textbf{ParaView}. \textbf{32} is the physical ID of manual tills.}
    \label{Fig:MTill}
\end{figure}

A manual till (keyword: \texttt{MTill}) is a rectangular box object. A manual till is defined, in this order, by:
\begin{itemize}
    \item its keyword \texttt{MTill};
    \item its location $(x_{0}, y_{0}, z_{0})$ and
    \item its size $(\ell_{x}, \ell_{y}, \ell_{z})$,
\end{itemize}

as shown in Code~\ref{Lst:InputFurniture_MTill} and in Figure~\ref{Fig:MTill}. The physical surface ID of every  manual till is \textbf{32}.

%-- Code - MTill
\begin{Code}[language=Python, firstnumber=1, caption={How to define manual till in \textit{IGG\_Furniture.dat}.}, label={Lst:InputFurniture_MTill}]
# FURNITURE 1
MTill
x0 y0 z0
lx ly lz
\end{Code}

%-- Figure - MTill

%--------------------------------------------------
%-----------------------
%-- Automatic Till
%-----------------------
\subsection{Automatic Till}\label{Sec:Feature_ATill}
\begin{figure}
    \centering
    \includegraphics[width=0.4\textwidth]{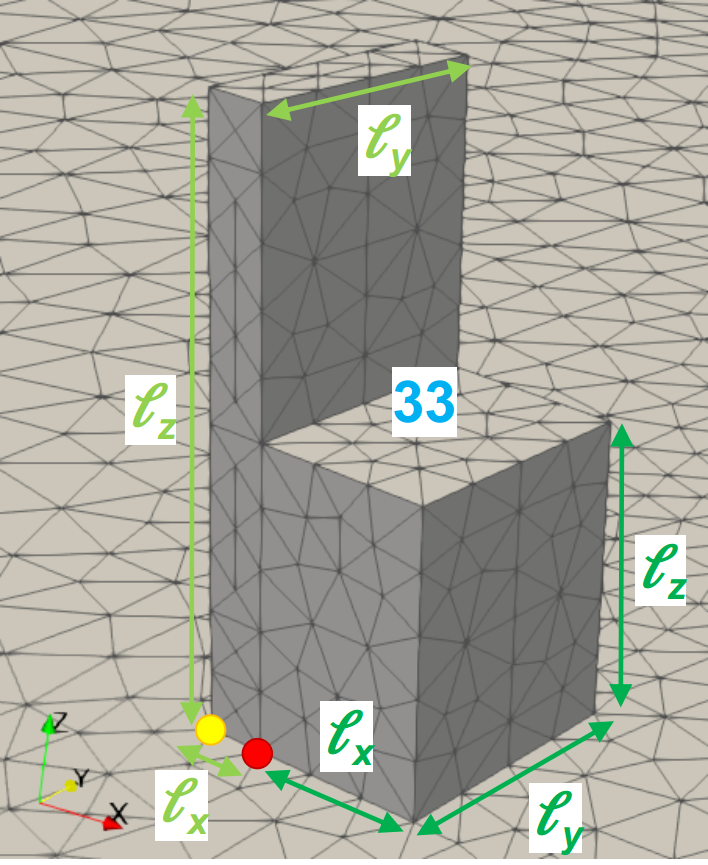}
    \caption{Visualisation of an automatic till \texttt{ATill} in \textbf{ParaView}. \textbf{33} is the physical ID of automatic tills.}
    \label{Fig:ATill}
\end{figure}

An automatic till (keyword: \texttt{ATill}) is a L-shape object: in other words, automatic till is a compound of 2 rectangular boxes. A automatic till is defined, in this order, by:
\begin{itemize}
    \item its keyword \texttt{MTill};
    \item its lower part location $(x_{0,lower}, y_{0,lower}, z_{0,lower})$,
    \item its lower part size $(\ell_{x,lower}, \ell_{y,lower}, \ell_{z,lower})$,
    \item its upper part location $(x_{0,upper}, y_{0,upper}, z_{0,upper})$ and
    \item its upper part size $(\ell_{x,upper}, \ell_{y,upper}, \ell_{z,upper})$,
\end{itemize}

as shown in Code~\ref{Lst:InputFurniture_ATill} and in Figure~\ref{Fig:ATill}. It exists two ways to represent automatic till, and more generally L-shape object, as shown in Figure~\ref{Fig:RShelf_Notation}: both are supported by \textbf{IGG}, and the user can use one or the other. The physical surface ID of every automatic till is \textbf{33}.

\textbf{Remark:} It is intended to implement a generic L-shape object in future version of \textbf{IGG}. For the time being, automatic till can be used as a generic L-shape object.

%-- Code - ATill
\begin{Code}[language=Python, firstnumber=1, caption={How to define automatic till in \textit{IGG\_Furniture.dat}.}, label={Lst:InputFurniture_ATill}]
# FURNITURE 1
ATill
x0_lower y0_lower z0_lower
lx_lower ly_lower lz_lower
x0_upper y0_upper z0_upper
lx_upper ly_upper lz_upper
\end{Code}

% Figure -- ATill

%--------------------------------------------------
%-----------------------
%-- Table
%-----------------------
\subsection{Table}\label{Sec:Feature_Table}
\begin{figure}
    \centering
    \begin{subfigure}[t]{0.45\textwidth}
        \includegraphics[width=\textwidth]{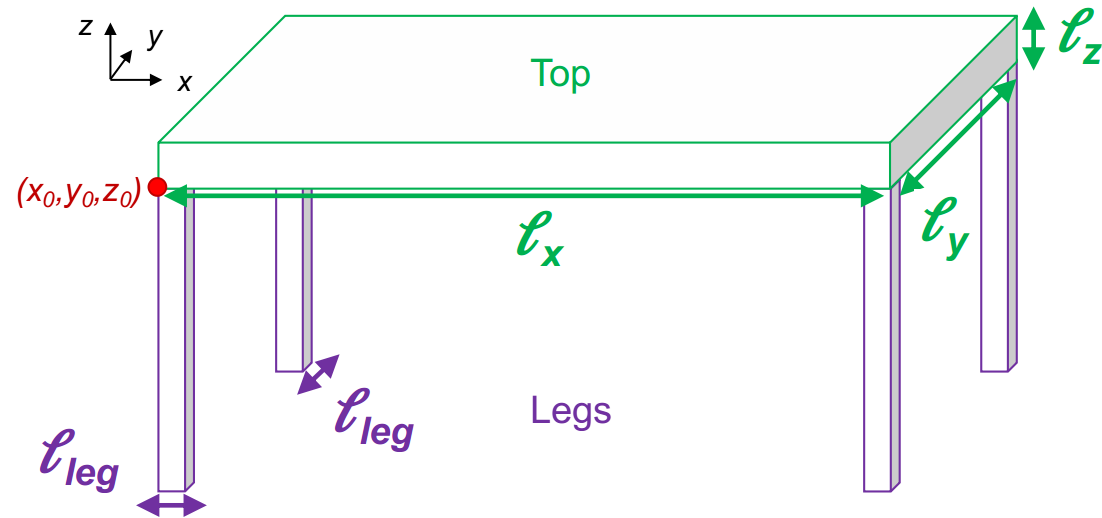}
        \caption{}
        \label{Fig:Table_Notation}
    \end{subfigure}
    \begin{subfigure}[t]{0.4\textwidth}
        \includegraphics[width=\textwidth]{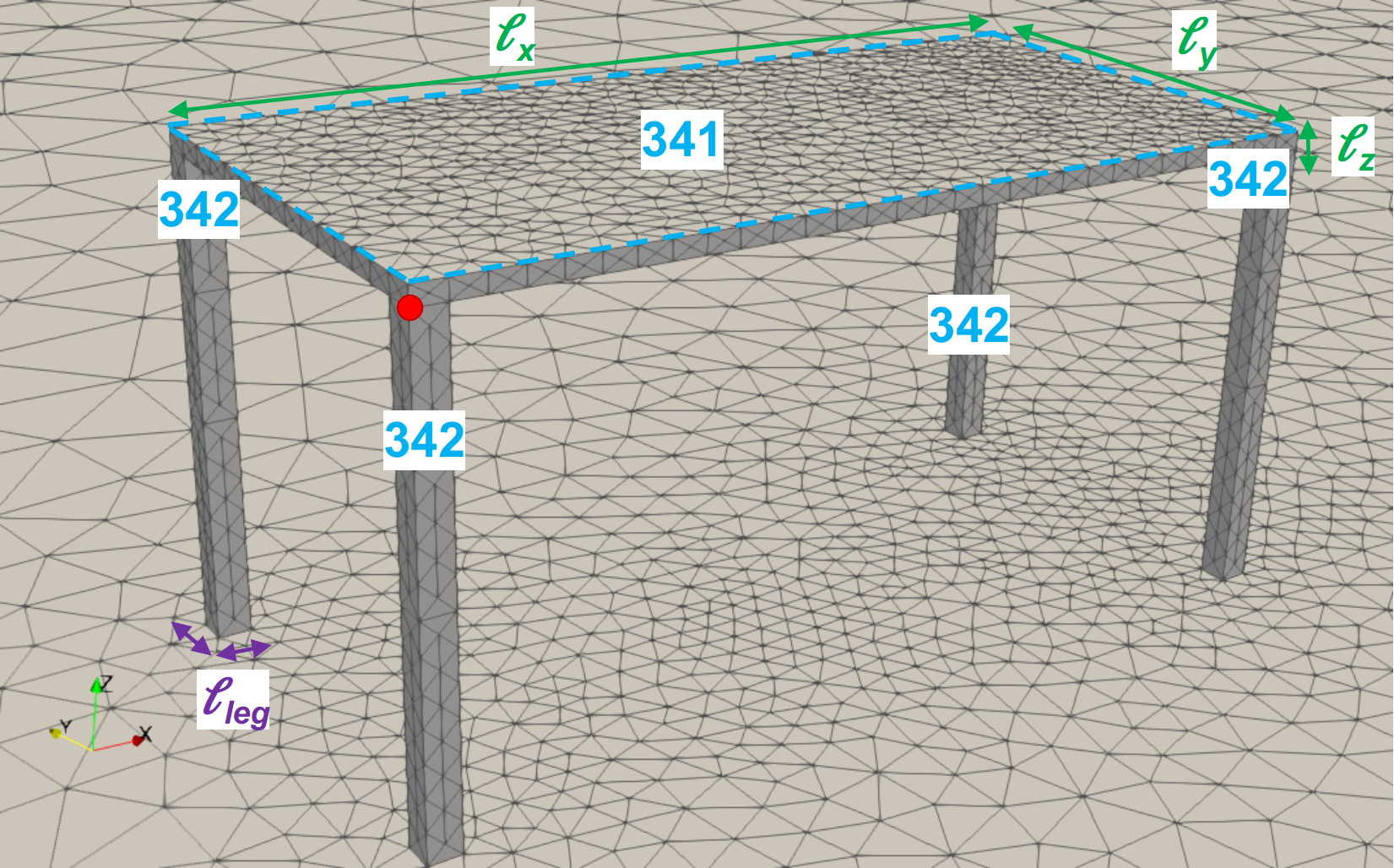}
        \caption{}
        \label{Fig:Table_ParaView}
    \end{subfigure}
    \caption{(a) Representation of a table in \textbf{IGG} and (b) visualisation of a table \texttt{Table} in \textbf{ParaView}. \textbf{341} is the physical ID of the top surface of the top part and \textbf{342} is the physical ID of the other surfaces.}
    \label{Fig:Table}
\end{figure}

A table (keyword: \texttt{Table}) is a compound of 5 rectangular boxes: 4 legs and 1 top as shown in Figure~\ref{Fig:Table_Notation}. 

\textbf{Remarks about the table geometry:} The 4 legs are assumed to have a square section having the same size $\ell_{leg} \times \ell_{leg}$ as shown in Figure~\ref{Fig:Table_Notation}. The legs are set at the four corner of the top part. 

A table is defined, in this order, by:
\begin{itemize}
    \item its keyword \texttt{Table};
    \item its top location $(x_{0}, y_{0}, z_{0})$,
    \item its top size $(\ell_{x}, \ell_{y}, \ell_{z})$ and
    \item its legs size $\ell_{leg}$,
\end{itemize}

as shown in Code~\ref{Lst:InputFurniture_Table} and in Figure~\ref{Fig:Table}. The physical surface IDs of every table is \textbf{34x}, where \textbf{x} is equal to \textbf{1} for the top surface of the top part or \textbf{2} otherwise as shown in Figure~\ref{Fig:Table_ParaView}.

%-- Code - Table
\begin{Code}[language=Python, firstnumber=1, caption={How to define a table in \textit{IGG\_Furniture.dat}.}, label={Lst:InputFurniture_Table}]
# FURNITURE 1
Table
x0 y0 z0
lx ly lz
l_leg
\end{Code}

% Figure -- Table

%--------------------------------------------------
%-----------------------
%-- Chair
%-----------------------
\subsection{Chair}\label{Sec:Feature_Chair}
\begin{figure}
    \centering
    \begin{subfigure}[t]{0.48\textwidth}
        \includegraphics[width=\textwidth]{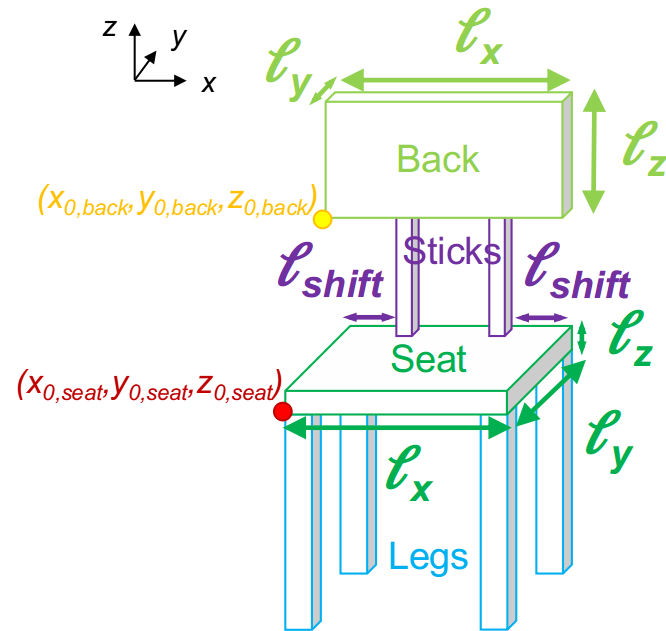}
        \caption{}
        \label{Fig:Chair_Notation}
    \end{subfigure}
    \begin{subfigure}[t]{0.35\textwidth}
        \includegraphics[width=\textwidth]{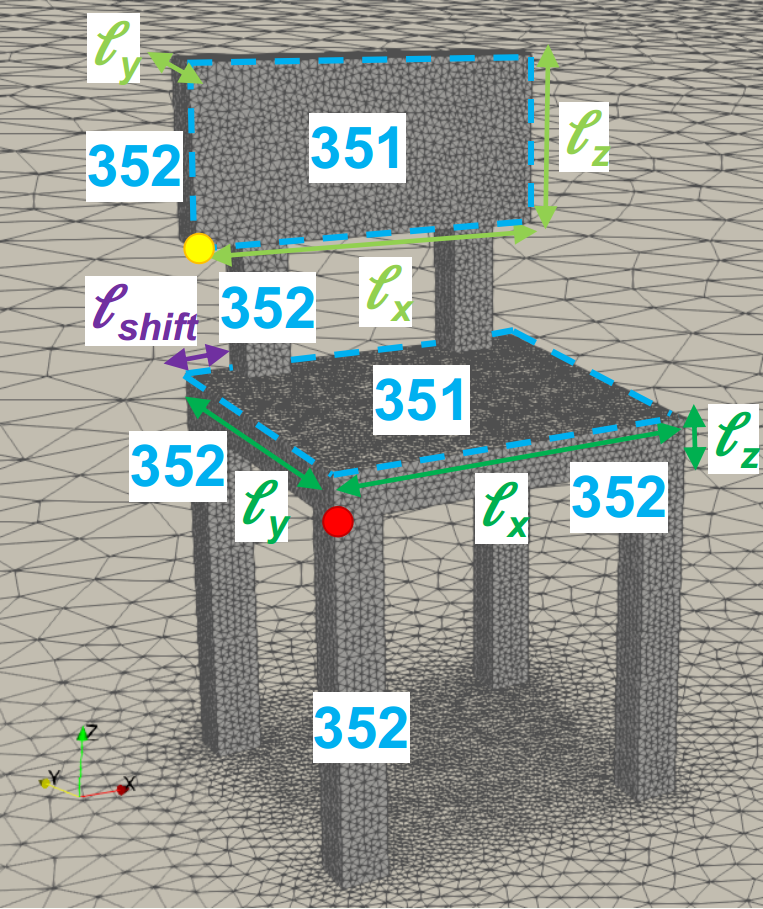}
        \caption{}
        \label{Fig:Chair_ParaView}
    \end{subfigure}
    \caption{(a) Representation of a chair in \textbf{IGG} and (b) visualisation of a chair \texttt{Chair} in \textbf{ParaView}. \textbf{351} is the physical ID of ``in-contact with people" surfaces and \textbf{352} is the physical ID of the other surfaces.}
    \label{Fig:Chair}
\end{figure}

A chair (keyword: \texttt{Chair}) is a compound of 8 rectangular boxes: 4 legs, 1 seat, 2 back sticks and 1 back as shown in Figure~\ref{Fig:Chair_Notation}. 

\textbf{Remarks about the chair geometry:} The 4 legs and the 2 back sticks are assumed to have a square section having the same size than the thickness of the seat and the back, respectively. The user does not have the control of the legs and stick dimensions and shape - they are determined automatically in \textbf{IGG}. The legs are set at the four corner of the seat part. The back size can be smaller or bigger than the seat size. In addition, the user has the possibility to have a shift in the location of the back sticks relative to the seat (see $\ell_{shift}$ in Figure~\ref{Fig:Chair_Notation}). It is believed that, for now, all these assumptions are fine as we are not intended to model Ikea catalogue! 

A chair is defined, in this order, by:
\begin{itemize}
    \item its keyword \texttt{Chair};
    \item its seat location $(x_{0,seat}, y_{0,seat}, z_{0,seat})$,
    \item its seat size $(\ell_{x,seat}, \ell_{y,seat}, \ell_{z,seat})$,
    \item its back location $(x_{0,back}, y_{0,back}, z_{0,back})$,
    \item its back size $(\ell_{x,back}, \ell_{y,back}, \ell_{z,back})$ and
    \item its length by which the back sticks are shifted compared to the seat $\ell_{shift}$,
\end{itemize}

as shown in Code~\ref{Lst:InputFurniture_Chair} and in Figure~\ref{Fig:Chair}. The physical surface IDs of every chair is \textbf{35x}, where \textbf{x} is equal to \textbf{1} for the surfaces in contact with potential people sitting on it or \textbf{2} otherwise. A chair has 2 ``in-contact with people" surfaces being the top surface of the seat and the adjacent lateral surface of the back as shown in Figure~\ref{Fig:Chair_ParaView}.

%-- Code - Chair
\begin{Code}[language=Python, firstnumber=1, caption={How to define a chair in \textit{IGG\_Furniture.dat}.}, label={Lst:InputFurniture_Chair}]
# FURNITURE 1
Chair
x0_seat y0_seat z0_seat
lx_seat ly_seat lz_seat
x0_back y0_back z0_back
lx_back ly_back lz_back
l_shift
\end{Code}

% Figure -- Chair

%--------------------------------------------------
%-----------------------
%-- Seat
%-----------------------
\subsection{Seat}\label{Sec:Feature_Seat}
\begin{figure}
    \centering
    \includegraphics[width=0.4\textwidth]{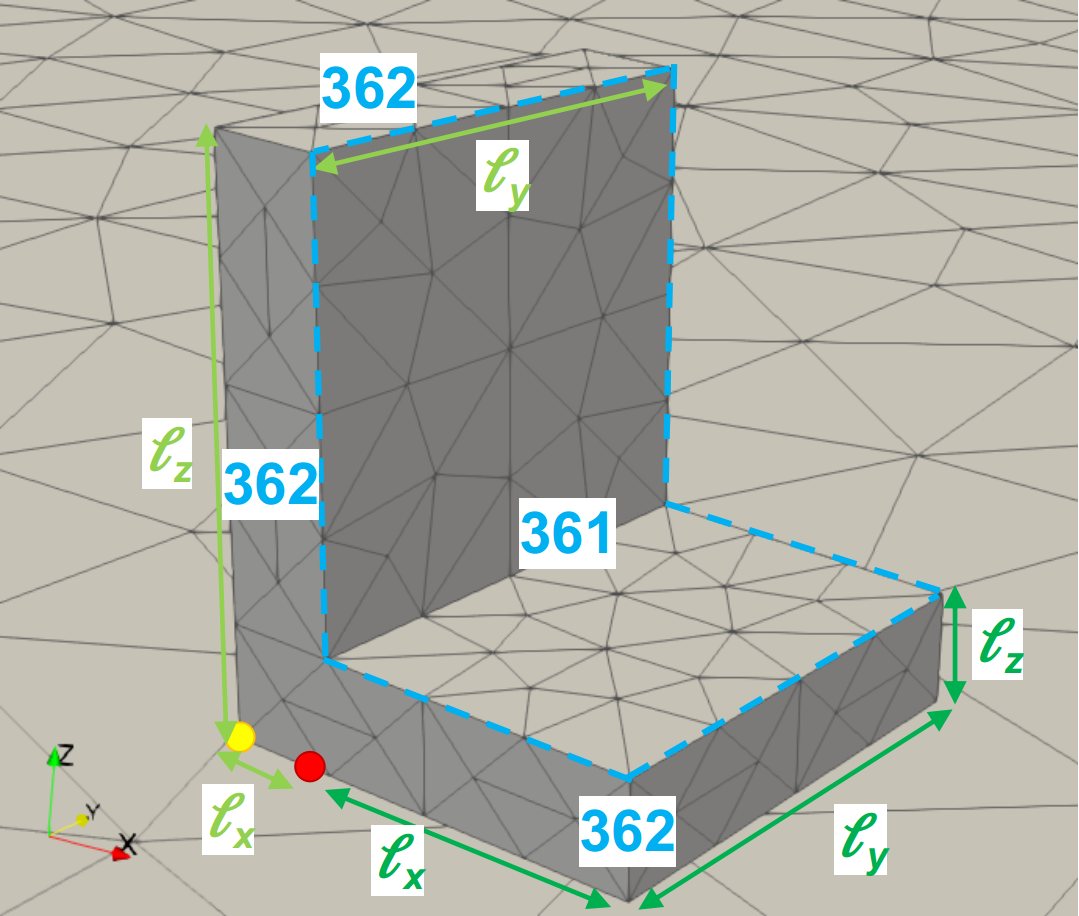}
    \caption{Visualisation of a seat \texttt{Seat} in \textbf{ParaView}. \textbf{361} is the physical ID of ``in-contact with people" surfaces and \textbf{362} is the physical ID of the other surfaces.}
    \label{Fig:Seat}
\end{figure}

A seat (keyword: \texttt{Seat}) is a L-shape object: in other words, a seat is a compound of 2 rectangular boxes. A seat is defined, in this order, by:
\begin{itemize}
    \item its keyword \texttt{Seat};
    \item its lower part location $(x_{0,lower}, y_{0,lower}, z_{0,lower})$,
    \item its lower part size $(\ell_{x,lower}, \ell_{y,lower}, \ell_{z,lower})$,
    \item its upper part location $(x_{0,upper}, y_{0,upper}, z_{0,upper})$ and
    \item its upper part size $(\ell_{x,upper}, \ell_{y,upper}, \ell_{z,upper})$,
\end{itemize}

as shown in Code~\ref{Lst:InputFurniture_Seat} and in Figure~\ref{Fig:Seat}. It exists two ways to represent seats, and more generally L-shape object, as shown in Figure~\ref{Fig:RShelf_Notation}: both are supported by \textbf{IGG}, and the user can use one or the other. The physical surface IDs of every seat is \textbf{36x}, where \textbf{x} is equal to \textbf{1} for the surfaces in contact with potential people sitting on it or \textbf{2} otherwise. A seat has 2 ``in-contact with people" surfaces being the top surface of the lower part and the adjacent lateral surface of the upper part as shown in Figure~\ref{Fig:Seat}.

%-- Code - RShelf
\begin{Code}[language=Python, firstnumber=1, caption={How to define a seat in \textit{IGG\_Furniture.dat}.}, label={Lst:InputFurniture_Seat}]
# FURNITURE 1
Seat
x0_lower y0_lower z0_lower
lx_lower ly_lower lz_lower
x0_upper y0_upper z0_upper
lx_upper ly_upper lz_upper
\end{Code}

% Figure -- Seat

%--------------------------------------------------
%-----------------------
%-- Stool
%-----------------------
\subsection{Stool}\label{Sec:Feature_Stool}
\begin{figure}
    \centering
    \begin{subfigure}[t]{0.4\textwidth}
        \includegraphics[width=\textwidth]{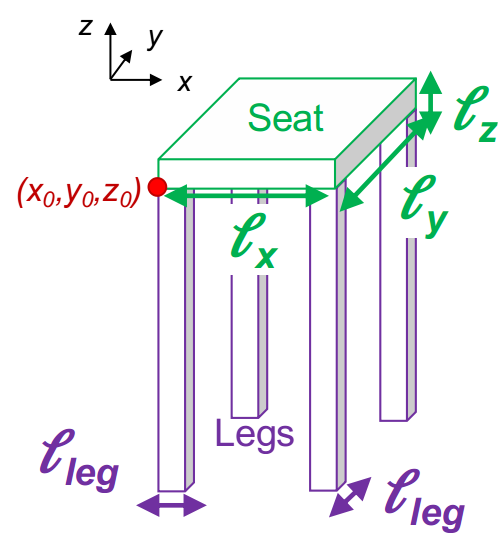}
        \caption{}
        \label{Fig:Stool_Notation}
    \end{subfigure}
    \begin{subfigure}[t]{0.4\textwidth}
        \includegraphics[width=\textwidth]{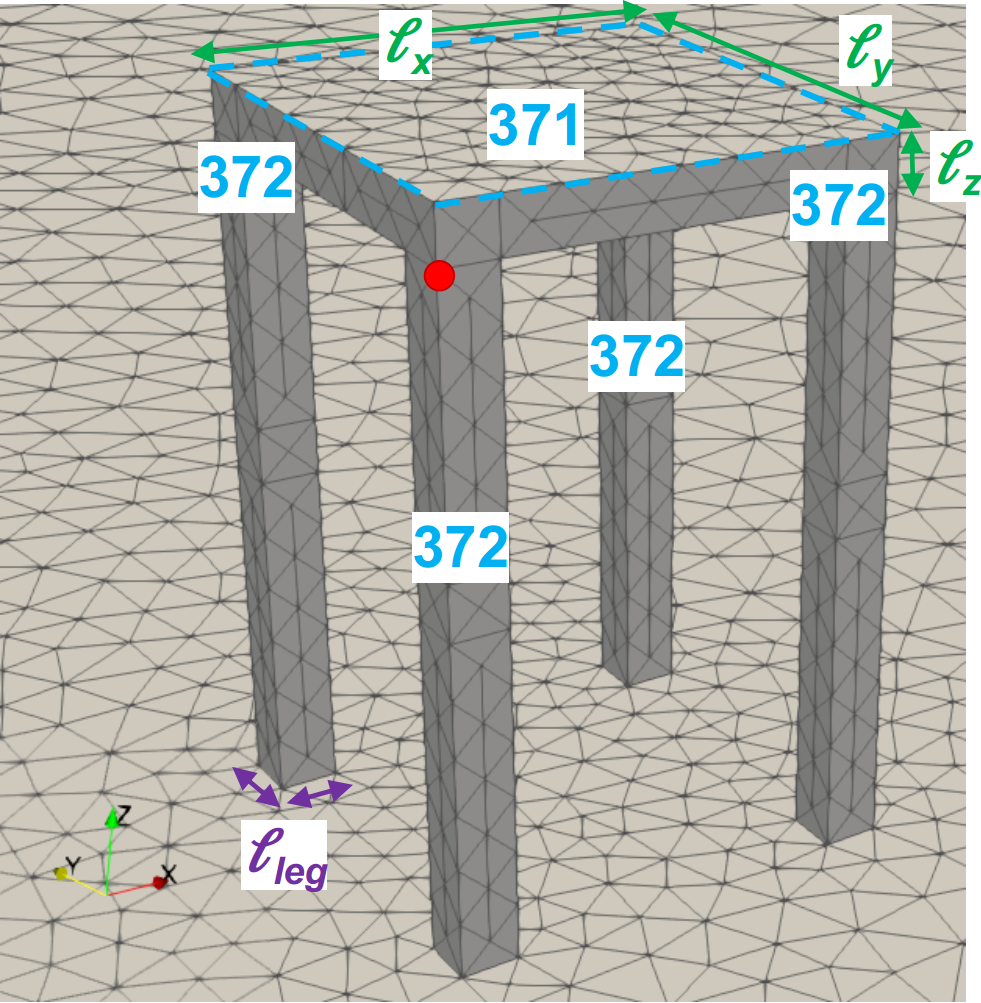}
        \caption{}
        \label{Fig:Stool_ParaView}
    \end{subfigure}
    \caption{(a) Representation of a stool in \textbf{IGG} and (b) visualisation of a stool \texttt{Stool} in \textbf{ParaView}. \textbf{371} is the physical ID of the top surface of the seat and \textbf{372} is the physical ID of the other surfaces.}
    \label{Fig:Stool}
\end{figure}

A stool (keyword: \texttt{Stool}) is a compound of 5 rectangular boxes: 4 legs and 1 seat as shown in Figure~\ref{Fig:Stool_Notation}. 

\textbf{Remarks about the stool geometry:} The 4 legs are assumed to have a square section having the same size $\ell_{leg} \times \ell_{leg}$ as shown in Figure~\ref{Fig:Stool_Notation}. The legs are set at the four corner of the seat. 

A stool is defined, in this order, by:
\begin{itemize}
    \item its keyword \texttt{Stool};
    \item its seat location $(x_{0}, y_{0}, z_{0})$,
    \item its seat size $(\ell_{x}, \ell_{y}, \ell_{z})$ and
    \item its legs size $\ell_{leg}$,
\end{itemize}

as shown in Code~\ref{Lst:InputFurniture_Stool} and in Figure~\ref{Fig:Stool}. The physical surface IDs of every stool is \textbf{37x}, where \textbf{x} is equal to \textbf{1} for the top surface of the seat or \textbf{2} otherwise as shown in Figure~\ref{Fig:Stool_ParaView}.

%-- Code - Stool
\begin{Code}[language=Python, firstnumber=1, caption={How to define a stool in \textit{IGG\_Furniture.dat}.}, label={Lst:InputFurniture_Stool}]
# FURNITURE 1
Stool
x0 y0 z0
lx ly lz
l_leg
\end{Code}

% Figure -- Stool

%--------------------------------------------------
%-----------------------
%-- Bed
%-----------------------
\subsection{Bed}\label{Sec:Feature_Bed}
\begin{figure}
    \centering
    \begin{subfigure}[t]{0.45\textwidth}
        \includegraphics[width=\textwidth]{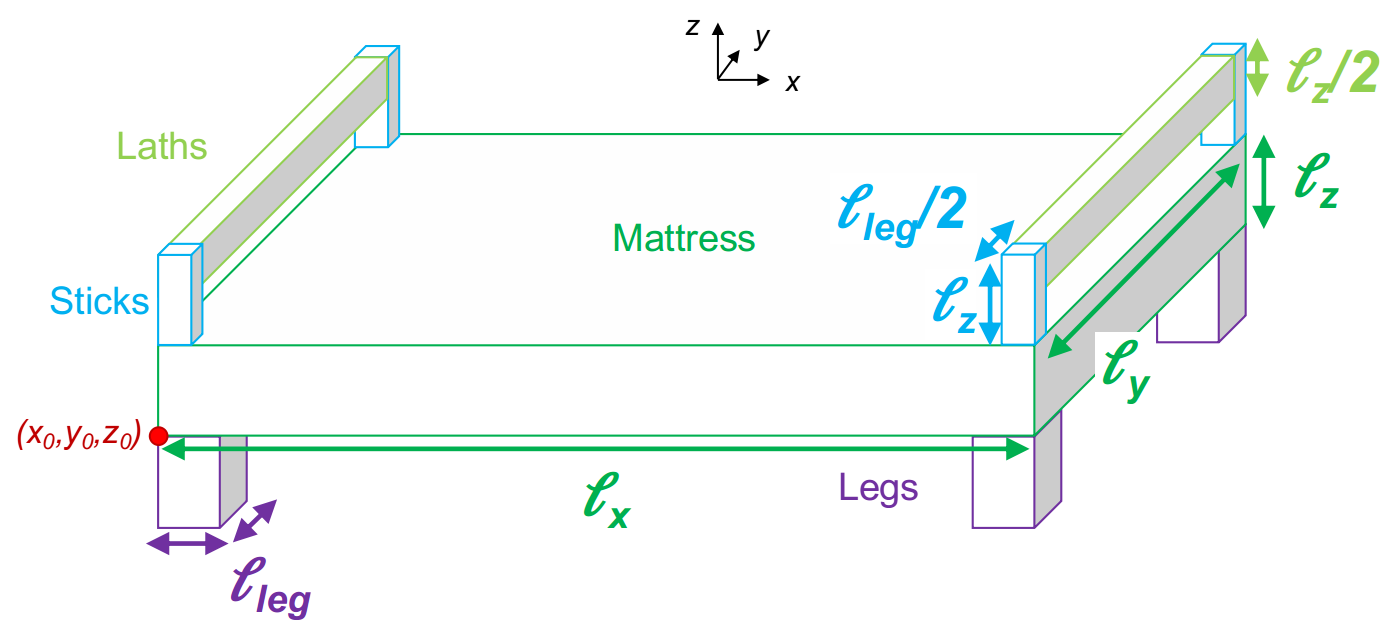}
        \caption{}
        \label{Fig:Bed_Notation}
    \end{subfigure}
    \begin{subfigure}[t]{0.38\textwidth}
        \includegraphics[width=\textwidth]{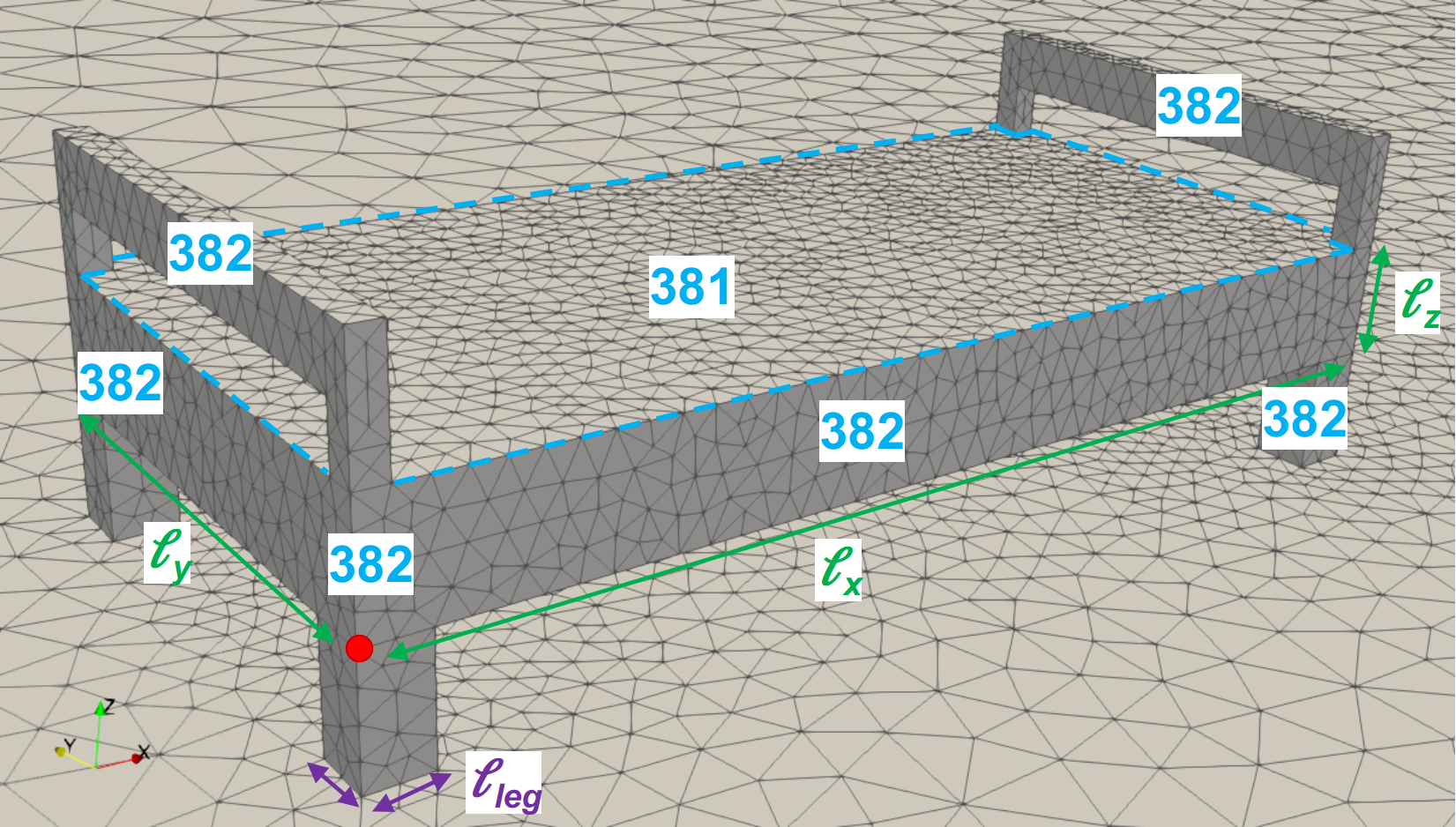}
        \caption{}
        \label{Fig:Bed_ParaView}
    \end{subfigure}
    \caption{(a) Representation of a bed in \textbf{IGG} and (b) visualisation of a bed \texttt{Bed} in \textbf{ParaView}. \textbf{381} is the physical ID of the top surface of the mattress and \textbf{372} is the physical ID of the other surfaces.}
    \label{Fig:Bed}
\end{figure}

A bed (keyword: \texttt{Bed}) is a compound of 11 rectangular boxes: 4 legs, 1 mattress, 4 sticks and 2 laths as shown in Figure~\ref{Fig:Bed_Notation}. 

\textbf{Remarks about the bed geometry:} The 4 legs are assumed to have a square section having the same size $\ell_{leg} \times \ell_{leg}$ as shown in Figure~\ref{Fig:Bed_Notation}. The legs are set at the four corner of the mattress. It is assumed that the 4 sticks, linking the laths, have square section twice smaller than the legs, i.e. $\ell_{leg}/2 \times \ell_{leg}/2 $ and have the same height than the mattress, i.e., $\ell_{z}$. Finally, the laths height is assumed to be twice smaller than the mattress height, i.e. $\ell_{z}/2$. The remaining dimensions are deduced naturally from the mattress and sticks sizes.

A bed is defined, in this order, by:
\begin{itemize}
    \item its keyword \texttt{Bed};
    \item its mattress location $(x_{0}, y_{0}, z_{0})$,
    \item its mattress size $(\ell_{x}, \ell_{y}, \ell_{z})$ and
    \item its legs size $\ell_{leg}$,
\end{itemize}

as shown in Code~\ref{Lst:InputFurniture_Bed} and in Figure~\ref{Fig:Bed}. The physical surface IDs of every bed is \textbf{38x}, where \textbf{x} is equal to \textbf{1} for the top surface of the mattress or \textbf{2} otherwise as shown in Figure~\ref{Fig:Bed_ParaView}.

%-- Code - Bed
\begin{Code}[language=Python, firstnumber=1, caption={How to define a bed in \textit{IGG\_Furniture.dat}.}, label={Lst:InputFurniture_Bed}]
# FURNITURE 1
Bed
x0 y0 z0
lx ly lz
l_leg
\end{Code}

% Figure -- Bed

%--------------------------------------------------
%-----------------------
%-- Bevel
%-----------------------
\subsection{Bevel}\label{Sec:Feature_Bevel}
\begin{figure}
    \centering
    \begin{subfigure}[t]{0.45\textwidth}
        \includegraphics[width=\textwidth]{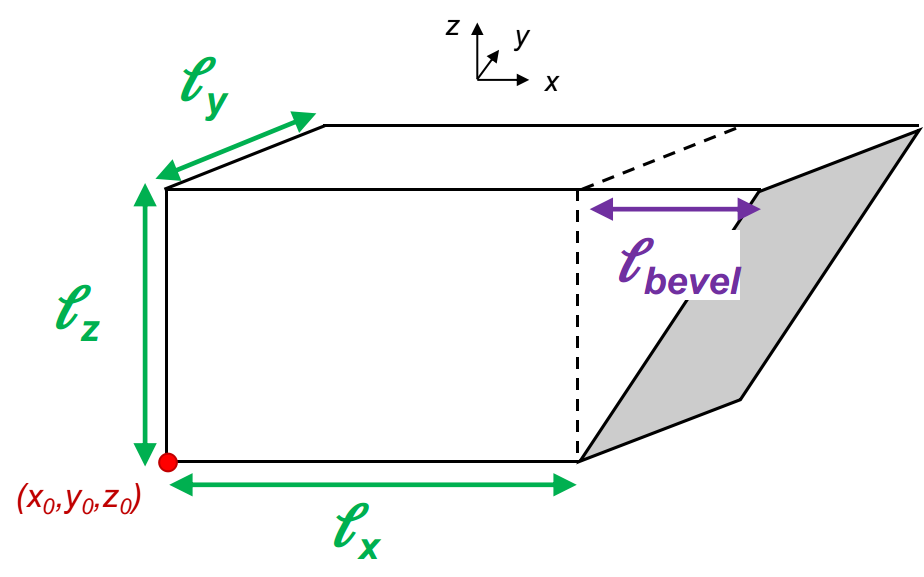}
        \caption{}
        \label{Fig:Bevel_Notation}
    \end{subfigure}
    \begin{subfigure}[t]{0.4\textwidth}
        \includegraphics[width=\textwidth]{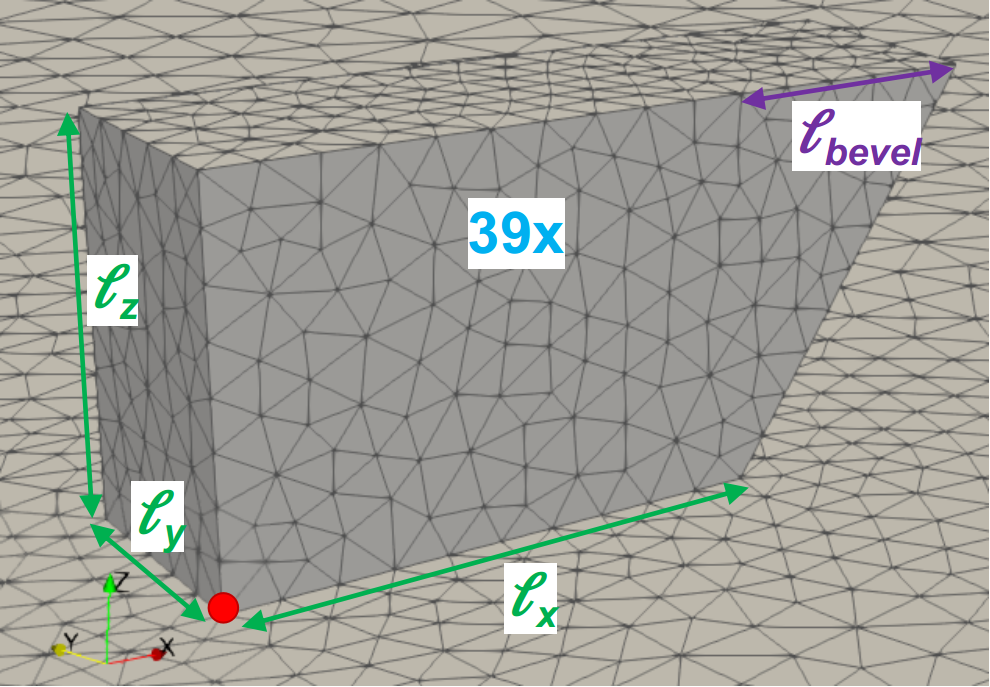}
        \caption{}
        \label{Fig:Bevel_ParaView}
    \end{subfigure}
    \caption{(a) Representation of a bevel in \textbf{IGG} and (b) visualisation of a bevel \texttt{Bevel} in \textbf{ParaView}. \textbf{39x} is the physical ID of bevels, where \textbf{x} if the bevel ID.}
    \label{Fig:Bevel}
\end{figure}

A bevel (keyword: \texttt{Bevel}) is a rectangular box with one of its surface having an inclination that meets the other surfaces at any angle but 90\degree, i.e. a rectangular box with one inclined surface, as shown in Figure~\ref{Fig:Bevel_Notation}. 

A bevel is defined, in this order, by:
\begin{itemize}
    \item its keyword \texttt{Bevel};
    \item its bevel orientation and direction \texttt{Surface1 Surface2};
    \item its location $(x_{0}, y_{0}, z_{0})$,
    \item its size $(\ell_{x}, \ell_{y}, \ell_{z})$ and
    \item its bevel size $\ell_{bevel}$,
\end{itemize}

as shown in Code~\ref{Lst:InputFurniture_Bevel} and in Figure~\ref{Fig:Bevel}. The physical surface IDs of every bevel is \textbf{39x}, where \textbf{x} is the ID of the bevel. For example, if there are 2 bevels into the domain, they will have the physical IDs \textbf{391} and \textbf{392}. In future version of \textbf{IGG}, bevels will be encapsulated under one unique physical ID.

When defining a bevel, the location and direction of the bevel need to be specified. \texttt{Surface1} designates the surface of the rectangular box where the inclined surface is located, while \texttt{Surface2} designates which edge needs to be ``moved" to create the bevel. The user can refer to Figure~\ref{Fig:Bevels_Direction} which summarises all the different bevel possibilities and the \texttt{Surface1 Surface2} (in purple) that need to be used.

%-- Code - Bed
\begin{Code}[language=Python, firstnumber=1, caption={How to define a bevel in \textit{IGG\_Furniture.dat}.}, label={Lst:InputFurniture_Bevel}]
# FURNITURE 1
Bevel
Surface1 Surface2
x0 y0 z0
lx ly lz
l_bevel
\end{Code}

% Figure -- Bevel

% Figure -- Bevels Direction
\begin{figure}
    \centering
    \begin{subfigure}[t]{\textwidth}
        \includegraphics[width=\textwidth]{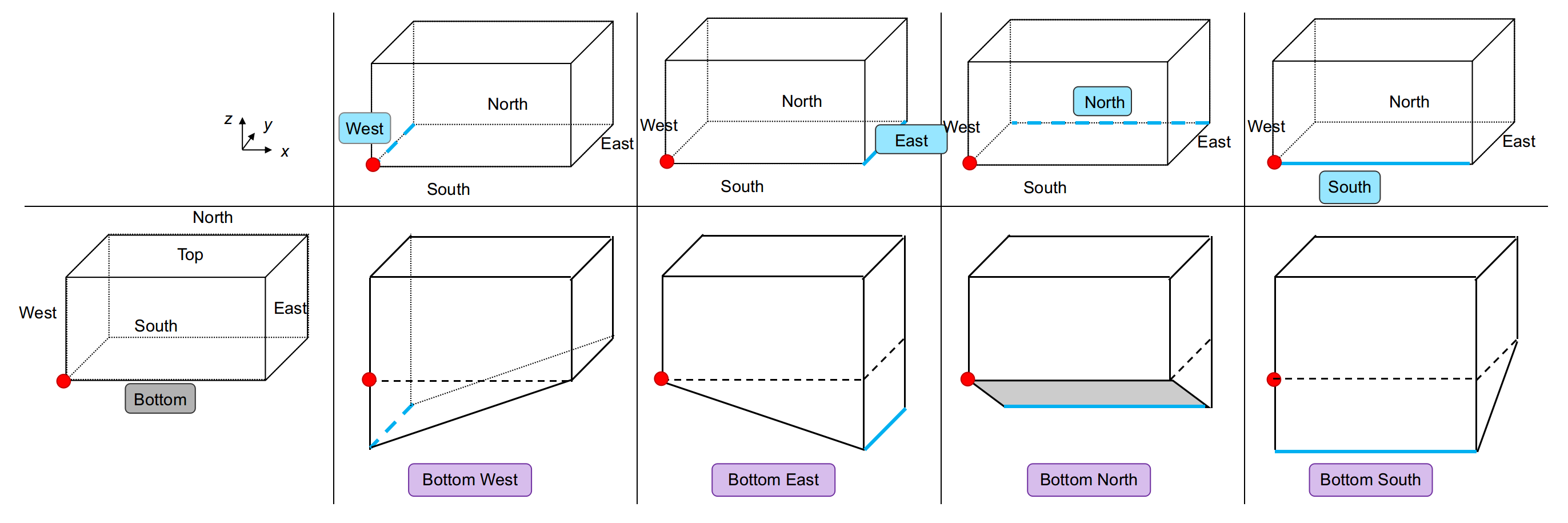}
        \caption{Bevel part located on the \textbf{Bottom} surface of the rectangular box.}
    \end{subfigure}
    \begin{subfigure}[t]{\textwidth}
        \includegraphics[width=\textwidth]{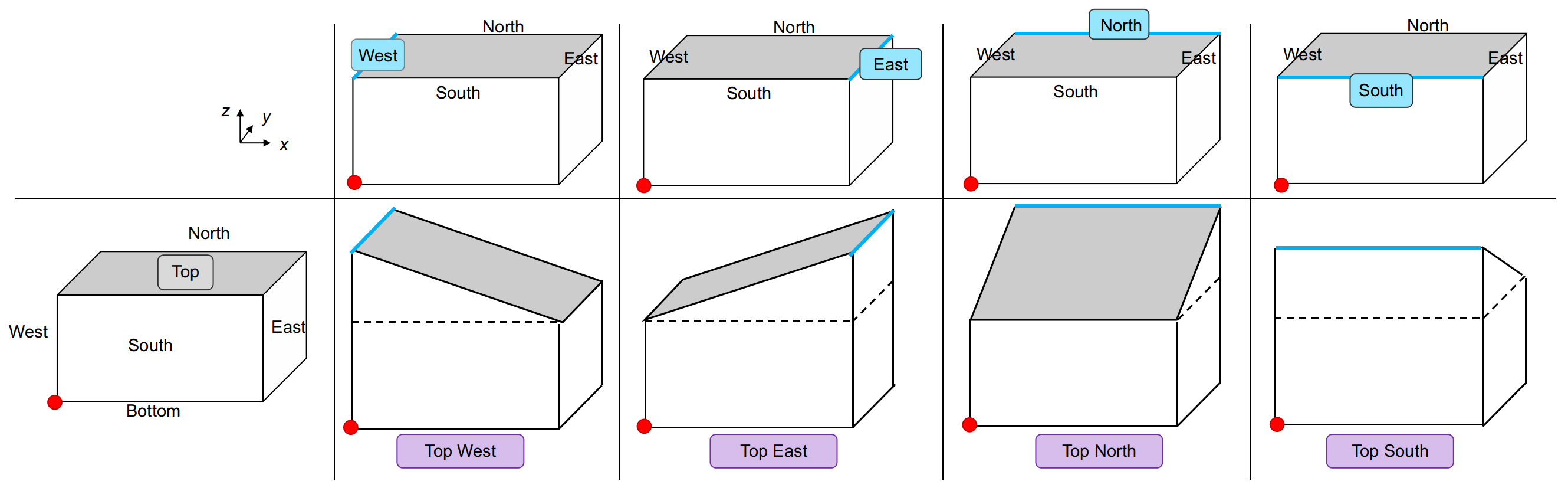}
        \caption{Bevel part located on the \textbf{Top} surface of the rectangular box.}
    \end{subfigure}
    \begin{subfigure}[t]{\textwidth}
        \includegraphics[width=\textwidth]{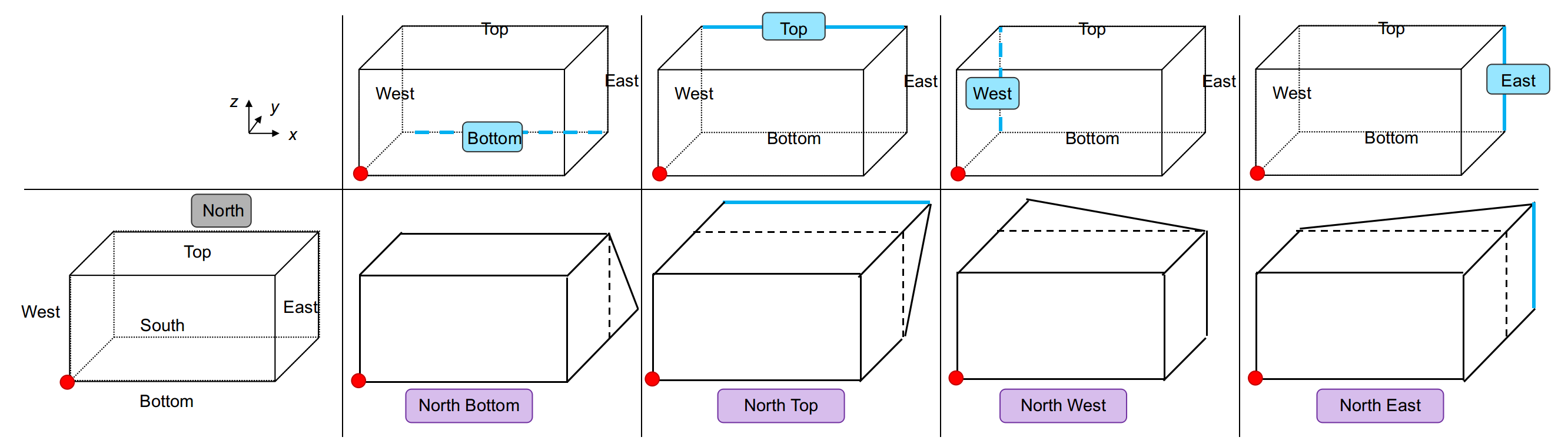}
        \caption{Bevel part located on the \textbf{North} surface of the rectangular box.}
    \end{subfigure}
    \caption{See next page...}
\end{figure}

\begin{figure}
    \ContinuedFloat
    \centering
    \begin{subfigure}[t]{\textwidth}
        \includegraphics[width=\textwidth]{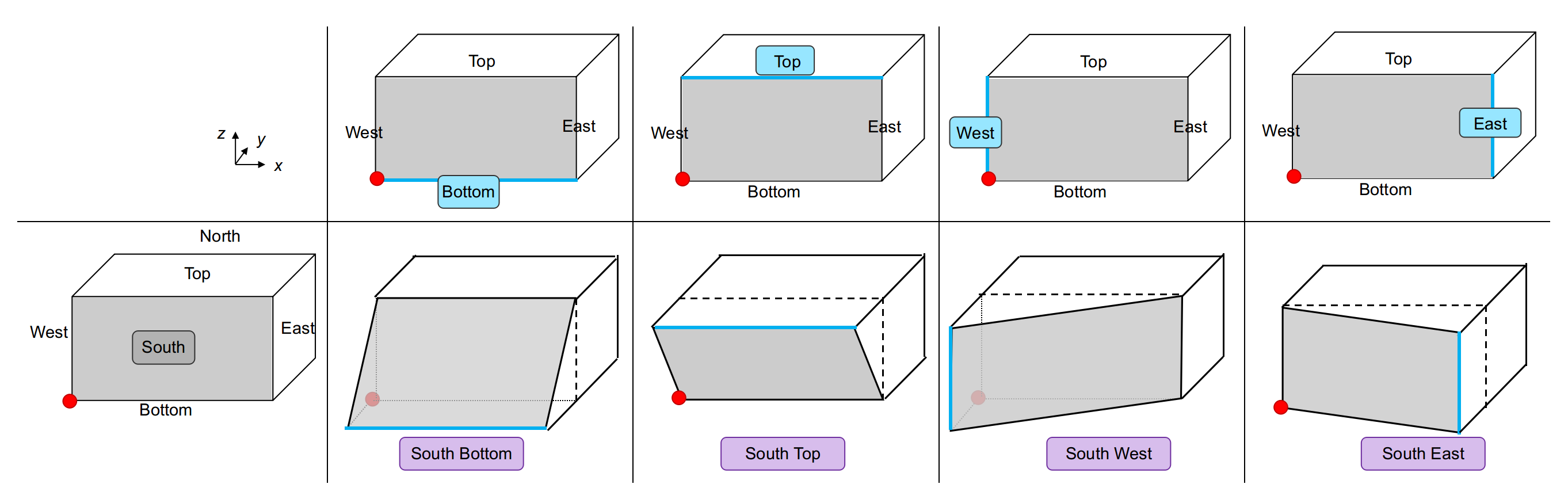}
        \caption{Bevel part located on the \textbf{South} surface of the rectangular box.}
    \end{subfigure}
    \begin{subfigure}[t]{\textwidth}
        \includegraphics[width=\textwidth]{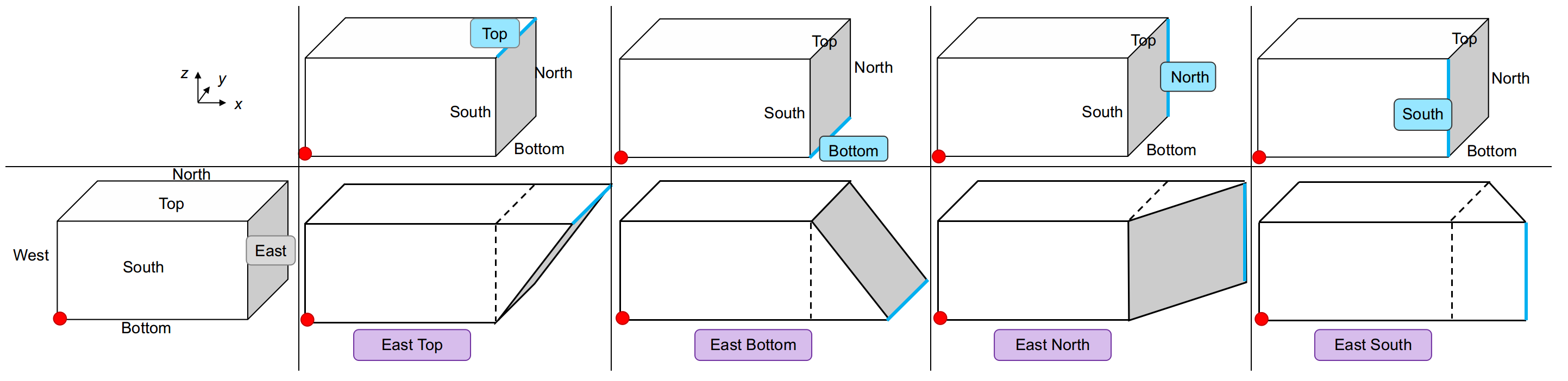}
        \caption{Bevel part located on the \textbf{East} surface of the rectangular box.}
    \end{subfigure}
    \begin{subfigure}[t]{\textwidth}
        \includegraphics[width=\textwidth]{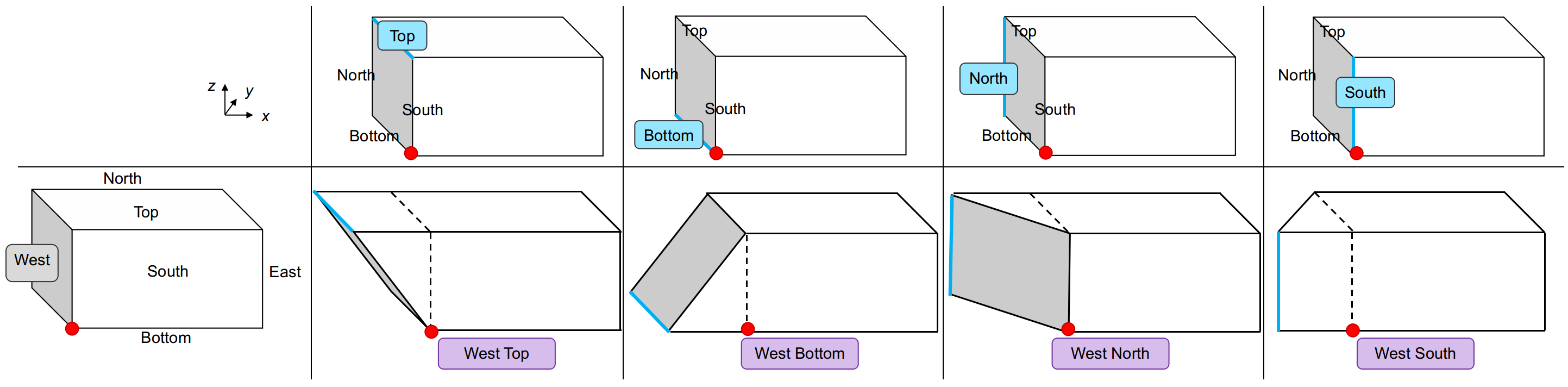}
        \caption{Bevel part located on the \textbf{West} surface of the rectangular box.}
    \end{subfigure}
    \caption{Representation of the different bevel orientation/direction supported by \textbf{IGG}. The grey represents \texttt{Surface1} where the bevel is located. The blue lines depicts the location \texttt{Surface2} of the edge being ``moved" to create the bevel. In purple is given the keyword \texttt{Surface1 Surface2} needed in \textit{IGG\_Furniture.dat} to generate the bevel.}
    \label{Fig:Bevels_Direction}
\end{figure}
%%%%%%%%%%%%%%%%%%%%%%%%%%%%%%%%%%%
%-- Accessories
%%%%%%%%%%%%%%%%%%%%%%%%%%%%%%%%%%%
\section{Accessories}\label{Sec:Feature_Accessories}
The accessories, which are defined by the user in \textit{IGG\_Accessories.dat}, supported by \textbf{IGG} are summarised in Table~\ref{Tab:Feature_Accessories}.

\begin{table}
    \centering
    \begin{tabular}{ |p{3.5cm}||p{3.5cm}|p{2.0cm}|p{4.0cm}|p{1.5cm}|  }
         \hline
         \textbf{Accessories type} & \textbf{Shape} & \textbf{Keyword} & \textbf{Physical ID} & \textbf{Section}  \\\hline
         Separator      & Rectangular box     & \texttt{Separator} & 20  & \ref{Sec:Feature_Separator} \\\hline
         Laptop         & L-shape             & \texttt{Laptop}    & 601 hot surfaces \newline 602 otherwise & \ref{Sec:Feature_Laptop} \\\hline
         Screen         & 3 rectangular boxes & \texttt{Screen}    & 611 hot screen surface \newline 612 otherwise & \ref{Sec:Feature_Screen} \\\hline
         Computer tower & Rectangular box     & \texttt{Tower}     & 621 & \ref{Sec:Feature_Tower} \\\hline
         Diffuser       & Rectangular boxes \newline Triangles & \texttt{Diffuser}    & 631 South  \newline 632 East \newline 633 North \newline 634 West & \ref{Sec:Feature_Diffuser} \\\hline
         Extractor      & Rectangular boxes \newline Triangles & \texttt{Extractor}   & 641 South  \newline 642 East \newline 643 North \newline 644 West & \ref{Sec:Feature_Extractor} \\\hline
    \end{tabular}
    \caption{\label{Tab:Feature_Accessories}Summary of the accessories supported by \textbf{IGG}.}
\end{table}

%--------------------------------------------------
%-----------------------
%-- Separator
%-----------------------
\subsection{Separator}\label{Sec:Feature_Separator}
\begin{figure}
    \centering
    \includegraphics[width=0.5\textwidth]{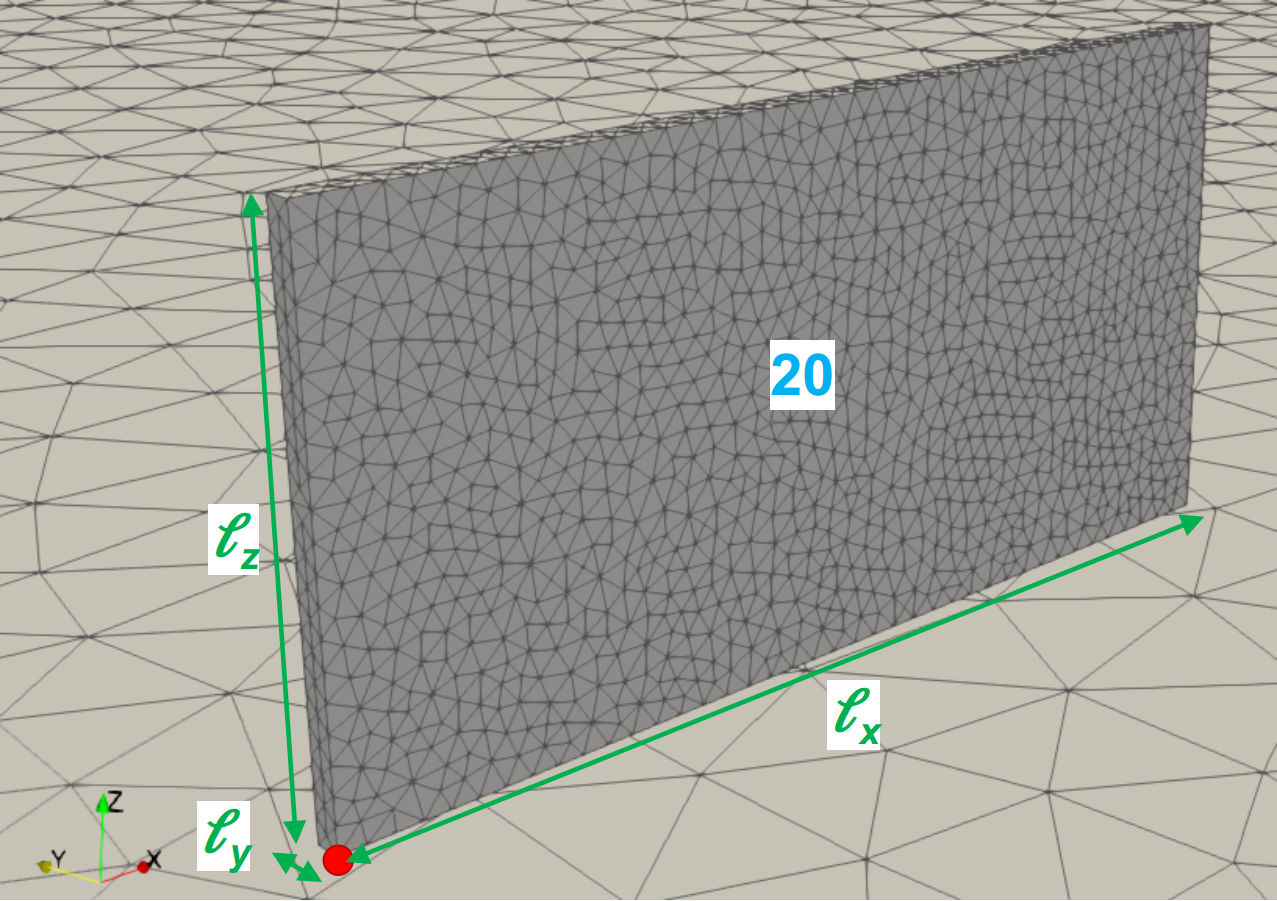}
    \caption{Visualisation of a plastic separator \texttt{Separator} in \textbf{ParaView}. \textbf{20} is the physical ID of separators.}
    \label{Fig:Separator}
\end{figure}

A plastic separator (keyword: \texttt{Separator}) is a rectangular box object. A plastic separator is defined, in this order, by:
\begin{itemize}
    \item its keyword \texttt{Separator};
    \item its location $(x_{0}, y_{0}, z_{0})$ and
    \item its size $(\ell_{x}, \ell_{y}, \ell_{z})$,
\end{itemize}

as shown in Code~\ref{Lst:InputAccessories_Separator} and in Figure~\ref{Fig:Separator}. The physical surface ID of every plastic separator is \textbf{20}.

%-- Code - Separator
\begin{Code}[language=Python, firstnumber=1, caption={How to define plastic separator in \textit{IGG\_Accessories.dat}.}, label={Lst:InputAccessories_Separator}]
# ACCESSORY 1
Separator
x0 y0 z0
lx ly lz
\end{Code}

%-- Figure - Separator

%--------------------------------------------------
%-----------------------
%-- Laptop
%-----------------------
\subsection{Laptop}\label{Sec:Feature_Laptop}
\begin{figure}
    \centering
    \includegraphics[width=0.4\textwidth]{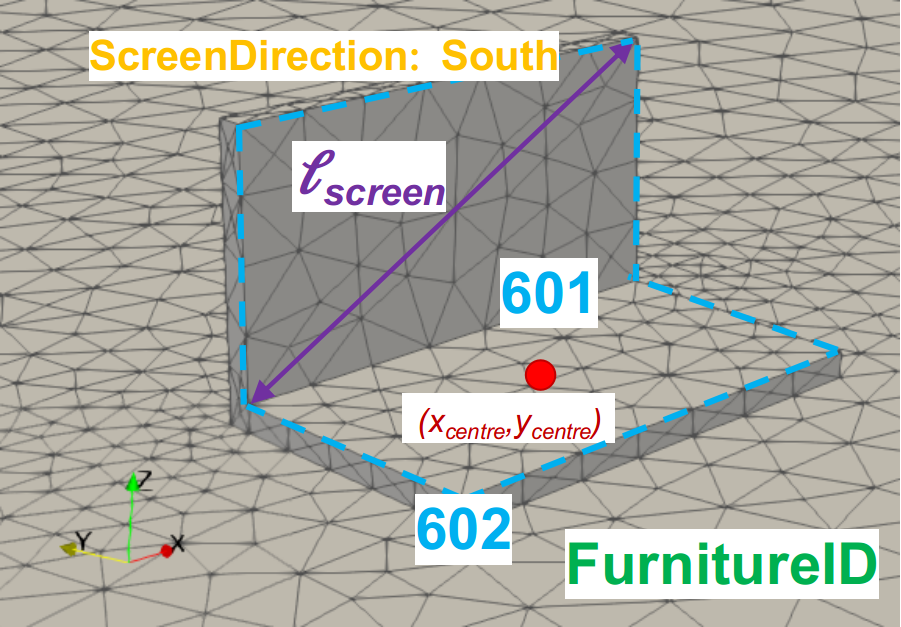}
    \caption{Visualisation of a laptop \texttt{Laptop} in \textbf{ParaView}. \textbf{601} is the physical ID of ``hot" surfaces and \textbf{602} is the physical ID of the other surfaces.}
    \label{Fig:Laptop}
\end{figure}

A laptop (keyword: \texttt{Laptop}) is a L-shape object: in other words, a laptop is a compound of 2 rectangular boxes. A laptop is defined, in this order, by:
\begin{itemize}
    \item its keyword \texttt{Laptop};
    \item its location in space $(x_{centre}, y_{centre})$,
    \item its screen size in inch $\ell_{screen}$,
    \item its furniture on which it stands \texttt{FurnitureID} and
    \item its screen direction \texttt{ScreenDirection}, i.e. the direction/surface of the domain the laptop screen is facing: \texttt{East}, \texttt{West}, \texttt{South} or \texttt{North};
\end{itemize}

as shown in Code~\ref{Lst:InputAccessories_Laptop} and in Figure~\ref{Fig:Laptop}. The physical surface IDs of every laptop is \textbf{60x}, where \textbf{x} is equal to \textbf{1} for the ``hot" surfaces or \textbf{2} otherwise. A laptop has 2 ``hot" surfaces being the screen and the keyboard as shown in Figure~\ref{Fig:Laptop}.

\textbf{Remark 1:} Be careful, here the centre of the laptop is required.

\textbf{Remark 2:} The screen size need to be given in inches.

\textbf{Remark 3:} The furniture ID is an integer and is the ID of the furniture as it appears in \textit{IGG\_Furniture.dat}. If the laptop is on the ground, use the ID 0.

%-- Code - Laptop
\begin{Code}[language=Python, firstnumber=1, caption={How to define laptop in \textit{IGG\_Accessories.dat}.}, label={Lst:InputAccessories_Laptop}]
# ACCESSORY 1
Laptop
Centre          xcentre ycentre
ScreenSize      l_screen
Furniture       FurnitureID
ScreenDirection ScreenDirection
\end{Code}

% Figure -- Laptop

%--------------------------------------------------
%-----------------------
%-- Screen
%-----------------------
\subsection{Screen}\label{Sec:Feature_Screen}
\begin{figure}
    \centering
    \begin{subfigure}[t]{0.4\textwidth}
        \includegraphics[width=\textwidth]{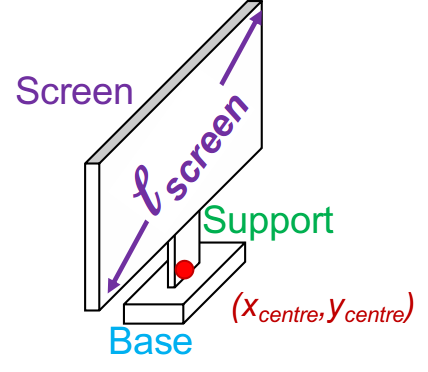}
        \caption{}
        \label{Fig:Screen_Notation}
    \end{subfigure}
    \begin{subfigure}[t]{0.4\textwidth}
        \includegraphics[width=\textwidth]{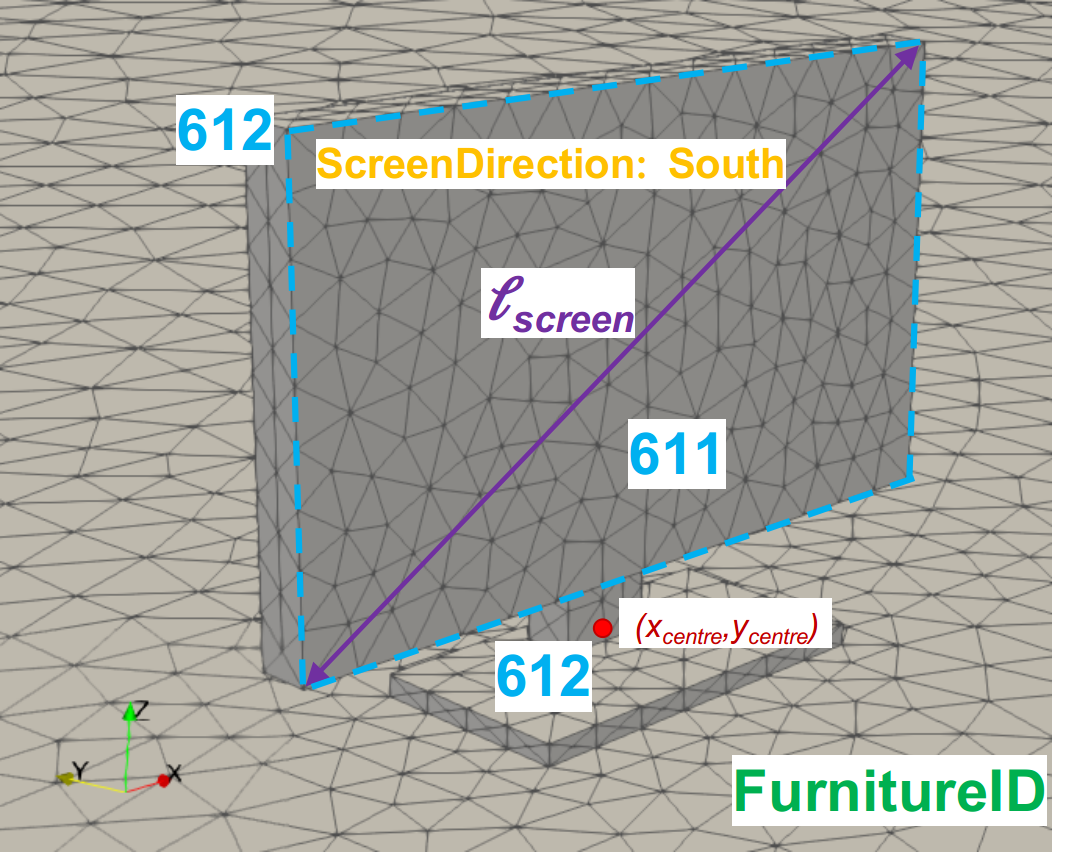}
        \caption{}
        \label{Fig:Screen_ParaView}
    \end{subfigure}
    \caption{(a) Representation of a screen in \textbf{IGG} and (b) visualisation of a screen \texttt{Screen} in \textbf{ParaView}. \textbf{611} is the physical ID of the hot surface of the screen and \textbf{612} is the physical ID of the other surfaces.}
    \label{Fig:Screen}
\end{figure}

A screen (keyword: \texttt{Screen}) is is a compound of 3 rectangular boxes: 1 screen, 1 base and 1 support as shown in Figure~\ref{Fig:Screen_Notation}. 

A screen is defined, in this order, by:
\begin{itemize}
    \item its keyword \texttt{Screen};
    \item its location in space $(x_{centre}, y_{centre})$,
    \item its screen size in inch $\ell_{screen}$,
    \item its furniture on which it stands \texttt{FurnitureID} and
    \item its screen direction \texttt{ScreenDirection}, i.e. the direction/surface of the domain the screen is facing: \texttt{East}, \texttt{West}, \texttt{South} or \texttt{North};
\end{itemize}

as shown in Code~\ref{Lst:InputAccessories_Screen} and in Figure~\ref{Fig:Screen}. The physical surface IDs of every screen is \textbf{61x}, where \textbf{x} is equal to \textbf{1} for the ``hot" screen surface or \textbf{2} otherwise as shown in Figure~\ref{Fig:Screen_ParaView}.

\textbf{Remark 1:} Be careful, here the centre of the screen is required.

\textbf{Remark 2:} The screen size need to be given in inches.

\textbf{Remark 3:} The furniture ID is an integer and is the ID of the furniture as it appears in \textit{IGG\_Furniture.dat}. If the laptop is on the ground, use the ID 0.

%-- Code - Screen
\begin{Code}[language=Python, firstnumber=1, caption={How to define a screen in \textit{IGG\_Accessories.dat}.}, label={Lst:InputAccessories_Screen}]
# ACCESSORY 1
Screen
Centre          xcentre ycentre
ScreenSize      l_screen
Furniture       FurnitureID
ScreenDirection ScreenDirection
\end{Code}

% Figure -- Screen

%--------------------------------------------------
%-----------------------
%-- Tower
%-----------------------
\subsection{Computer Tower}\label{Sec:Feature_Tower}
\begin{figure}
    \centering
    \includegraphics[width=0.3\textwidth]{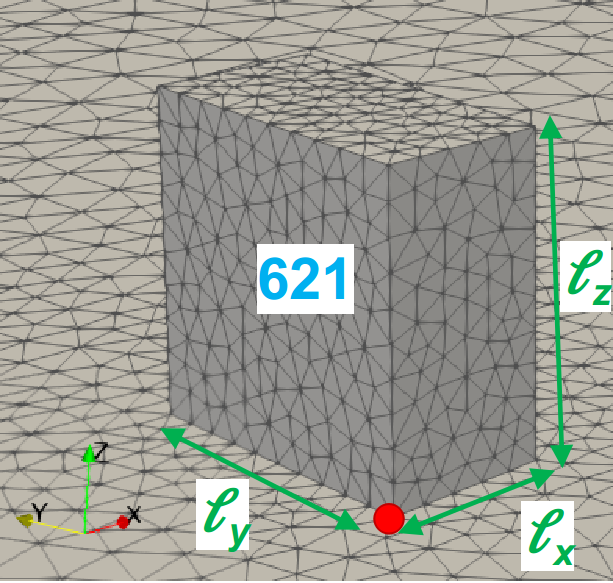}
    \caption{Visualisation of a plastic separator \texttt{Separator} in \textbf{ParaView}. \textbf{621} is the physical ID of computer tower.}
    \label{Fig:Tower}
\end{figure}

A computer tower (keyword: \texttt{Tower}) is a rectangular box object. A computer tower is defined, in this order, by:
\begin{itemize}
    \item its keyword \texttt{Tower};
    \item its location $(x_{0}, y_{0}, z_{0})$ and
    \item its size $(\ell_{x}, \ell_{y}, \ell_{z})$,
\end{itemize}

as shown in Code~\ref{Lst:InputAccessories_Tower} and in Figure~\ref{Fig:Tower}. The physical surface ID of every computer tower is \textbf{621}.

%-- Code - Tower
\begin{Code}[language=Python, firstnumber=1, caption={How to define a computer tower in \textit{IGG\_Accessories.dat}.}, label={Lst:InputAccessories_Tower}]
# ACCESSORY 1
Tower
x0 y0 z0
lx ly lz
\end{Code}

%-- Figure - Tower

%--------------------------------------------------
%-----------------------
%-- Ceiling Diffuser
%-----------------------
\subsection{Ceiling Diffuser}\label{Sec:Feature_Diffuser}
\begin{figure}
    \centering
    \includegraphics[width=0.5\textwidth]{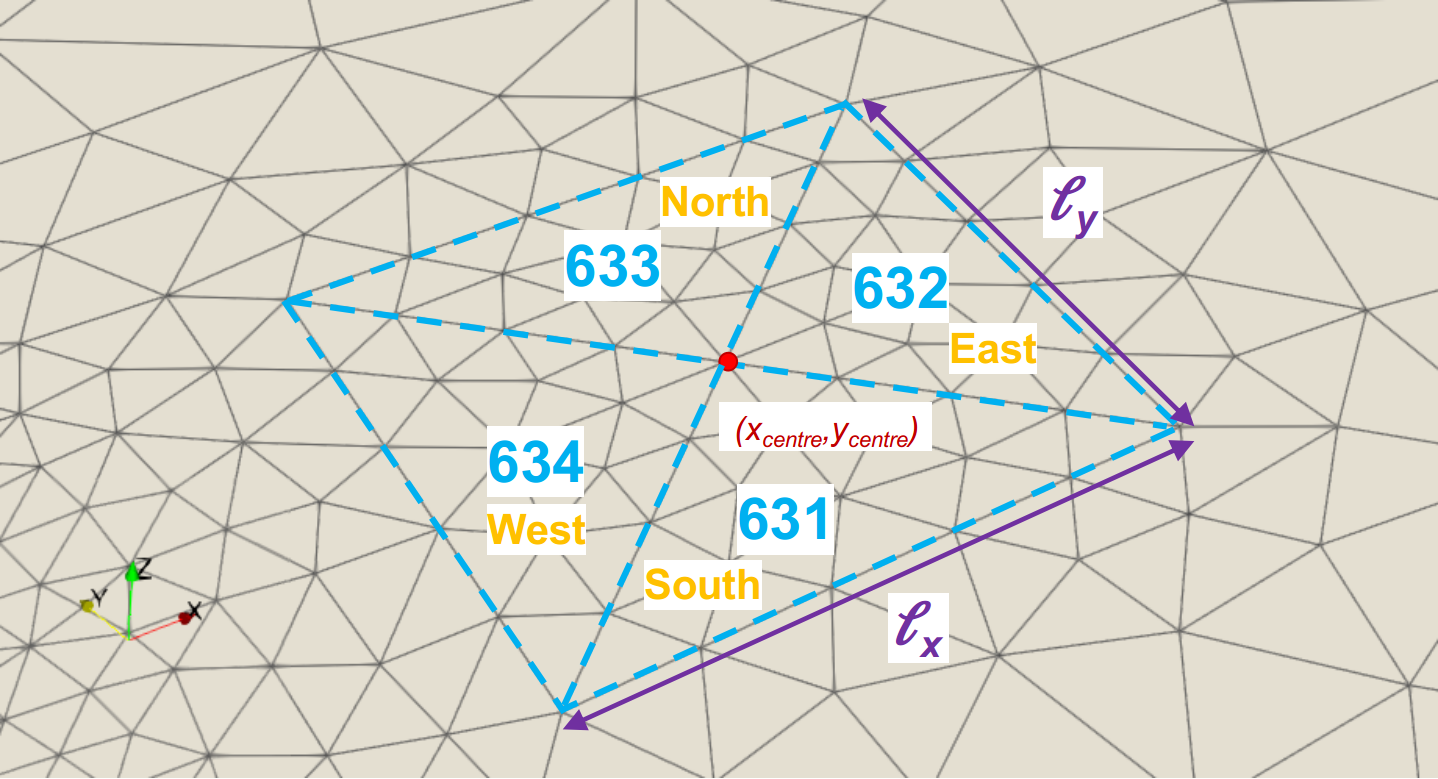}
    \caption{Visualisation of a 4-way diffuser \texttt{Diffuser} in \textbf{ParaView}. \textbf{631}, \textbf{632}, \textbf{633} and \textbf{634} are the physical IDs of the surfaces where air is supplied.}
    \label{Fig:Diffuser}
\end{figure}

A diffuser (keyword: \texttt{Diffuser}) is a 2D rectangular box object divided into smaller rectangles or triangles depending the diffuser type (see Section~\ref{Sec:Feature_2D}). A diffuser is defined, in this order, by:
\begin{itemize}
    \item its keyword \texttt{Diffuser};
    \item its ventilation direction \texttt{North}, \texttt{South}, \texttt{East} and/or \texttt{West};
    \item its location $(x_{centre}, y_{centre})$ and
    \item its size $(\ell_{x}, \ell_{y})$.
\end{itemize}

as shown in Code~\ref{Lst:InputAccessories_Diffuser} and in Figure~\ref{Fig:Diffuser}. The physical IDs defining the diffuser are:
\begin{itemize}
    \item \textbf{631} for the surface where the air is released towards the \texttt{South}
    \item \textbf{632} for the surface where the air is released towards the \texttt{East}
    \item \textbf{633} for the surface where the air is released towards the \texttt{North}
    \item \textbf{634} for the surface where the air is released towards the \texttt{West}
\end{itemize}

\textbf{Remark 1:} Be careful, here the centre of the diffuser is required.

\textbf{Remark 2:} Only diffusers at ceiling height are supported.

\textbf{Remark 3:} If 1 ventilation direction is provided, it is a 1-way diffuser; if 2 are provided, it is a 2-way diffuser... Refer to Figure~\ref{Fig:Diffuser} to look at all the possibilities supported.

\textbf{Remark 4:} A diffuser can only be an inlet. It cannot be both an inlet/outlet. For example, if the diffuser is a 2-way diffuser, the two surfaces defining it behave as inlets.

%-- Code - Diffuser
\begin{Code}[language=Python, firstnumber=1, caption={How to define a ceiling diffuser in \textit{IGG\_Accessories.dat}.}, label={Lst:InputAccessories_Diffuser}]
# ACCESSORY 1
Diffuser
Direction  Direction1 Direction2 ...
Centre     xcentre ycentre
Dimension  lx ly
\end{Code}

%-- Figure - Diffuser

%--------------------------------------------------
%-----------------------
%-- Ceiling Extractor
%-----------------------
\subsection{Ceiling Extractor}\label{Sec:Feature_Extractor}
\begin{figure}
    \centering
    \includegraphics[width=0.5\textwidth]{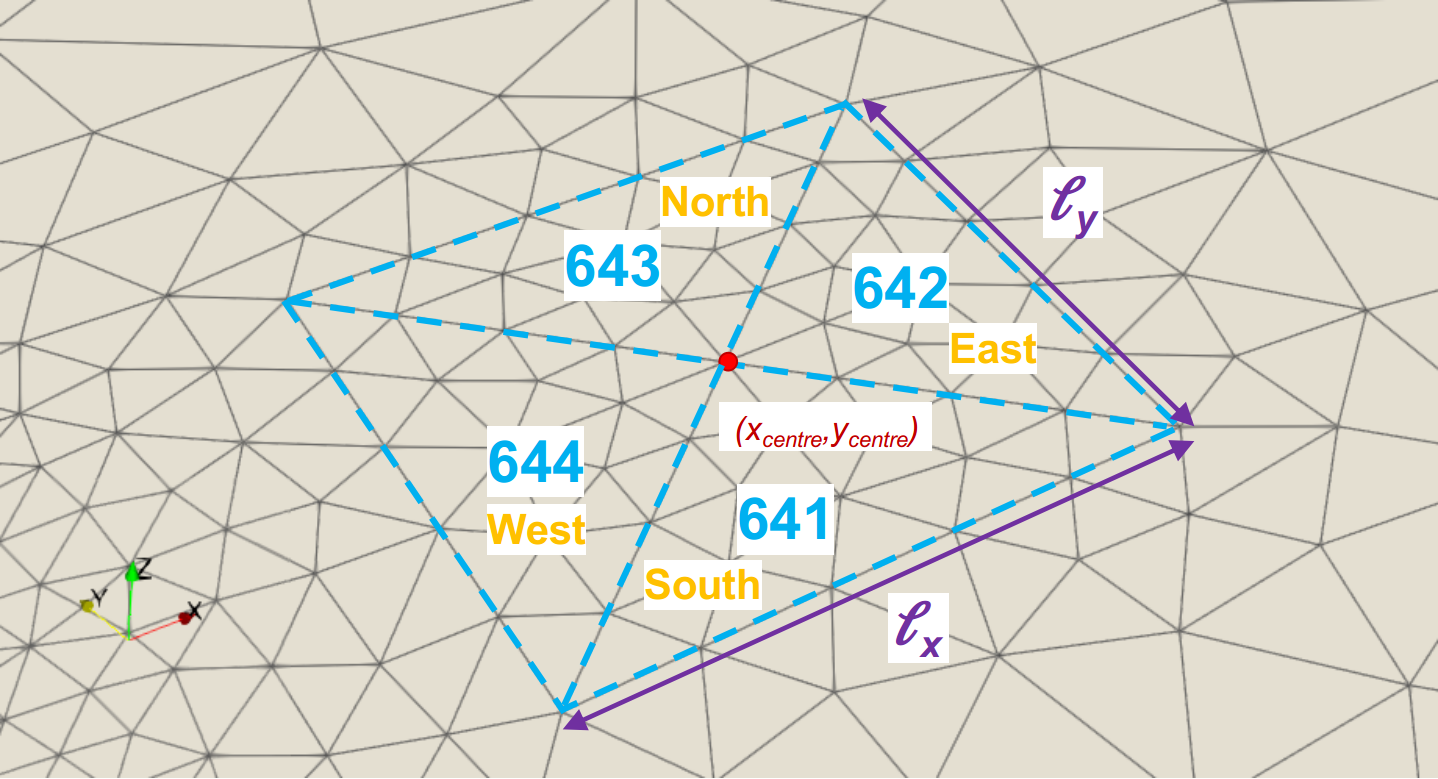}
    \caption{Visualisation of a 4-way extractor \texttt{Extractor} in \textbf{ParaView}. \textbf{641}, \textbf{642}, \textbf{643} and \textbf{644} are the physical IDs of the surfaces where air is extracted.}
    \label{Fig:Extractor}
\end{figure}

An extractor (keyword: \texttt{Extractor}) is a 2D rectangular box object divided into smaller rectangles or triangles depending the diffuser type (see Section~\ref{Sec:Feature_2D}). An extractor is defined, in this order, by:
\begin{itemize}
    \item its keyword \texttt{extractor};
    \item its extract direction \texttt{North}, \texttt{South}, \texttt{East} and/or \texttt{West};
    \item its location $(x_{centre}, y_{centre})$ and
    \item its size $(\ell_{x}, \ell_{y})$.
\end{itemize}

as shown in Code~\ref{Lst:InputAccessories_Extractor} and in Figure~\ref{Fig:Extractor}. The physical IDs defining the extractor are:
\begin{itemize}
    \item \textbf{641} for the surface where the air is extracted towards the \texttt{South}
    \item \textbf{642} for the surface where the air is extracted towards the \texttt{East}
    \item \textbf{643} for the surface where the air is extracted towards the \texttt{North}
    \item \textbf{644} for the surface where the air is extracted towards the \texttt{West}
\end{itemize}

\textbf{Remark 1:} Be careful, here the centre of the extractor is required.

\textbf{Remark 2:} Only extractors at ceiling height are supported.

\textbf{Remark 3:} If 1 extract direction is provided, it is a 1-way extractor; if 2 are provided, it is a 2-way extractor... Refer to Figure~\ref{Fig:Diffuser_Notation} to look at all the possibilities supported.

\textbf{Remark 4:} An extractor can only be an outlet. It cannot be both an inlet/outlet. For example, if the extractor is a 2-way extractor, the two surfaces defining it behave as outlets.

%-- Code - Extractor
\begin{Code}[language=Python, firstnumber=1, caption={How to define a ceiling extractor in \textit{IGG\_Accessories.dat}.}, label={Lst:InputAccessories_Extractor}]
# ACCESSORY 1
Extractor
Direction Direction1 Direction2 ...
Centre    xcentre ycentre
Dimension lx ly
\end{Code}

%-- Figure - Extractor
        %----------------------------------------------------------%
%-- Description of humans                                --%
%----------------------------------------------------------%
\chapter{Humans}\label{Sec:Humans}
Humans can be added into the geometry and parametrised into the input file \textit{IGG\_Humans.dat}. Assuming that the \texttt{File Name} defined in \textit{IGG\_Numerics.dat} is \texttt{MyGeom}, a summary of the human states is given in \textit{MyGeom\_HumansStates.dat} and can be open with a file editor. As the user can let \textbf{IGG} uses the default values for, amongst other, the \texttt{Height}, the \texttt{Weight}..., \textit{MyGeom\_HumansStates.dat} summarises the values and characteristics used for each human as well as the area of each body part.

%-----------------------
%-- Proportion
%-----------------------
\section{Human geometry and body proportion}\label{Sec:Humans_Geometry}
A human is a compound of 16 rectangular boxes for the body, 1 bevel for the nose and 1 2D feature for the mouth. The body parts are: 
1 Head, 1 Neck, 1 Trunk, 2 Arms, 2 Forearms, 2 Hands, 1 Pelvis, 2 Thighs, 2 Legs and 2 Feet.

The body proportion of the human are following standards based on the head unit (hu) defined as in equation~\eqref{Eq:HeadUnit}:
\begin{equation}
    hu = H/8.
    \label{Eq:HeadUnit}
\end{equation}

where $H$ is the human height in meters. It is assumed that every human has a height of $8hu$.

\begin{figure}
    \centering
    \begin{subfigure}[t]{0.45\textwidth}
        \includegraphics[width=\textwidth]{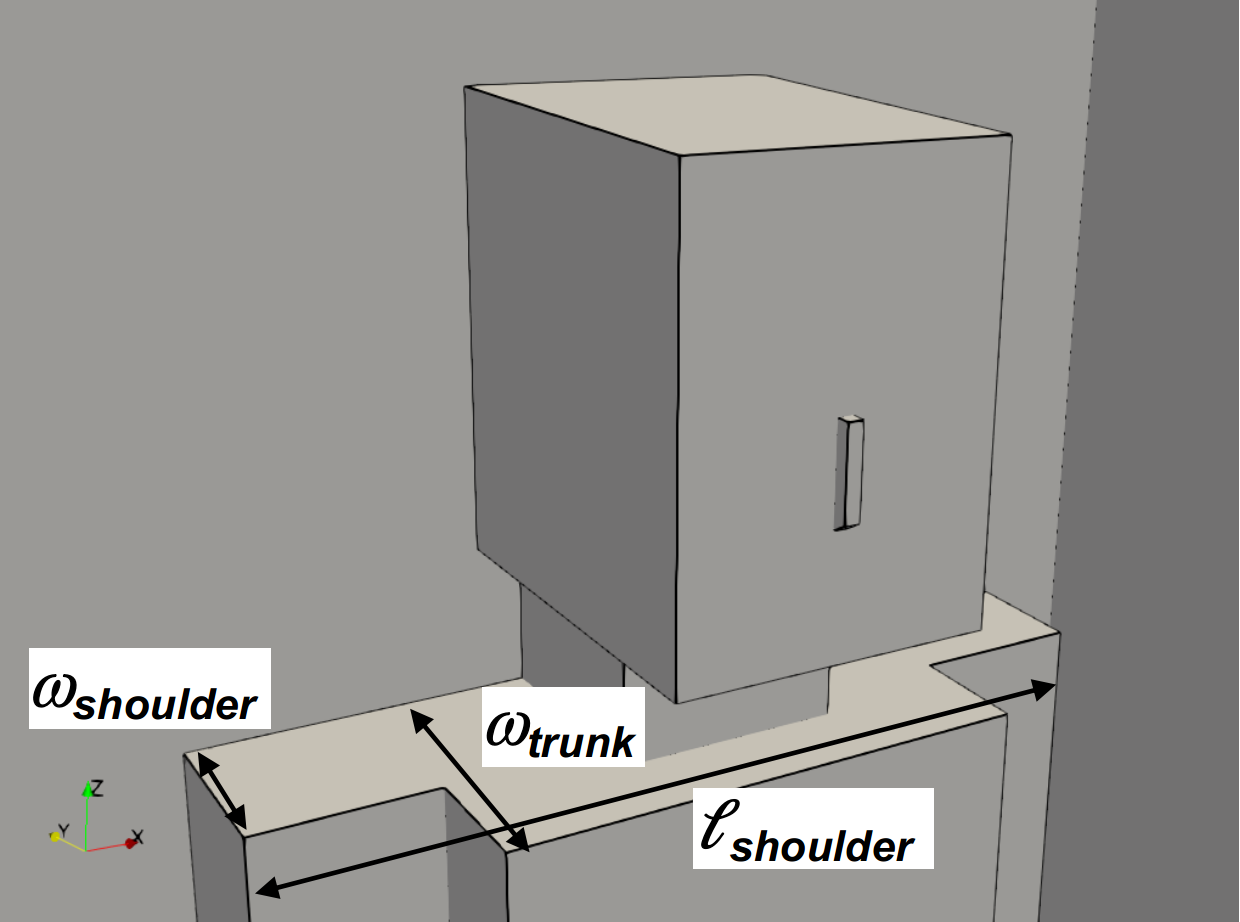}
        \caption{Length between shoulder $\ell_{shoulder}$. Width of shoulders $w_{shoulder}$. Width of trunk $w_{trunk}$.}
        \label{Fig:Human_ProportionShoulder}
    \end{subfigure}
    \begin{subfigure}[t]{0.6\textwidth}
        \includegraphics[width=\textwidth]{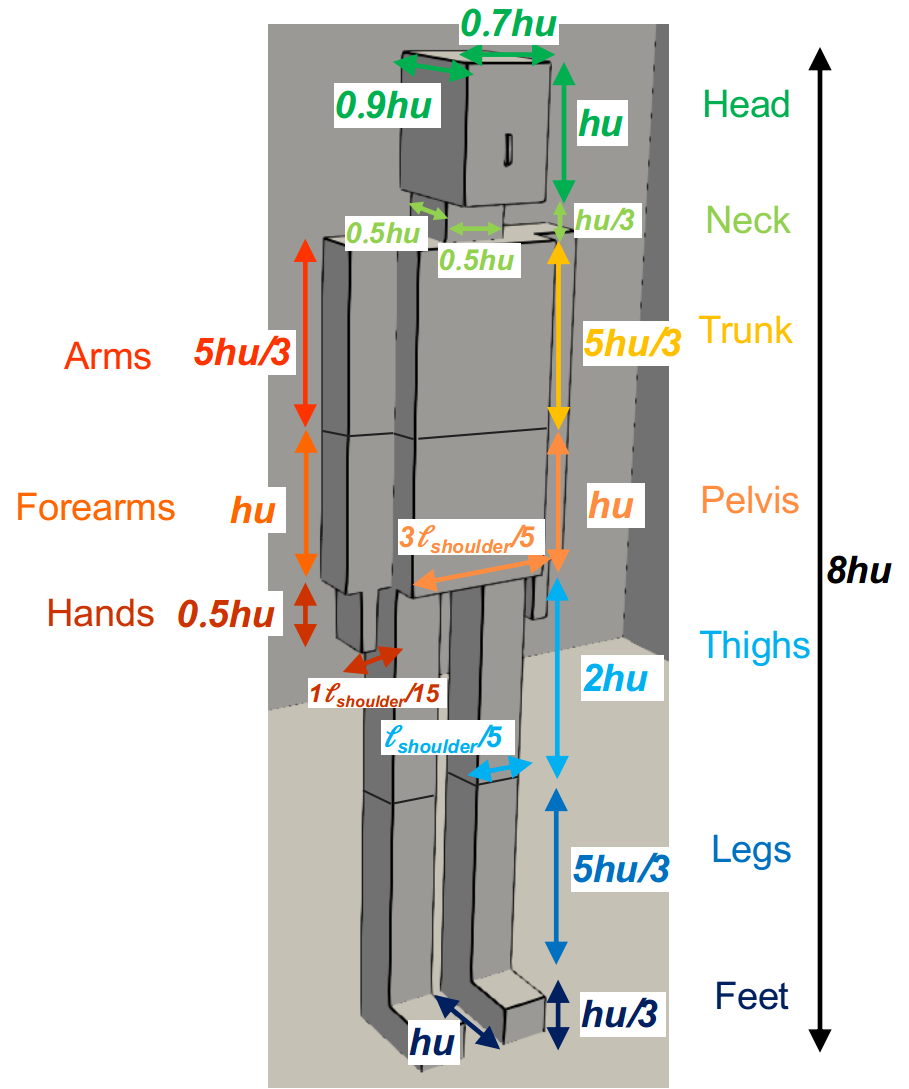}
        \caption{}
        \label{Fig:Human_ProportionBody}
    \end{subfigure}
    \caption{Human proportions.}
    \label{Fig:Human_Proportion}
\end{figure}

Depending the weight profile of the human, the length between shoulders, the width of the trunk and the width of the shoulders are defined as in Tale~\ref{Tab:Proportion_Weight}. See Figure~\ref{Fig:Human_ProportionShoulder} to visualise what these dimensions mean.

%-- Table - Basic proportion
\begin{table}
    \centering
    \begin{tabular}{ |p{3.0cm}||p{1.5cm}|p{1.5cm}|p{1.5cm}|  }
         \hline
         \textbf{Weight profile} & \textbf{$\ell_{shoulder}$} & \textbf{$w_{shoulder}$} & \textbf{$w_{trunk}$}  \\
         \hline
         \texttt{Thin}   & $1.8hu$   & $0.3hu$ & $0.5hu$  \\\hline
         \texttt{Normal} & $2.0hu$   & $0.4hu$ & $0.7hu$  \\\hline
         \texttt{Fat}    & $2.2hu$   & $0.6hu$ & $0.9hu$  \\\hline
    \end{tabular}
    \caption{\label{Tab:Proportion_Weight}Summary of the proportion for length between shoulder and width of shoulder/trunk.}
\end{table}

% Figure -- Basic proportion

%-----------------------
%-- Options
%-----------------------
\section{Human characteristics}\label{Sec:Humans_Characteristics}
Humans are divided into 10 parts: \texttt{Head}, \texttt{Neck}, \texttt{Arms}, \texttt{Forearms}, \texttt{Hands}, \texttt{Trunk}, \texttt{Pelvis}, \texttt{Thighs}, \texttt{Legs} and \texttt{Feet}. At the beginning of \textit{IGG\_Humans.dat}, the user can decide to represent only some part of the body if wanted.

Humans can have 3 different activities \texttt{Activity}: \texttt{Standing}, \texttt{Sitting} or \texttt{Lying}. It is \textbf{mandatory} to specify the human activity to each human defined in \textit{IGG\_Humans.dat}.

\subsection{General options}
Whatever the human activity, the user can specify several optional characteristics such as:
\begin{itemize}
    \item The human height (keyword: \texttt{Height}): float in centimetres. By default 170 $cm$ is used.
    \item The human weight profile (keyword: \texttt{Weight}): \texttt{Thin}, \texttt{Normal} or \texttt{Fat}. By default \texttt{Normal} is used.
    \item The human forearms position (keyword: \texttt{Forearm}): \texttt{Up} or \texttt{Down}. By default \texttt{Down} is used.
    \item The nose model (keyword: \texttt{NoseModel}): \texttt{T} or \texttt{F}. By default \texttt{F} is used.
    \item The mouth model (keyword: \texttt{MouthModel}): \texttt{T} or \texttt{F}. By default \texttt{F} is used.
    \item The mouth area (keyword: \texttt{MouthArea}): float in $cm^{2}$ or \texttt{Auto}. By default the mouth area is 1.18 $cm^{2}$. If \texttt{MouthArea} is specified but not \texttt{MouthModel}, then \texttt{MouthModel} is set to \texttt{T}.
    \item The nose area (keyword: \texttt{NoseArea}): float in $cm^{2}$ or \texttt{Auto}. By default the nose area is 0.65 $cm^{2}$. If \texttt{NoseArea} is specified but not \texttt{NoseModel}, then \texttt{NoseModel} is set to \texttt{T}.
    \item The nose angle (keyword: \texttt{NoseAngle}): float in degree or \texttt{Auto}. By default the nose angle is $60\degree$. If \texttt{NoseAngle} is specified but not \texttt{NoseModel}, then \texttt{NoseModel} is set to \texttt{T}.
    \item The body part releasing heat (keyword: \texttt{HeatingPart}): \texttt{All}, \texttt{Head}, \texttt{Neck}, \texttt{Arms}, \texttt{Forearms}, \texttt{Hands}, \texttt{Trunk}, \texttt{Pelvis}, \texttt{Thighs}, \texttt{Legs} and/or \texttt{Feet}. By default \texttt{All} is used.
\end{itemize}

\textbf{Remark:} The keywords can be used in any order, there is no specific order.

\subsection{Standing}
If the \texttt{Activity} is \texttt{Standing}, then the following is \textbf{mandatory}:
\begin{itemize}
    \item The face direction (keyword: \texttt{FaceDirection}) of the human, i.e. which surface of the domain the human is looking: \texttt{South}, \texttt{North}, \texttt{East} or \texttt{West}.
    \item The location, i.e. centre, of the human (keyword: \texttt{Centre}): a 3-tuple of float $(x_{centre}, y_{centre}, z_{centre})$.
\end{itemize}
In the case the human is standing on a furniture, then the following is needed:
\begin{itemize}
    \item The furniture on which the human is standing (keyword: \texttt{Furniture}): an integer corresponding to the ID of the furniture as it appears in \textit{IGG\_Furniture.dat}. In this case, $z_{centre}$ provided by the keywork \texttt{Centre} is ignored.
\end{itemize}

\subsection{Sitting}
If the \texttt{Activity} is \texttt{Sitting}, then the following is \textbf{mandatory}:

\begin{itemize}
    \item The furniture on which the human is sit (keyword: \texttt{Furniture}): an integer corresponding to the ID of the furniture as it appears in \textit{IGG\_Furniture.dat}.
    \item If the furniture is not a \texttt{Chair} or a \texttt{Seat}, then the face direction of the human (keyword: \texttt{FaceDirection}) is mandatory too, i.e. which surface of the domain the human is looking: \texttt{South}, \texttt{North}, \texttt{East} or \texttt{West}.
\end{itemize}

\textbf{Remark:} When a human is sitting on a \texttt{Chair} or a \texttt{Seat}: the back of the human always touches the back of the seat; if the human is too tall and its feet are going under the ground, then the legs are shorten to touch perfectly the ground; if the human is too short and its thighs are not long enough compare to the seat size, then the thighs are elongated to fit the seat.

\subsection{Lying}
If the \texttt{Activity} is \texttt{Lying}, then the following is \textbf{mandatory}:
\begin{itemize}
    \item The furniture on which the human is lying (keyword: \texttt{Furniture}): an integer corresponding to the ID of the furniture as it appears in \textit{IGG\_Furniture.dat}.
    \item The face direction of the human (keyword: \texttt{FaceDirection}), i.e. which surface of the domain the human is looking: \texttt{South}, \texttt{North}, \texttt{East}, \texttt{West} or \texttt{Ceiling}.
    \item The lying position (keyword: \texttt{LyingPosition}) of the human, i.e. towards which surface of the domain the human head is: \texttt{South}, \texttt{North}, \texttt{East} or \texttt{West}.
\end{itemize}

\textbf{Remark:} When a human are lying on one side , i.e \texttt{FaceDirection} not being \texttt{Ceiling}, the arm, forearm and hand of this side are not modelled in order to have the human perfectly lying on the mattress. 

\section{Physical IDs}\label{Sec:Humans_Physical IDs}
Body part (except mouths and noses) physical IDs have the form "9xyz" where::
\begin{itemize}
    \item "x":: human ID 
    \item "y":: body part ID: 0: head; 1: neck; 2: arms; 3: forearms; 4: hands; 5: trunk; 6: pelvis; 7: thighs; 8: legs; 9: feet
    \item "z":: 1 for "heat" BCs; 2 for "no heat" BCs 
\end{itemize}

Mouths and noses physical IDs have the form "8xyz" where::
\begin{itemize}
    \item "x":: human ID 
    \item "y":: body part ID: 0: mouth; 1: nose
    \item "z":: 1 for "exit" BCs; 2 for "other" BCs 
\end{itemize}

Ex: "9201" is human 2 head generating heat; "8111" is human 1 exit of nose. 
    "92142" refers to the non-heating hands of human 21. 
NB: "8x02" does not exist - mouth is a surface and is an exit only.

        %----------------------------------------------------------%
%-- Coarsen and refinement of the mesh                   --%
%----------------------------------------------------------%
\chapter{Mesh size constraints}\label{Sec:Mesh}

The input file \textit{IGG\_Mesh.dat} (see Code~\ref{Lst:InputMesh}) allows the user to specify the mesh characteristics.
\begin{itemize}
    \item \textbf{General mesh sizes} is firstly defined. \texttt{NMinElement} is the minimum number of elements on one surface. \texttt{MinLength} and \texttt{MaxLength} are the minimum and maximum edge length of the mesh, respectively.
    \item \textbf{Constraints} can be added to the mesh. First, the number of constraints is defined. In Code~\ref{Lst:InputMesh}, 7 constraints are specified by the user for example. Then the constraints are defined. See Section~\ref{Sec:Mesh} for more details.
    \begin{itemize}
        \item Feature-specific constraint: the feature type to refine is specified, then the edge length wanted given.
        \item Non-feature-specific constraint: the user can refined/coarsen the mesh in some areas if wanted. The area to be refined/coarsen can be defined by a box or a sphere.
    \end{itemize}
\end{itemize}

The mesh can refined locally by the user. The user can also decide to refine specific part of some features.

\section{Refinement of rectangular boxes}
The following features are 2D features or one single rectangular boxes and will be always refined/coarsen everywhere - the option \texttt{All} should always be specified and is the only one supported by \textbf{IGG}:
\begin{itemize}
    \item Box
    \item Normal Shelf
    \item Manual Till
    \item Separator
    \item Computer Tower
    \item Inlet of ventilation
    \item Outlet of ventilation
    \item Windows
    \item Doors
    \item Diffusers
    \item Extractors
\end{itemize}

%-- Code - General
\begin{Code}[language=Python, firstnumber=1, caption={How to define a constraint in \textit{IGG\_Mesh.dat}.}, label={Lst:InputMeshe_ConstraintFeature}]
# CONSTRAINT 1
Furniture 1
All
m_size1

# CONSTRAINT 2
Accessory 1
All
m_size2

# CONSTRAINT 3
Inlet 1
All
m_size3
\end{Code}

\section{Refinement of complex objects}
The following features can be refined/coarsen in some part:
\begin{itemize}
    \item Refrigerated Shelf, Automatic Till, Seat
        \begin{itemize}
            \item \texttt{All}: refine the whole feature
            \item \texttt{Lower}: the smallest, i.e. lower, part of the L-shape
            \item \texttt{Upper}: the tallest, i.e. upper, part of the L-shape.
        \end{itemize}
    \item Table, Stool
        \begin{itemize}
            \item \texttt{All}: refine the whole feature
            \item \texttt{Top}:  the top, i.e. upper, part of the feature
            \item \texttt{Legs}: the four legs of the feature.
        \end{itemize}
    \item Chair
        \begin{itemize}
            \item \texttt{All}: refine the whole feature
            \item \texttt{Seat}: the seat part of the chair
            \item \texttt{Legs}: the four legs of the chair.
            \item \texttt{Back}: the back part of the chair.
        \end{itemize}
    \item Bed: note that constraining upper and lower legs is equivalent to constraint the whole bed...
        \begin{itemize}
            \item \texttt{All}: refine the whole feature
            \item \texttt{Mattress}: the mattress part of the bed
            \item \texttt{Legs}: the four upper and lower legs of the bed.
            \item \texttt{Laths}: the laths upper part of the bed.
        \end{itemize}
\end{itemize}

The rectangular box defining the domain can also be refined using the following keywords:
\begin{itemize}
    \item \texttt{All}: refine all the surfaces of the domain
    \item \texttt{Bottom}: refine the Bottom surface of the domain
    \item \texttt{Top}: refine the Top surface of the domain
    \item \texttt{East}: refine the East surface of the domain
    \item \texttt{West}: refine the West surface of the domain
    \item \texttt{South}: refine the South surface of the domain
    \item \texttt{North}: refine the North surface of the domain
\end{itemize}

Humans can be also refined/coarsen. The keywords can be \texttt{All} or any specific part of the body. Note that for the arms, forearms, thighs, hands, legs and feet, the keywords need to be in singular form. In short, the keywords are: \texttt{All}, \texttt{Head}, \texttt{Mouth}, \texttt{Nose}, \texttt{Neck}, \texttt{Trunk}, \texttt{Pelvis}, \texttt{Arm}, \texttt{Forearm}, \texttt{Hand}, \texttt{Thigh}, \texttt{Leg}, \texttt{Foot}.  

\begin{Code}[language=Python, firstnumber=1, caption={How to define a constraint in \textit{IGG\_Mesh.dat}.}, label={Lst:InputMeshe_ConstraintComplex}]
# CONSTRAINT 1
Furniture 1
Mattress
m_size1

# CONSTRAINT 2
Human 1
Hand
m_size2

# CONSTRAINT 3
Human 2
All
m_size3
\end{Code}

\section{Refinement in the volume}
The user can refined/coarsen the mesh in some areas within the domain volume if wanted. The area to be refined/coarsen can be defined by a box or a sphere.
To refine in a box (keyword: \texttt{Box}) area: the user needs to define the centre of the box $(x_{C}$, $y_{C}$, $z_{C})$ and its size $(l_{x}$, $l_{y}$, $l_{z})$. To refine in a sphere (keyword: \texttt{Sphere}) area: the user needs to define the centre of the sphere $(x_{C}$, $y_{C}$, $z_{C})$ and the radius $\mathcal{R}_{C}$. 

\begin{Code}[language=Python, firstnumber=1, caption={How to define a constraint in \textit{IGG\_Mesh.dat}.}, label={Lst:InputMeshe_ConstraintArea}]
# CONSTRAINT 1
Box
xcentre ycentre zcentre
lx ly lz
m_size1

# CONSTRAINT 2
Sphere
xcentre ycentre zcentre
radius
m_size2
\end{Code}

        %----------------------------------------------------------%
%-- Physical IDs                                         --%
%----------------------------------------------------------%
\chapter{Physical IDs for Boundary Conditions}\label{Sec:PhysicalIDsBC}
Assuming that the \texttt{File Name} defined in \textit{IGG\_Numerics.dat} is \texttt{MyGeom}, \textit{MyGeom\_PhysicalIDs.dat} summarises the Physical IDs assigned to surfaces. These Physical IDs are needed by \textbf{Fluidity} to apply boundary conditions. In the first part of the file, the Physical IDs by feature types (furniture, accessories, humans...) are summaries. In a second part of the file, an attempt is done to group the Physical IDs by boundary conditions types, i.e. walls, heating, inlet, outlet, heat generation... The Physical IDs are grouped by common boundary conditions used in CFD - but the user can decide to use different boundary conditions than the ones suggested if wanted.

\section{Physical IDs summary}\label{Sec:PhysicalIDsBC_ID}
\subsection{Features}\label{Sec:PhysicalIDsBC_IDFeature}
The physical surface IDs are:
\begin{itemize}
    \item 1: Domain surface South
    \item 2: Domain surface East
    \item 3: Domain surface North
    \item 4: Domain surface West
    \item 5: Domain surface Bottom
    \item 6: Domain surface Top
    \item The physical IDs of 2D features have the form \textbf{xyy}, where:
    \begin{itemize}
        \item \textbf{x} is the integer:
        \begin{itemize}
            \item \textbf{1} for an inlet
            \item \textbf{2} for an outlet
            \item \textbf{3} for a door
            \item \textbf{4} for a window
        \end{itemize}
        \item \textbf{yy} is an integer referring to the feature number from \textbf{01} to \textbf{99} as defined in \textit{IGG\_Domain.dat}.
    \end{itemize}
    \item 40: Generic Box
    \item 30: Normal Shelf
    \item 311: Refrigerated shelf cold surfaces
    \item 312: Refrigerated shelf other surfaces
    \item 32: Manual Till
    \item 33: Automatic Till
    \item 341: Table top surface
    \item 342: Table other surfaces
    \item 351: Chair in-contact w/ people surfaces
    \item 352: Chair other surfaces
    \item 361: Seat in-contact w/ people surfaces
    \item 362: Seat other surfaces
    \item 371: Stool top surface
    \item 372: Stool other surfaces
    \item 381: Bed mattress top surface
    \item 382: Bed other surfaces
    \item 39x: Bevel x=bevel ID 
    \item 20: Separator
    \item 601: Laptop hot surfaces
    \item 602: Laptop other surfaces
    \item 611: Screen hot surfaces
    \item 622: Screen other surfaces
    \item 621: Computer Tower
    \item 631: Diffuser releasing air towards South
    \item 632: Diffuser releasing air towards East
    \item 633: Diffuser releasing air towards North
    \item 634: Diffuser releasing air towards West
    \item 641: Extractor extracting air towards South
    \item 642: Extractor extracting air towards East
    \item 643: Extractor extracting air towards North
    \item 644: Extractor extracting air towards West
\end{itemize}

\subsection{Human}\label{Sec:PhysicalIDsBC_IDHuman}
Body part (except mouths and noses) physical IDs have the form "9xyz" where::
\begin{itemize}
    \item "x":: human ID 
    \item "y":: body part ID: 0: head; 1: neck; 2: arms; 3: forearms; 4: hands; 5: trunk; 6: pelvis; 7: thighs; 8: legs; 9: feet
    \item "z":: 1 for "heat" BCs; 2 for "no heat" BCs 
\end{itemize}

Mouths and noses physical IDs have the form "8xyz" where::
\begin{itemize}
    \item "x":: human ID 
    \item "y":: body part ID: 0: mouth; 1: nose
    \item "z":: 1 for "exit" BCs; 2 for "other" BCs 
\end{itemize}

Ex: "9201" is human 2 head generating heat; "8111" is human 1 exit of nose. 
    "92142" refers to the non-heating hands of human 21. 
NB: "8x02" does not exist - mouth is a surface and is an exit only.

\section{Boundary Conditions}\label{Sec:PhysicalIDsBC_BCs}
Boundary conditions suggested are as follows:
\begin{itemize}
    \item For Velocity/Pressure:
    \begin{itemize}
        \item No Slip/Slip (walls)
        \item Inlets
        \item Outlets
        \item Doors
        \item Windows
        \item Diffuser
        \item Extractor
    \end{itemize}
    \item For Temperature/Tracer
    \begin{itemize}
        \item Heat flux (for human, screen...)
        \item Adiabatic
        \item Fixed temperature (for refrigerated shelves for ex.)
    \end{itemize}
\end{itemize}

        %------------------------------------%
        %-- Acknowledgements               --%
        %------------------------------------% 
        \section*{Acknowledgements}
        This work is supported by the EPSRC Grand Challenge grant ‘Managing Air for Green Inner Cities (MAGIC) EP/N010221/1; EPSRC RELIANT, Risk EvaLuatIon fAst iNtel-ligent Tool for COVID19 (EP/V036777/1); EPSRC INHALE, Health assessment across biological length scales (EP/T003189/1); CO-TRACE COvid-19 Transmission Risk Assessment Case studies - Education Establishments EP/W001411/1; PROTECT COVID-19 National Core Study on transmission and environment and . This work has been undertaken,in part, as a contribution to ‘Rapid Assistance in Modelling the Pandemic’ (RAMP), initiated by the Royal Society.
        TRACK... 
        
        %------------------------------------%
        %-- References                     --%
        %------------------------------------% 
        \bibliographystyle{unsrt}
        \bibliography{biblio.bib}
    \end{sloppypar}
\end{document}